# Insights into the behavior of certain optical systems gleaned from Feynman's approach to quantum electrodynamics

Masud Mansuripur
James C. Wyant College of Optical Sciences, The University of Arizona, Tucson

**Abstract**. Richard Feynman's method of path integrals is based on the fundamental assumption that a system starting at a point $A$ and arriving at a point $B$ takes all possible paths from $A$ to $B$, with each path contributing its own (complex) probability amplitude. The sum of the amplitudes over all these paths then yields the overall probability amplitude that the system starting at $A$ would end up at $B$. We apply Feynman's method to several optical systems of practical interest and discuss the nuances of the method as well as instances where the predicted outcomes agree or disagree with those of classical optical theory. Examples include the properties of beam-splitters, passage of single photons through Mach-Zehnder and Sagnac interferometers, electric and magnetic dipole scattering, reciprocity, time-reversal symmetry, the optical theorem, the Ewald-Oseen extinction theorem, far field diffraction, and the two-photon interference phenomenon known as the Hong-Ou-Mandel effect.

**1. Introduction**. Many problems in quantum optics have their classical counterparts.[1,2] For example, Thomas Young's famous double-slit experiment, when carried out with single photons, continues to exhibit the iconic interference fringes that were so crucial in understanding the wave nature of light against the Newtonian corpuscular ideas. Similarly, the passage of a single photon through a Mach-Zehnder interferometer is governed by principles that have an uncanny resemblance to those that guide the behavior of classical coherent light. And Glauber's coherent state mimics the characteristic features of classical coherent light in many consequential ways.[3,4] In fact, there are numerous instances in classical optics where the observed behavior of light can be described in the language of quantum optics (or quantum electrodynamics) in ways that deepen our understanding of light and help bridge the gap created by its mysterious wave-particle duality.

It is a goal of the present paper to showcase a few situations where well-known classical optical phenomena submit to elementary descriptions in terms of single photons, following a methodology pioneered by Richard Feynman.[5] In a nutshell, Feynman asserts that the photon (a Bose particle) should be assumed to take all the allowed paths through a system, each path having its own probability amplitude — a complex number. When the various paths taken by the photon are physically indistinguishable, one must add all the probability amplitudes that lead from a specific initial condition to a specific final condition along different paths, in order to find the overall probability amplitude of the corresponding event. The probability of occurrence of the event is then the squared absolute value of the probability amplitude thus computed.

There exist situations, of course, where the predictions of quantum optics, while agreeing with experimental findings, contradict those of the classical theory. Here, once again, Feynman's method proves its usefulness when one tries to understand (or at least clearly explain) the nature of the observed phenomena. A rather trivial example is provided by a single-photon wavepacket arriving at a $50/50$ beam-splitter. According to the classical theory, the splitter should divide the incident optical energy between its two exit ports, with detectors placed in these ports each receiving one-half of the incident energy. In contrast, the quantum theory of light assigns equal probabilities to each of the two distinct photodetection events, guaranteeing that, in every instance the experiment is carried out, the incident photon (in its entirety) will be picked up by only one of the two detectors.[5]

Even more interesting perhaps is the case of two identical single-photon wavepackets arriving simultaneously at the two entrance ports of a $50/50$ beam-splitter. While the classical theory maintains that detectors placed in the two exit ports will each pick up a single photon, the correct prediction is provided by the quantum theory, which asserts that both photons will arrive together at one detector or the other (with equal probability), but are never divided between the two detectors.[6] These and other examples that reveal the profound differences between the predictions of the classical and quantum theories of light are discussed in the second half of the paper.

The organization of the paper is as follows. Section 2 contains a detailed analysis of the lossless beam-splitter, which plays an important role in numerous optical systems. Here, we derive the fundamental relations between the Fresnel reflection and transmission coefficients of beam-splitters that will be needed in some of the subsequent sections. In Sec.3, we examine the passage of a single photon through a Mach-Zehnder interferometer and derive the conditions under which the photon could emerge from one or the other of its exit channels. A similar analysis of a single photon going through a Sagnac interferometer is the subject of Sec.4. Section 5 is devoted to the problem of electromagnetic (EM) scattering from a small spherical particle, where we relate the susceptibility of the host material to the polarizability of the particle in the presence of an external EM field. The scattering amplitude of an electric dipole in various directions (relative to that of the incident photon) is subsequently discussed in Sec.6, with the polarization state of the scattered photon properly taken into account. Here, we also demonstrate the underlying principle of reciprocity in EM systems using the inherent symmetries of single-photon scattering. The results of Sec.6 are then generalized in Sec.7 to cover EM scattering from magnetic dipoles. Section 8 is a brief description of diffraction from a blazed grating as an elementary example of the principle of reciprocity. Another example is provided in Sec.9 in the context of optical transmission through slabs and multilayer stacks involving multiple internal reflections. Section 10 describes a thought experiment that involves the scattering of a single photon from a pair of identical particles, then raises a question as to whether quantum interference is still viable when a certain amount of angular momentum is transferred from the incident photon to its scatterer.

In Sec.11, we examine the reflection and transmission coefficients of an EM plane-wave arriving at a thin dielectric sheet, then use the results to provide an elementary demonstration of the principle of time-reversal symmetry in classical optics. The results of Sec.11 are also used in Sec.12 to elucidate the fundamental argument behind the Ewald-Oseen extinction theorem. Section 13 is devoted to yet another important theorem of classical electrodynamics; here, we use the notion of single-photon scattering to straightforwardly prove the so-called optical theorem, revealing the intimate relation between the scattering cross-section of a material body and its forward scattering amplitude.

The classical problems of scalar and vector diffraction are visited in Sec.14 from the viewpoint of single-photon scattering. We derive the standard formulas of far field diffraction from an object, then invoke the properties of beam-splitters to argue that the results should apply not only to single photons but also to any coherent state of the incident beam. Section 15 is a concise description of the quintessential quantum-optical phenomenon known as the Hong-Ou-Mandel effect. This section provides the motivation for, and a segue to, our subsequent discussions of multi-photon states.

We return to the problem of beam-splitter in Sec.16, this time examining two wavepackets, one in the number state $|n_1\rangle$, the other in $|n_2\rangle$, that simultaneously arrive at the entrance ports 1 and 2 of a lossless beam-splitter. Here, we explain how the splitter distributes these $n_1 + n_2$ photons between its two output ports, thus arriving at the probability of detecting $m$ photons in port 3 and the remaining $n_1 + n_2 - m$ photons in port 4.

Up to this point, we have relied primarily on the principle of superposition and the notions of distinguishability and indistinguishability of events to reach conclusions that either reaffirm the well-known results of classical electrodynamics or provide plausibility arguments for decidedly non-classical phenomena. To proceed further, we must employ some of the more formal tools and techniques of quantum electrodynamics. This we do in the remainder of the paper, starting with Sec.17, where the orthonormal eigenmodes of the EM field in free space are formally defined. The annihilation and creation operators, $\hat{a}$ and $\hat{a}^\dagger$, are then introduced in Sec.18 and used to describe Glauber's coherent state $|\gamma\rangle$ and its properties. Next, we introduce in Sec.19 the EM field operators pertaining to the electric field $\boldsymbol{E}(\boldsymbol{r},t)$, magnetic field $\boldsymbol{B}(\boldsymbol{r},t)$, vector potential $\boldsymbol{A}(\boldsymbol{r},t)$, Poynting vector $\boldsymbol{S}(\boldsymbol{r},t)$, and the overall energy of EM modes. This is followed by an examination of certain physical characteristics of the number states, the coherent state, and the (mixed) thermal state.

The operator algebra is deployed in Sec.20 to re-examine the characteristic behavior of beam-splitters previously studied in Sec.16, and to confirm the results obtained in that earlier section. We rely on the operator algebra once again in Sec.21 to demonstrate that a pair of coherent beams, $|\gamma_1\rangle$ and $|\gamma_2\rangle$, arriving at the input ports of a beam-splitter give rise to another pair of coherent beams, $|\gamma_3\rangle$ and $|\gamma_4\rangle$, that emerge from the corresponding output ports. The passage of thermal light through a beam-splitter is the subject of Sec.22. Section 23 is a brief foray into the Sudarshan-Glauber P-representation. The paper closes with a summary and a few concluding remarks in Sec.24.

**2. Characteristics of beam-splitters**. We consider a beam-splitter constructed from thin-film layers of homogeneous, isotropic, and transparent materials of differing thicknesses and refractive indices, as depicted in Fig.1. A monochromatic plane-wave of frequency $\omega$, $k$-vector $\boldsymbol{k} = (\omega/c)\hat{\boldsymbol{k}}$, and linear polarization (either $p$- or $s$-polarized) arrives at oblique incidence at the front-facet of the splitter. The Fresnel reflection and transmission coefficients at this front-facet are $(\rho_p, \tau_p)$ and $(\rho_s, \tau_s)$ for $p$- and $s$-polarized light, respectively.[7,8] In what follows, we examine the properties of $\rho = |\rho|e^{\mathrm{i}\varphi_\rho}$ and $\tau = |\tau|e^{\mathrm{i}\varphi_\tau}$ without specifying the $p$ or $s$ subscript; we expect the reader to understand that our arguments apply to both $p$- and $s$-polarized light separately and independently of each other. Let us mention in passing that $\rho$ and $\tau$ represent the complex probability amplitudes for reflection and transmission of a single photon occupying a wave-packet of frequency $\omega$, $k$-vector $\boldsymbol{k} = (\omega/c)\hat{\boldsymbol{k}}$, and linear polarization (either $p$ or $s$) that arrives at the splitter along the direction of the unit-vector $\hat{\boldsymbol{k}}$ and in the number state $|1\rangle$.

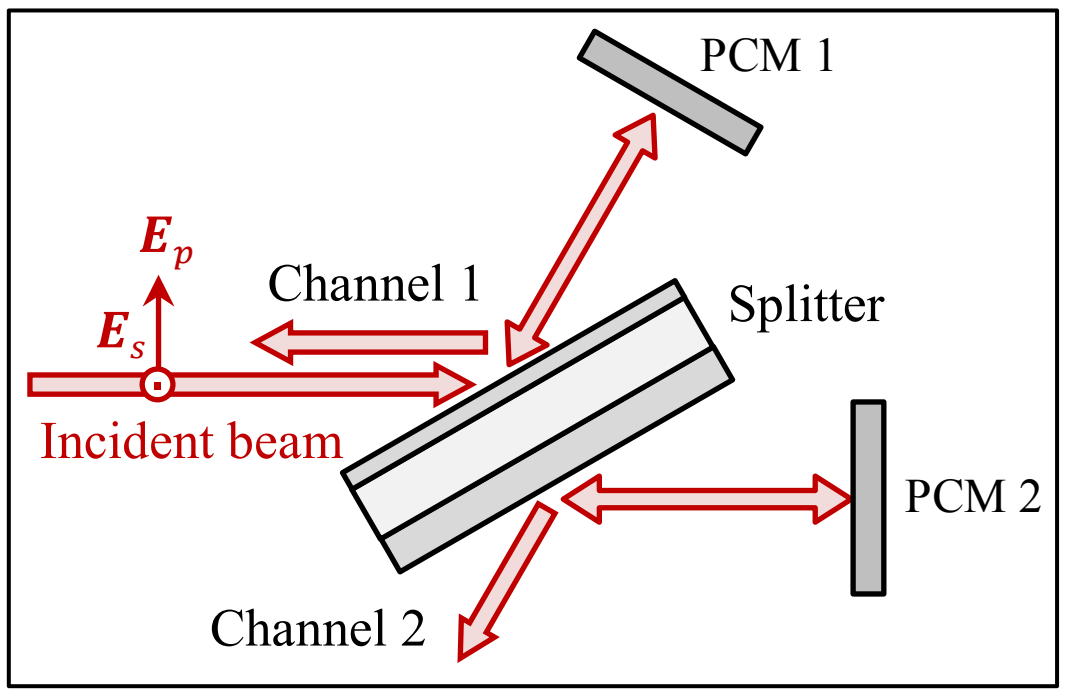

**Fig.1**. The Fresnel reflection and transmission coefficients at the front facet of the beam-splitter are $(\rho_p, \tau_p)$ and $(\rho_s, \tau_s)$ for the $p$- and $s$-polarized incident light, respectively. The lateral symmetry of the multilayer dielectric stack guarantees that, for both polarization states, $(\rho, \tau)$ are the same for incidence from the left- and the right-hand sides on the front-facet. In contrast, unless the splitter has structural symmetry in the front-to-back direction, the reflection and transmission coefficients $(\rho_p', \tau_p')$ and $(\rho_s', \tau_s')$ at the back-side of the splitter might differ from the corresponding ones at the front-side.

The lateral symmetry of the multilayer stack comprising the beam-splitter ensures that the Fresnel coefficients $(\rho, \tau)$ are the same for light arriving at the front facet from either the left- or the right-hand side, so long as the incidence angle and the polarization state of the beam remain the same. However, unless the splitter has structural symmetry in the front-to-back direction, one cannot say that its Fresnel coefficients $(\rho', \tau')$ for incidence from the back-side are the same as those from the front-side — again for a fixed incidence angle and a fixed polarization state. Our goal in the present section is to demonstrate that, for a lossless beam-splitter, $|\rho| = |\rho'|$, $\tau = \tau'$, and $\varphi_\tau = ½(\varphi_\rho + \varphi'_\rho) \pm \pi/2$. Our analysis does *not* provide any information on the relation between $\varphi_\rho$ and $\varphi'_\rho$, nor does it specify the sign of $\pi/2$ in the preceding expression that relates $\varphi_\tau$ to $\varphi_\rho + \varphi'_\rho$.

We begin by recognizing that, in the absence of optical absorption, conservation of photon number (or conservation of energy in the language of classical optics) imposes the following constraints on the various reflection and transmission coefficients:

$$|\rho|^2 + |\tau|^2 = 1, \tag{2.1}$$

$$|\rho'|^2 + |\tau'|^2 = 1. \tag{2.2}$$

Other constraints are imposed by the requirement of time-reversal symmetry (see Sec.11 for a brief discussion of time-reversal symmetry). The phase-conjugate mirrors PCM1 and PCM2 shown in Fig.1 return the reflected and transmitted waves to the front and rear facets of the splitter, albeit with the complex amplitudes of the returning waves conjugated. The returning waves must reconstruct the incident beam (now conjugated and propagating backward) in channel 1; they must also cancel each other out in the direction identified in Fig.1 as channel 2. We thus have

reconstruction in channel 1 →
$$\rho\rho^* + \tau'\tau^* = 1, \tag{2.3}$$

cancellation in channel 2 →
$$\tau\rho^* + \rho'\tau^* = 0. \tag{2.4}$$

Equation (2.3) in conjunction with Eqs.(2.1) and (2.2) now yields

$$\tau' = \tau, \tag{2.5}$$

$$|\rho'| = \sqrt{1 - |\tau'|^2} = \sqrt{1 - |\tau|^2} = |\rho|. \tag{2.6}$$

Finally, upon combining Eq.(2.4) and (2.6), we arrive at

$$\rho'/\rho^* = -\tau/\tau^* \quad \rightarrow \quad \varphi'_\rho + \varphi_\rho = 2\varphi_\tau \pm \pi \quad \rightarrow \quad \varphi_\tau = ½(\varphi_\rho + \varphi'_\rho) \pm \pi/2. \tag{2.7}$$

This concludes our analysis of the reflection and transmission coefficients of lossless splitters with lateral symmetry such as those constructed in the form of multilayer stacks with isotropic, homogeneous, and transparent dielectric layers. In those special cases when the stack has structural symmetry in the front-to-back direction, we will have $\varphi_\rho = \varphi'_\rho$ and, consequently, $\varphi_\tau = \varphi_\rho \pm \pi/2$.

An alternative analysis of lossless splitters starts by assuming that two plane-waves with $E$-field amplitudes $E_1$ and $E_2$ (both $p$-polarized or both $s$-polarized) enter through channels 1 and 2 of the splitter depicted in Fig.1. With the phase-conjugate mirrors now removed, energy conservation requires that the sum of the emergent beam intensities be equal to that of the entering beams; that is,

$$|\rho E_1 + \tau' E_2|^2 + |\tau E_1 + \rho' E_2|^2 = |E_1|^2 + |E_2|^2. \tag{2.8}$$

Expanding and rearranging the terms on the left-hand side of the above equation, we arrive at

$$(|\rho|^2 + |\tau|^2)|E_1|^2 + (|\rho'|^2 + |\tau'|^2)|E_2|^2 + 2Re[(\rho\tau'^* + \tau\rho'^*)E_1E_2^*] = |E_1|^2 + |E_2|^2. \tag{2.9}$$

We now invoke Eqs.(2.1) and (2.2) to eliminate the terms containing $|E_1|^2$ and $|E_2|^2$. Since the remaining term contains $E_1E_2^*$, which can have an arbitrary amplitude and phase, we conclude that the satisfaction of Eq.(2.9) requires that $\rho\tau'^* + \tau\rho'^* = 0$. Consequently,

$$\rho\tau'^* = -\tau\rho'^* \rightarrow \begin{cases} |\rho||\tau'| = |\tau||\rho'| \rightarrow |\rho|^2(1-|\rho'|^2) = (1-|\rho|^2)|\rho'|^2 \rightarrow |\rho| = |\rho'|, \\ \varphi_\rho - \varphi'_\tau = \varphi_\tau - \varphi'_\rho \pm \pi \quad \rightarrow \quad \varphi_\rho + \varphi'_\rho = \varphi_\tau + \varphi'_\tau \pm \pi. \end{cases} \tag{2.10}$$

As before, it is seen that $|\rho| = |\rho'|$, which also implies that $|\tau| = |\tau'|$. However, unlike the previous analysis that was informed by time-reversal symmetry, the present method cannot establish the equality of $\varphi_\tau$ and $\varphi'_\tau$. The best one can do in this case is to conclude that $\varphi_\rho + \varphi'_\rho$ differs from $\varphi_\tau + \varphi'_\tau$ by $\pi$. Only when the splitter has front-to-back structural symmetry will we have $\rho = \rho'$, $\tau = \tau'$, and $\varphi_\tau = \varphi_\rho \pm \pi/2$.

**3. The Mach-Zehnder interferometer.**[2,7,8] A wavepacket in the single-photon state $|1\rangle$ enters the Mach-Zehnder interferometer depicted in Fig.2, where, for the sake of simplicity, both $50/50$ beam-splitters are assumed to be symmetric, having reflection and transmission coefficients $\rho = 1/\sqrt{2}$ and $\tau = \mathrm{i}/\sqrt{2}$, irrespective of the direction of incidence. Depending on the path-length difference between the two arms of the device, either constructive or destructive interference can take place at the second splitter, in which case the photon consistently emerges from one or the other exit channel of the interferometer. Denoting the phase acquired upon propagation in the upper path by $\varphi_1$ and that in the lower path by $\varphi_2$, one can adjust the phase difference $\Delta\varphi = \varphi_2 - \varphi_1$ by properly positioning the retroreflector.

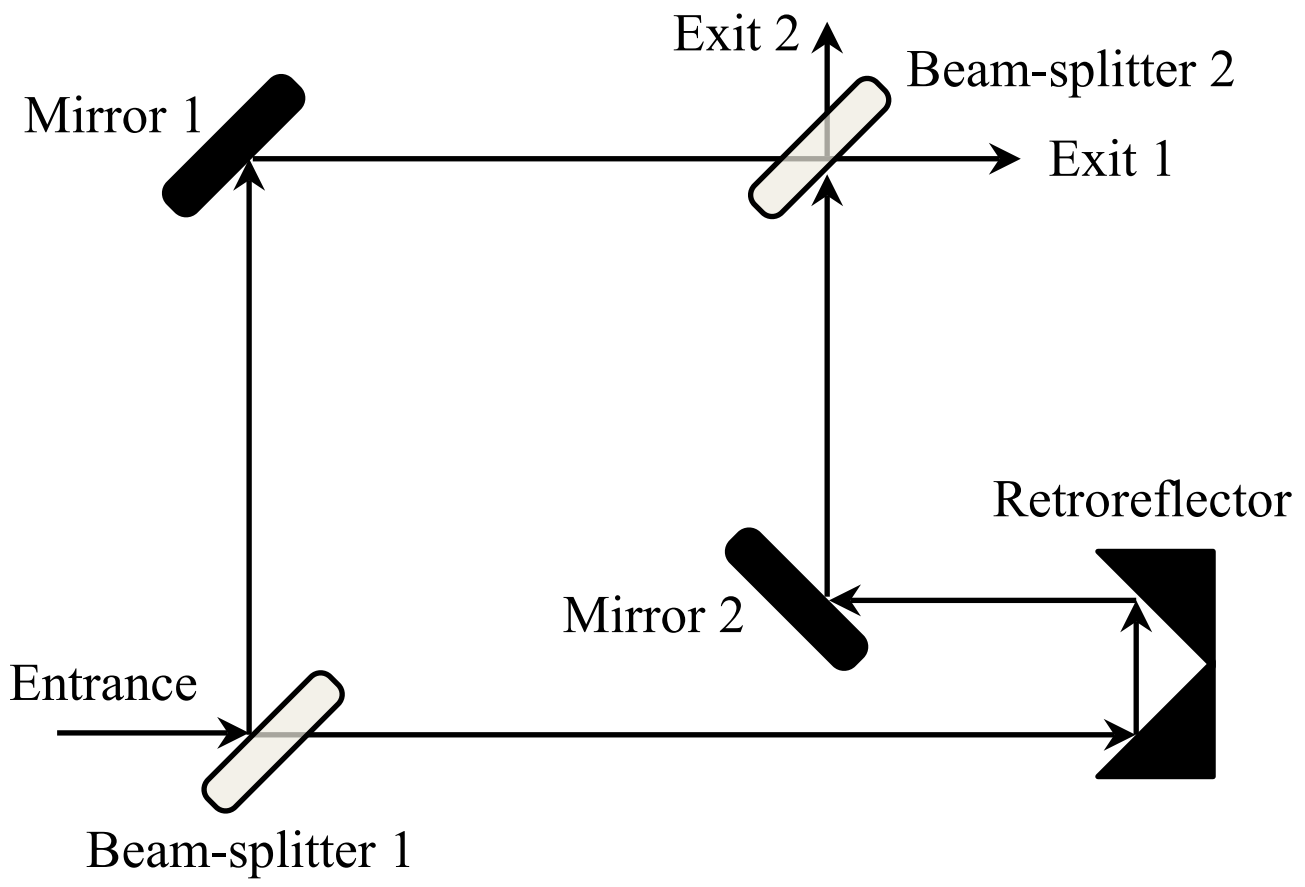

**Fig.2**. A wavepacket in the single-photon state $|1\rangle$ arrives at the entrance facet of a Mach-Zehnder interferometer whose identical beam-splitters are symmetric and have reflection and transmission coefficients $\rho = 1/\sqrt{2}$ and $\tau = \mathrm{i}/\sqrt{2}$, irrespective of the direction of incidence. Repositioning the retroreflector allows for changing the optical path-length of the lower arm of the device, hence adjusting the phase difference $\Delta\varphi = \varphi_2 - \varphi_1$ between the two paths that the photon can take in going from splitter 1 to splitter 2. The photon will consistently emerge at Exit 1 (2) if $\Delta\varphi$ is an even (odd) multiple of $\pi$.

Both alternative paths leading to Exit 1 involve one reflection and one transmission at the splitters, so the corresponding photon amplitudes will be $\rho\tau e^{\mathrm{i}\varphi_1} = \tfrac{1}{2}\mathrm{i}e^{\mathrm{i}\varphi_1}$ and $\rho\tau e^{\mathrm{i}\varphi_2} = \tfrac{1}{2}\mathrm{i}e^{\mathrm{i}\varphi_2}$. At Exit 2, however, the upper path requires reflections at both splitters whereas the lower path requires two transmissions; therefore, the corresponding amplitudes are $\rho^2 e^{\mathrm{i}\varphi_1} = \tfrac{1}{2}e^{\mathrm{i}\varphi_1}$ and $\tau^2 e^{\mathrm{i}\varphi_2} = -\tfrac{1}{2}e^{\mathrm{i}\varphi_2}$. If $\Delta\varphi = 2n\pi$ for some integer $n$, then the photon emerges at Exit 1 with probability 1. However, if $\Delta\varphi = (2n+1)\pi$, the amplitudes add up to zero at Exit 1 and the photon emerges at Exit 2, where the sum of its corresponding amplitudes equals 1. This is the fundamental principle of operation of the Mach-Zehnder interferometer.

In general, the two beam-splitters need not be identical, nor symmetric, although both must be $50/50$ splitters. For the first splitter having $\rho_1 = e^{\mathrm{i}\varphi_\rho}/\sqrt{2}$ and $\tau_1 = e^{\mathrm{i}\varphi_\tau}/\sqrt{2}$, the actual values of $\varphi_\rho$ and $\varphi_\tau$ are inconsequential. As for the second splitter having $\rho_2 = e^{\mathrm{i}\varphi_\rho}/\sqrt{2}$ and $\tau_2 = e^{\mathrm{i}\varphi_\tau}/\sqrt{2}$ for incidence from the left, and $\rho_2'' = e^{\mathrm{i}\varphi_\rho''}/\sqrt{2}$ and $\tau_2' = e^{\mathrm{i}\varphi_\tau'}/\sqrt{2}$ for incidence from below (see the analysis of Sec.2), it is only necessary to have $\varphi_\rho + \varphi_\rho'' = \varphi_\tau + \varphi_\tau' \pm \pi$, which is always true for such beam-splitters.

We note in passing that the incident photon arriving at the first splitter is ideally in a single mode specified in terms of its frequency $\omega$, $k$-vector $\boldsymbol{k}$, and polarization state $\hat{\boldsymbol{e}}$. To a good approximation, one can take this single-photon (in the number state $|1\rangle$) to be a collimated wave-packet of large (but finite) cross-sectional area and long (but finite) duration of total energy $\hbar\omega$. The duration $\Delta t$ of the wavepacket is typically on the order of the inverse linewidth of the photon's source; that is, $\Delta t \sim (\Delta\omega)^{-1}$. The path-length difference between the two arms of the interferometer must be much less than $c\Delta t$ (where $c$ is the speed of light in free space) to enable the photon to interfere with itself at the second splitter.

**4. The Sagnac interferometer**.[9,10] Figure 3 shows the diagram of a Sagnac interferometer consisting of a $50/50$ beam-splitter $S$, a first mirror $M_1$, and a second mirror $M_2$ that jointly form the triangular loop $SM_1M_2$ around which light can travel clockwise or counterclockwise. A single photon arriving from the source at the beam-splitter has a probability amplitude for going clockwise around the loop and arriving at the observation plane, and also an amplitude for traveling counter-clockwise and reaching the same observation plane. Since there is no way in the end to tell which path the photon has taken, these two amplitudes must be added together.†

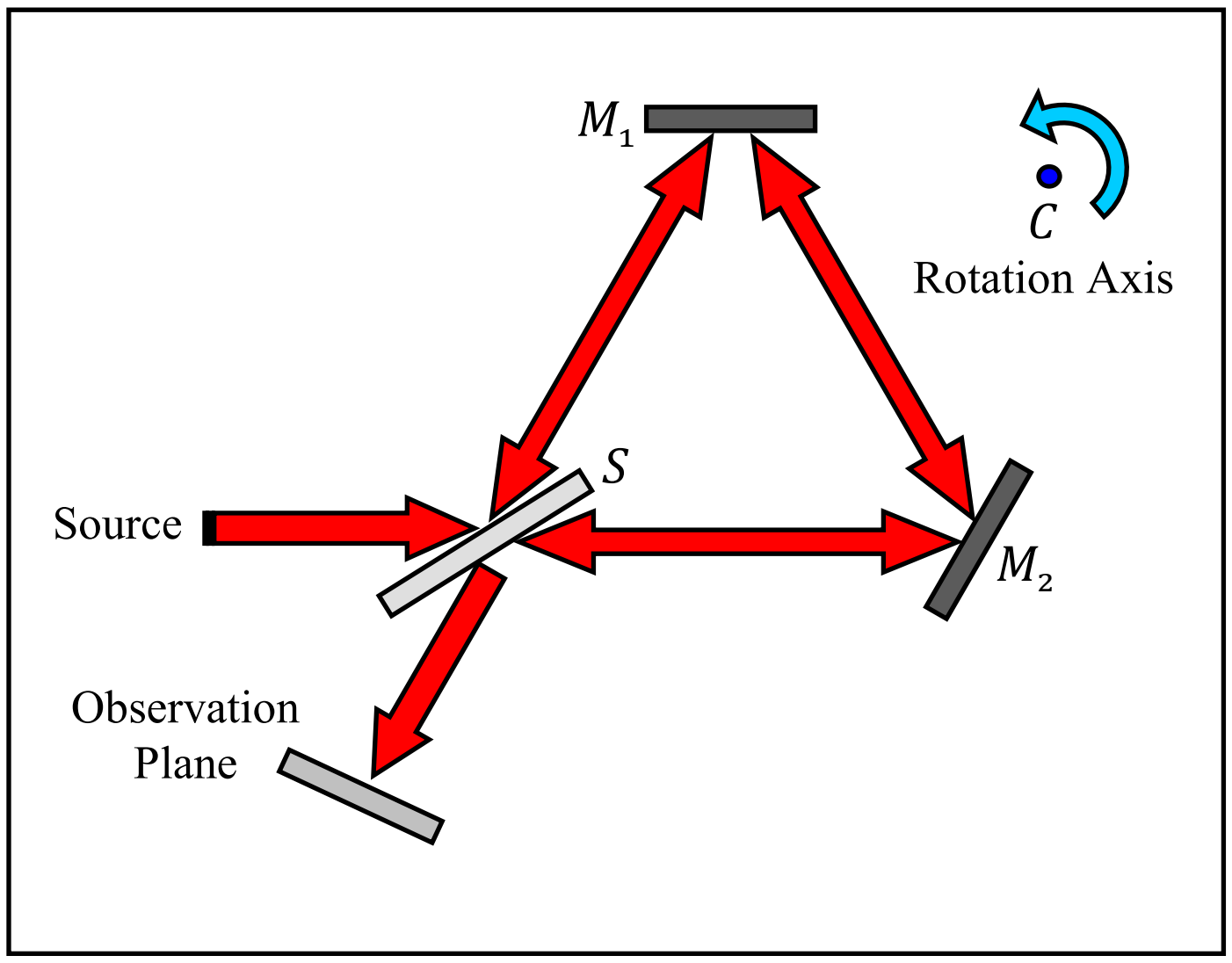

**Fig.3**. In a Sagnac interferometer, the light from the source is split at a $50/50$ beam-splitter $S$ to travel in opposite directions around a loop formed by the splitter and the mirrors $M_1$ and $M_2$. Upon returning to the splitter, one beam is reflected and the other one transmitted at $S$, so that a superposition of the two arrives at the observation plane. With the interferometer standing still, the counter-propagating beams around the $SM_1M_2$ loop remain in phase; however, one beam suffers two reflections at the splitter while the other one gets transmitted twice; the net relative phase between the two beams will then be 180° and, therefore, no light reaches the observation plane. In contrast, when the interferometer rotates around an axis (say, one that crosses the $SM_1M_2$ plane at $C$), the counter-propagating beams acquire a relative phase that allows a fraction of the circulating light to reach the observation plane.

†From the perspective of an observer in an inertial reference frame, each time the photon is reflected at $M_1$ or $M_2$ or $S$, its frequency is Doppler shifted due to the motion of these reflectors.[8] It can be shown, however, that the overall Doppler shifts for counter-propagating photons are precisely the same, so that one cannot rely on the photon frequency at the observation plane to determine the path that it has taken through the interferometer.

When the interferometer is at rest (i.e., not rotating around some axis), the accumulated phase angles for the two counter-propagating paths around the loop will be identical. However, the clockwise path involves two reflections at the splitter, corresponding to a probability amplitude $\rho\rho'' = ½e^{\mathrm{i}(\varphi_\rho+\varphi''_\rho)}$, while the counterclockwise path involves two transmissions, corresponding to an amplitude $\tau\tau' = ½e^{\mathrm{i}(\varphi_\tau+\varphi'_\tau)}$. These two amplitudes differ by a 180° phase (see Sec.2), indicating that the photon cannot reach the observation plane when the interferometer is standing still. In contrast, when the interferometer rotates around an axis that crosses its plane at some arbitrary location, say, $C$, there will be an optical path-length difference associated with the two propagation directions, in which case the photon will have a nonzero probability of being detected at the observation plane. Below, we show that the total phase difference acquired by the counter-propagating beams equals $(4k_0/c)\boldsymbol{A}\cdot\boldsymbol{\Omega}$, where $k_0 = 2\pi/\lambda_0$ is the photon's wave-number, $c$ is the speed of light in vacuum, $\boldsymbol{A}$ is the perpendicular vector to the plane of the interferometer whose magnitude $A$ equals the area of the $SM_1M_2$ triangle, and $\boldsymbol{\Omega}$ is a vector along the rotation axis whose magnitude $\Omega$ is the angular velocity of the stage on which the interferometer is mounted.

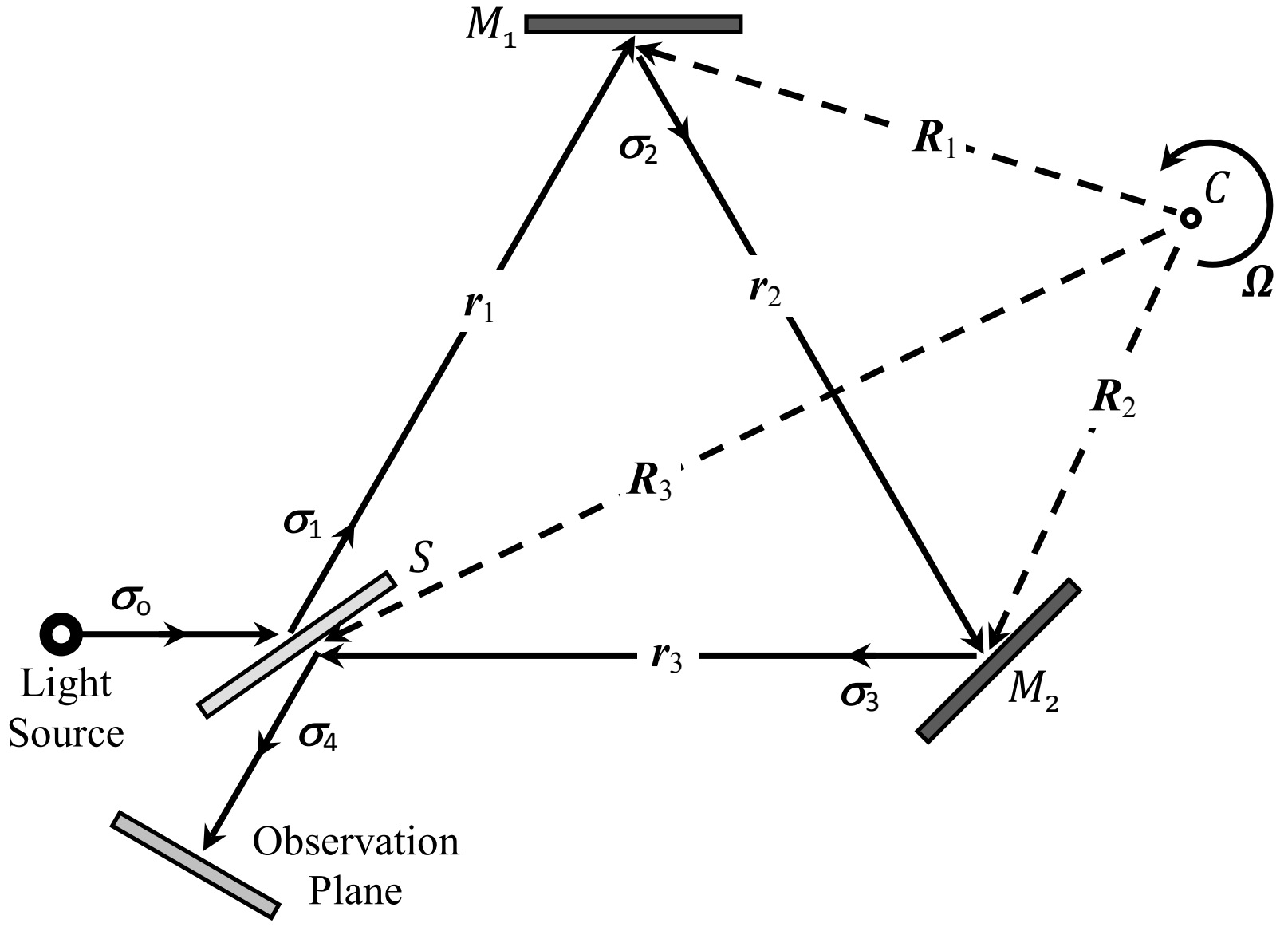

**Fig.4**. A single photon of frequency $\omega_0$, vacuum wavelength $\lambda_0 = 2\pi c/\omega_0$, and wave-number $k_0 = 2\pi/\lambda_0$ emerges from the light source in the direction of the unit-vector $\boldsymbol{\sigma}_0$. In its clockwise path around the Sagnac interferometer, the photon travels from $S$ to $M_1$ along the unit-vector $\boldsymbol{\sigma}_1 = \boldsymbol{r}_1/r_1$, then from $M_1$ to $M_2$ along the unit-vector $\boldsymbol{\sigma}_2 = \boldsymbol{r}_2/r_2$, and finally from $M_2$ back to $S$ along the unit-vector $\boldsymbol{\sigma}_3 = \boldsymbol{r}_3/r_3$. (The directions of $\boldsymbol{\sigma}_1, \boldsymbol{\sigma}_2, \boldsymbol{\sigma}_3$ are reversed when the photon takes the counterclockwise path.) The rotation axis $\boldsymbol{\Omega}$ crosses the plane $SM_1M_2$ of the interferometer at some arbitrary point $C$, with the vectors $\boldsymbol{R}_1, \boldsymbol{R}_2, \boldsymbol{R}_3$ connecting $C$ to the center points of $M_1$, $M_2$, and $S$, respectively. The area $A$ of the $r_1r_2r_3$ triangle is the magnitude of the vector $\boldsymbol{A}$ that is perpendicular to the plane of the interferometer. Note that the rotation axis $\boldsymbol{\Omega}$ is not necessarily aligned with the area vector $\boldsymbol{A}$.

With reference to Fig.4, we note that, upon its first reflection at the splitter $S$, the photon travels to the mirror $M_1$, taking $\Delta t = r_1/c$ seconds to reach the center of the mirror. During this time, the center of $M_1$ moves by $\Delta\boldsymbol{r}_1 = (\boldsymbol{\Omega}\times\boldsymbol{R}_1)\Delta t$, so that the photon's incremental phase-shift (due to the rotation of the stage) will be[8]

$$\Delta\varphi_1 = \boldsymbol{k}_1\cdot\Delta\boldsymbol{r}_1 = k_0\boldsymbol{\sigma}_1\cdot(\boldsymbol{\Omega}\times\boldsymbol{R}_1)\,r_1/c = (k_0/c)(\boldsymbol{R}_1\times\boldsymbol{r}_1)\cdot\boldsymbol{\Omega}. \tag{4.1}$$

The vector $\boldsymbol{R}_1 \times \boldsymbol{r}_1$ is perpendicular to the plane of the $r_1R_1R_3$ triangle, with a magnitude that is twice the area of this triangle. A similar argument applied to the second leg of the interferometer (from $M_1$ to $M_2$) yields $\Delta\varphi_2 = (k_0/c)(\boldsymbol{R}_2 \times \boldsymbol{r}_2) \cdot \boldsymbol{\Omega}$, with the vector $\boldsymbol{R}_2 \times \boldsymbol{r}_2$ being perpendicular to the $R_2R_1r_2$ triangle and having a magnitude twice the area of the triangle. And, finally, the incremental phase-shift along the leg between $M_2$ and $S$ is $\Delta\varphi_3 = (k_0/c)(\boldsymbol{R}_3 \times \boldsymbol{r}_3) \cdot \boldsymbol{\Omega}$, with the vector $\boldsymbol{R}_3 \times \boldsymbol{r}_3$ being perpendicular to the $R_3R_2r_3$ triangle and having a magnitude twice the area of the triangle. Although these three triangles are co-planar, their normal vectors are oriented such that those of $r_1R_1R_3$ and $R_3R_2r_3$ are parallel to each other and antiparallel to that of $R_2R_1r_2$. Therefore, the total incremental phase accumulated during a single round trip around the $r_1r_2r_3$ loop is given by

$$\Delta\varphi = \Delta\varphi_1 + \Delta\varphi_2 + \Delta\varphi_3 = (k_0/c)(\boldsymbol{R}_1 \times \boldsymbol{r}_1 + \boldsymbol{R}_2 \times \boldsymbol{r}_2 + \boldsymbol{R}_3 \times \boldsymbol{r}_3) \cdot \boldsymbol{\Omega} = (2k_0/c)\boldsymbol{A} \cdot \boldsymbol{\Omega}. \quad (4.2)$$

If the photon is initially transmitted through the splitter $S$, it would go around the $r_1r_2r_3$ triangle in the reverse direction, accumulating an incremental phase shift of $-\Delta\varphi$. All in all, the cumulative phase difference between the two directions of travel around the interferometer is going to be $2\Delta\varphi = (4k_0/c)\boldsymbol{A} \cdot \boldsymbol{\Omega}$. (Note that the angular velocity vector $\boldsymbol{\Omega}$ need not be perpendicular to the $r_1r_2r_3$ loop.) The $2\Delta\varphi$ phase-shift brought about by the rotation of the plane of the interferometer thus alters the overall probability of the incident photon reaching the observation plane, assuring the certainty of its detection within that plane whenever $2\Delta\varphi$ happens to be an odd multiple of $\pi$.

**5. Scattering of light from small particles**. In preparation for an analysis of reciprocity in dipole scattering in the next section, we provide here an overview of the classical theory of EM scattering from a small polarizable particle. Let a monochromatic plane-wave of frequency $\omega$ propagate along the $y$-axis. A small spherical particle placed at the origin of the coordinate system, having volume $v$ and polarizability $\alpha(\omega) = |\alpha|e^{\mathrm{i}\phi_\alpha}$, scatters the incident light. On a spherical surface of (large) radius $r$ centered at the origin, the $\boldsymbol{E}$ and $\boldsymbol{H}$ fields of the incident plane-wave may be expressed as[11]

$$\boldsymbol{E}_{\mathrm{inc}}(\boldsymbol{r},t) = E_0\hat{\boldsymbol{z}}\exp[\mathrm{i}(\omega/c)(y-ct)] = E_0(\cos\theta\,\hat{\boldsymbol{r}} - \sin\theta\,\hat{\boldsymbol{\theta}})\exp[\mathrm{i}(\omega/c)(r\sin\theta\sin\varphi - ct)]. \quad (5.1)$$

$$\boldsymbol{H}_{\mathrm{inc}}(\boldsymbol{r},t) = (E_0/Z_0)\hat{\boldsymbol{x}}\exp[\mathrm{i}(\omega/c)(y-ct)] = (E_0/Z_0)(\sin\theta\cos\varphi\,\hat{\boldsymbol{r}} + \cos\theta\cos\varphi\,\hat{\boldsymbol{\theta}} - \sin\varphi\,\hat{\boldsymbol{\varphi}}) \times \exp[\mathrm{i}(\omega/c)(r\sin\theta\sin\varphi - ct)]. \quad (5.2)$$

Here, $c = (\mu_0\varepsilon_0)^{-½}$ is the speed of light in vacuum, $Z_0 = (\mu_0/\varepsilon_0)^{½}$ is the impedance of free space, and $(\theta, \varphi)$ are the polar and azimuthal angles in our spherical coordinate system. On the spherical surface of radius $r$, the EM fields radiated by the dipole (i.e., the far fields) are given by[12]

$$\boldsymbol{E}_{\mathrm{rad}}(\boldsymbol{r},t) = -\alpha E_0(\omega/c)^2\sin\theta\,\hat{\boldsymbol{\theta}}\exp[\mathrm{i}(\omega/c)(r-ct)]/(4\pi\varepsilon_0 r). \quad (5.3)$$

$$\boldsymbol{H}_{\mathrm{rad}}(\boldsymbol{r},t) = -\alpha E_0(\omega/c)^2\sin\theta\,\hat{\boldsymbol{\varphi}}\exp[\mathrm{i}(\omega/c)(r-ct)]/(4\pi\varepsilon_0 Z_0 r). \quad (5.4)$$

The total optical power radiated by the dipole is computed via the Poynting vector $\boldsymbol{S}_{\mathrm{rad}}$, as follows:

$$P_{\mathrm{out}} = |\alpha|^2E_0^2(\omega/c)^4\int_0^\pi \sin^3\theta\,\mathrm{d}\theta/(16\pi\varepsilon_0^2Z_0) = |\alpha|^2E_0^2(\omega/c)^4/(12\pi\varepsilon_0^2Z_0). \quad (5.5)$$

As for the EM energy density entering the spherical surface (due to the existence of cross-terms between the incident plane-wave and the dipole's radiated spherical wave), we have[11,12]

$$\langle\boldsymbol{S}_{\mathrm{in}}(\boldsymbol{r},t)\rangle = ½Re(\boldsymbol{E}_{\mathrm{inc}} \times \boldsymbol{H}_{\mathrm{rad}}^* + \boldsymbol{E}_{\mathrm{rad}} \times \boldsymbol{H}_{\mathrm{inc}}^*) = \frac{|\alpha|E_0^2(\omega/c)^2\hat{\boldsymbol{r}}}{8\pi\varepsilon_0 Z_0 r} \times Re\left[\sin^2\theta\, e^{\mathrm{i}\omega r\sin\theta\sin\varphi/c}e^{-\mathrm{i}(\phi_\alpha + r\omega/c)} + \sin\theta\sin\varphi\, e^{-\mathrm{i}\omega r\sin\theta\sin\varphi/c}e^{\mathrm{i}(\phi_\alpha + r\omega/c)}\right]. \quad (5.6)$$

Integrating the above energy flux density over the entire spherical surface yields

$$P_{\text{in}} = \frac{|\alpha|E_0^2(\omega/c)^2 r}{8\pi\varepsilon_0 Z_0} Re\left[e^{-\mathrm{i}(\phi_\alpha + r\omega/c)}\int_{\theta=0}^{\pi}\sin^3\theta\int_{\varphi=0}^{2\pi} e^{\mathrm{i}(\omega/c)r\sin\theta\sin\varphi}\mathrm{d}\varphi\mathrm{d}\theta\right.$$

$$\left.+e^{\mathrm{i}(\phi_\alpha + r\omega/c)}\int_{\theta=0}^{\pi}\sin^2\theta\int_{\varphi=0}^{2\pi}\sin\varphi\, e^{-\mathrm{i}\omega r\sin\theta\sin\varphi/c}\mathrm{d}\varphi\mathrm{d}\theta\right]$$

$$= \frac{|\alpha|E_0^2(\omega/c)^2 r}{4\varepsilon_0 Z_0} Re\left[e^{-\mathrm{i}(\phi_\alpha + r\omega/c)}\int_0^{\pi}\sin^3\theta\, J_0(\omega r\sin\theta/c)\mathrm{d}\theta\right.$$ (Gradshteyn & Ryzhik **3-915**-2)[13]

$$\left.-\mathrm{i}e^{\mathrm{i}(\phi_\alpha + r\omega/c)}\int_{\theta=0}^{\pi}\sin^2\theta\, J_1(\omega r\sin\theta/c)\mathrm{d}\theta\right]. \qquad (5.7)$$

The remaining integrals in Eq.(5.7) are evaluated as follows:

$$\int_0^{\pi}\sin(\theta)J_0(2a\sin\theta)\mathrm{d}\theta = \pi J_{-½}(a)J_{½}(a).$$ (G&R **6.681**-8)

$$\int_0^{\pi}\sin(3\theta)J_0(2a\sin\theta)\mathrm{d}\theta = -\pi J_{-3/2}(a)J_{3/2}(a).$$ (G&R **6.681**-8)

$$\rightarrow \int_0^{\pi}(3\sin\theta - 4\sin^3\theta)J_0(2a\sin\theta)\mathrm{d}\theta = -\pi J_{-3/2}(a)J_{3/2}(a)$$

$$\rightarrow \int_0^{\pi}\sin^3\theta\, J_0(2a\sin\theta)\mathrm{d}\theta = ¼\pi J_{-3/2}(a)J_{3/2}(a) + ¾\pi J_{-½}(a)J_{½}(a)$$ (G&R **8.464**-1, 2, 3, 4)

$$= -\frac{1}{2a}\left(\frac{\sin a}{a} - \cos a\right)\left(\frac{\cos a}{a} + \sin a\right) + \frac{3}{2a}\sin a\cos a$$

$$= \frac{\sin(2a)}{a} + \frac{\cos(2a)}{2a^2} - \frac{\sin(2a)}{4a^3}. \qquad (5.8)$$

Ignore, as these terms decline faster than $r^{-1}$.

$$\int_0^{\pi}J_1(2a\sin\theta)\mathrm{d}\theta = \pi J_{½}^2(a).$$ (G&R **6.681**-9)

$$\int_0^{\pi}\cos(2\theta)J_1(2a\sin\theta)\mathrm{d}\theta = -\pi J_{-½}(a)J_{3/2}(a).$$ (G&R **6.681**-9)

$$\rightarrow \int_0^{\pi}\sin^2\theta\, J_1(2a\sin\theta)\mathrm{d}\theta = ½\pi J_{½}^2(a) + ½\pi J_{-½}(a)J_{3/2}(a)$$ (G&R **8.464**-1, 2, 3, 4)

$$= \frac{\sin^2 a}{a} + \frac{\cos a}{a}\left(\frac{\sin a}{a} - \cos a\right) = \frac{\sin(2a)}{2a^2} - \frac{\cos(2a)}{a}. \qquad (5.9)$$

The total (time-averaged) optical power absorbed by the dipole is thus found to be

$$P_{\text{in}} = \frac{|\alpha|E_0^2(\omega/c)}{2\varepsilon_0 Z_0}[\cos(\phi_\alpha + r\omega/c)\sin(\omega r/c) - \sin(\phi_\alpha + r\omega/c)\cos(\omega r/c)] = -\frac{|\alpha|E_0^2(\omega/c)\sin\phi_\alpha}{2\varepsilon_0 Z_0}. \qquad (5.10)$$

This time-averaged power absorbed by the dipole may also be computed directly, as follows:[12]

$$P_{\text{abs}} = T^{-1}\int_{t=0}^{T}\boldsymbol{E}_{\text{inc}}(\boldsymbol{r}=0,t)\cdot\frac{\partial\boldsymbol{p}(t)}{\partial t}\mathrm{d}t = T^{-1}\int_{t=0}^{T}E_0\cos(\omega t)\frac{\partial}{\partial t}|\alpha|E_0\cos(\omega t - \phi_\alpha)\,\mathrm{d}t$$

$$= -T^{-1}|\alpha|E_0^2\omega\int_{t=0}^{T}\cos(\omega t)\sin(\omega t - \phi_\alpha)\,\mathrm{d}t = ½|\alpha|E_0^2\omega\sin\phi_\alpha. \qquad (5.11)$$

Note: $c = (\varepsilon_0 Z_0)^{-1}$

To ensure that the optical power entering the sphere of radius $r$ is greater than or equal to the power leaving that sphere, we must have

$$|\alpha| \le 3\varepsilon_0\lambda_0^3\sin\phi_\alpha/(2\pi)^2. \qquad (5.12)$$

Equality holds when there is no loss of energy during the scattering process. When the dipole scatters only a fraction of the incident optical energy and absorbs the remainder (converting the absorbed energy to other forms of energy, or to scattered light of frequencies other than $\omega$), the above inequality will be a strict one. (In our SI system of units, $\varepsilon_0 = 8.85418782 \cdots \times 10^{-12}$ farad/meter, and the free-space wavelength $\lambda_0$ of the incident and scattered waves is in meters.)

Let us denote the dipole density (i.e., the number of dipoles per unit volume) by $N$ and the (tiny) volume of each diploe by $v$. Then, in a densely-packed material medium consisting of such dipoles, we will have $N = v^{-1}$. Approximating a spherical dipole's self-field by[14]

$$\boldsymbol{E}_{\text{self}} \cong -\frac{(1 - \mathrm{i}4\pi^2 v/\lambda_0^3)\alpha E_0 \hat{\boldsymbol{z}}}{3\varepsilon_0 v}, \tag{5.13}$$

and writing for the electric susceptibility $\chi_e(\omega)$ of the bulk material $\varepsilon_0\chi_e(E_0\hat{\boldsymbol{z}} + \boldsymbol{E}_{\text{self}}) = \alpha E_0\hat{\boldsymbol{z}}/v$, we find

$$N\alpha/\varepsilon_0 \cong [\chi_e^{-1} + ⅓ - \mathrm{i}(4\pi^2/3N\lambda_0^3)]^{-1}. \tag{5.14}$$

Given that $N\lambda_0^3$ is typically a very large number, the dimensionless polarizability per unit volume, $N\alpha/\varepsilon_0$, should be of the same order of magnitude as $\chi_e$. Thus, upon multiplication by $N/\varepsilon_0$, the right-hand side of Eq.(5.12) should yield typical values of the (dimensionless) susceptibility of ordinary dielectric media (i.e., $\chi_e \sim 1$ at optical frequencies). In practice, for highly transparent material media, $\phi_\alpha$ is extremely small, e.g., on the order of $10^{-9}$ rad.

The above identities can be equivalently written as

$$|\chi_e^{-1} + ⅓ - \mathrm{i}(4\pi^2/3N\lambda_0^3)| \cong (N|\alpha|/\varepsilon_0)^{-1} \geq 4\pi^2/(3N\lambda_0^3 \sin\phi_\alpha). \tag{5.15}$$

Considering that, in accordance with Eq.(5.14), the phase of $\chi_e^{-1} + ⅓ - \mathrm{i}(4\pi^2/3N\lambda_0^3)$ equals $-\phi_\alpha$, one can infer from Eq.(5.15) that the only restriction on $\chi_e$ (for any material medium at any frequency $\omega$) should be $\mathrm{Im}(\chi_e) \geq 0$. The diagram in Fig.5 shows the region in the lower-half of the complex plane where $(N\alpha/\varepsilon_0)^{-1}$ is allowed to reside. Recalling that the susceptibility $\chi_e$ of ordinary transparent or partially absorptive media must be in the upper-half plane, its inverse, $\chi_e^{-1}$, should reside somewhere in the lower-half plane. Consequently, one expects $(N\alpha/\varepsilon_0)^{-1}$ to be located in the lower-half plane, as shown in Fig.5. Since the magnitude $(N|\alpha|/\varepsilon_0)^{-1}$ and phase $-\phi_\alpha$ of this complex number are constrained by Eq.(5.15), the number is *not* allowed to reside in the area between the red curve and the real axis of the complex plane.

Note that $4\pi^2/(3N\lambda_0^3)$ is typically a small number (on the order of $10^{-9}$), which means that the horizontal and vertical axes of the diagram in Fig.5 are depicted on vastly different scales. Although the red curve closely hugs the real axis, there clearly exists a region in the lower half-plane where the magnitude and phase of $(N\alpha/\varepsilon_0)^{-1}$ cannot satisfy the required inequality.

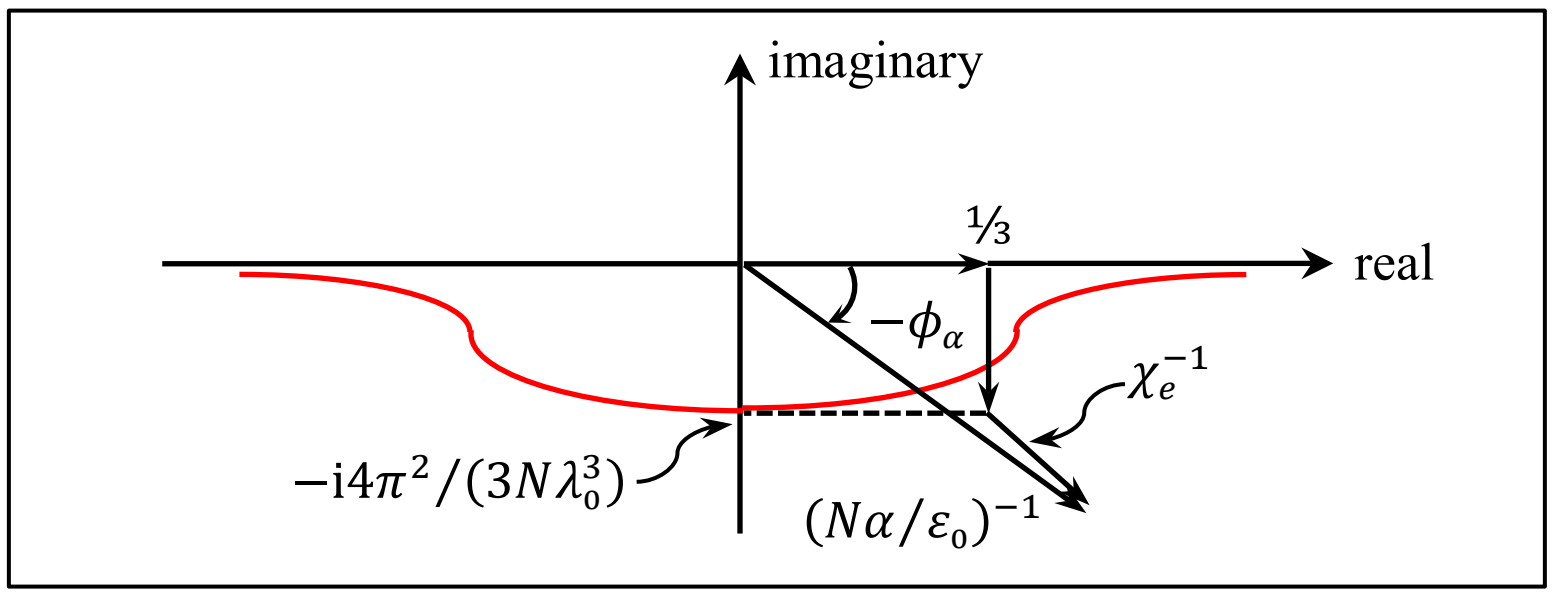

**Fig.5**. In the complex $\chi_e$ plane, the inverse susceptibility $\chi_e^{-1}$ plus $⅓ - \mathrm{i}(4\pi^2/3N\lambda_0^3)$ may, in principle, be anywhere in the lower-half plane except in the region between the red curve and the real axis. This implies that $\chi_e$ can be anywhere in the upper-half plane, where $\mathrm{Im}(\chi_e) \geq 0$. The various depicted entities are related to each other through Eq.(5.15).

In general, $\chi_e(\omega)$ obeys the Lorentz oscillator model,[11,12] and the sole limitation on its parameter values (i.e., plasma frequency $\omega_p$, resonance frequency $\omega_r$, damping coefficient $\gamma$) is that $\gamma \geq 0$. The constraint imposed by inequality (5.15) on the material medium's polarizability $N\alpha/\varepsilon_0$ is unrelated to the Kramers-Kronig relations,[11,12] which constrain the real and imaginary parts of $\chi_e(\omega)$ in relation to each other across the entire range of frequencies. In contrast, inequality (5.15) must hold at each and every frequency $\omega$, independently of the values of $\chi_e$ at other frequencies.

**6. Scattering amplitude of an electric dipole**. If the $k$-vector is specified as $\boldsymbol{k} = k_0\hat{\boldsymbol{r}} = (\omega/c)\hat{\boldsymbol{r}}$ within the spherical coordinate system $(r, \theta, \varphi)$ depicted in Fig.6, one can define the mutually orthogonal pair of unit-vectors $\boldsymbol{\mathcal{E}}'$ and $\boldsymbol{\mathcal{E}}''$ that are also perpendicular to $\boldsymbol{k}$ as follows:

$$\boldsymbol{\mathcal{E}}' = -\cos\theta\cos\varphi\,\hat{\boldsymbol{x}} - \cos\theta\sin\varphi\,\hat{\boldsymbol{y}} + \sin\theta\,\hat{\boldsymbol{z}}, \tag{6.1}$$

$$\boldsymbol{\mathcal{E}}'' = \sin\varphi\,\hat{\boldsymbol{x}} - \cos\varphi\,\hat{\boldsymbol{y}}. \tag{6.2}$$

One may then proceed to express the corresponding $E$-fields of a pair of right- and left-circularly polarized (RCP and LCP) monochromatic plane-waves as

$$\boldsymbol{E}_{R,L}(\boldsymbol{r}, t) = E^{(\pm)}(\boldsymbol{\mathcal{E}}' \pm \mathrm{i}\boldsymbol{\mathcal{E}}'')e^{\mathrm{i}(\boldsymbol{k}\cdot\boldsymbol{r} - \omega t)}. \tag{6.3}$$

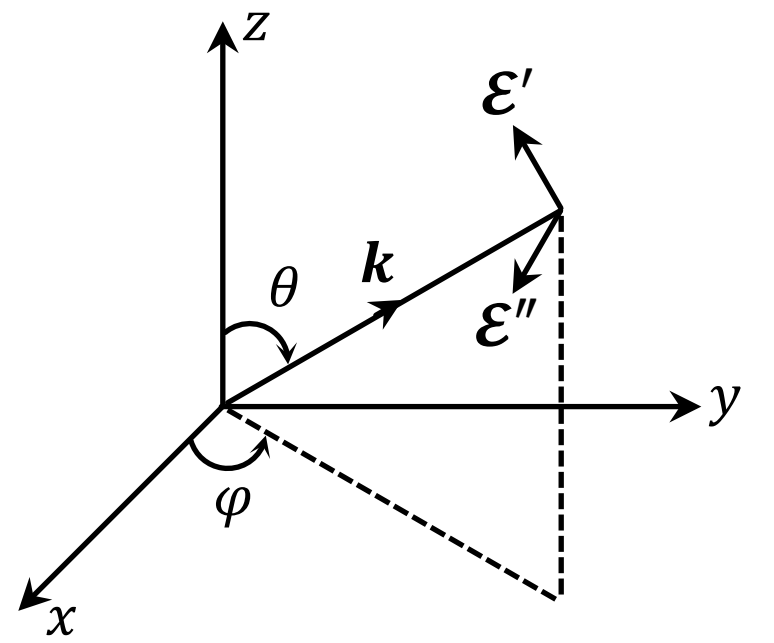

**Fig. 6**. The orthogonal pair of unit-vectors $(\boldsymbol{\mathcal{E}}', \boldsymbol{\mathcal{E}}'')$ is used to define the states of right and left circular polarization (RCP and LCP) of the scattered photon along the $k$-vector. While $\boldsymbol{\mathcal{E}}'$ is perpendicular to $\boldsymbol{k}$ within the $kz$-plane, $\boldsymbol{\mathcal{E}}''$, being aligned with $\boldsymbol{k} \times \boldsymbol{\mathcal{E}}'$, is perpendicular to the $kz$-plane.

Let an electric point-dipole $\boldsymbol{p}_s(t) = (p_{sx}\hat{\boldsymbol{x}} + p_{sy}\hat{\boldsymbol{y}} + p_{sz}\hat{\boldsymbol{z}})e^{-\mathrm{i}\omega t}$ sit at the origin of coordinates, with the subscript $s$ identifying it as a source of radiation. At a sufficiently distant observation point $\boldsymbol{r}$ away from the origin, in accordance with Eq.(5.3), the radiated $E$-field will be given by

$$\begin{pmatrix} E_x \\ E_y \\ E_z \end{pmatrix} = \frac{k_0^2 e^{\mathrm{i}k_0 r}}{4\pi\varepsilon_0 r}\begin{pmatrix} 1 - \sin^2\theta\cos^2\varphi & -\sin^2\theta\sin\varphi\cos\varphi & -\sin\theta\cos\theta\cos\varphi \\ -\sin^2\theta\sin\varphi\cos\varphi & 1 - \sin^2\theta\sin^2\varphi & -\sin\theta\cos\theta\sin\varphi \\ -\sin\theta\cos\theta\cos\varphi & -\sin\theta\cos\theta\sin\varphi & \sin^2\theta \end{pmatrix}\begin{pmatrix} p_{sx} \\ p_{sy} \\ p_{sz} \end{pmatrix}. \tag{6.4}$$

Writing this $E$-field as a sum of RCP and LCP modes, we will have

$$E^{(+)}(-\cos\theta\cos\varphi + \mathrm{i}\sin\varphi) + E^{(-)}(-\cos\theta\cos\varphi - \mathrm{i}\sin\varphi)$$
$$= (1 - \sin^2\theta\cos^2\varphi)p_{sx} - (\sin^2\theta\sin\varphi\cos\varphi)p_{sy} - (\sin\theta\cos\theta\cos\varphi)p_{sz}, \tag{6.5a}$$

$$E^{(+)}(-\cos\theta\sin\varphi - \mathrm{i}\cos\varphi) + E^{(-)}(-\cos\theta\sin\varphi + \mathrm{i}\cos\varphi)$$
$$= -(\sin^2\theta\sin\varphi\cos\varphi)p_{sx} + (1 - \sin^2\theta\sin^2\varphi)p_{sy} - (\sin\theta\cos\theta\sin\varphi)p_{sz}, \tag{6.5b}$$

$$E^{(+)} + E^{(-)} = -\cos\theta\cos\varphi\, p_{sx} - \cos\theta\sin\varphi\, p_{sy} + \sin\theta\, p_{sz}. \tag{6.5c}$$

The above equations are readily solved for the RCP and LCP probability amplitudes,‡ yielding

$$E_{\text{out}}^{(\pm)}=\frac{k_0^2}{8\pi\varepsilon_0}[(-\cos\theta_{\text{out}}\cos\varphi_{\text{out}}\mp \mathrm{i}\sin\varphi_{\text{out}})p_{sx}+(-\cos\theta_{\text{out}}\sin\varphi_{\text{out}}\pm \mathrm{i}\cos\varphi_{\text{out}})p_{sy}+\sin\theta_{\text{out}}p_{sz}]. \quad (6.6)$$

Here, a subscript is used to emphasize that $(\theta_{\text{out}},\varphi_{\text{out}})$ represent the polar and azimuthal angles of the outgoing photon. Note that the emission of a photon in the reverse direction would entail a change of these angles to $(\pi-\theta_{\text{out}},\varphi_{\text{out}}+\pi)$, which is tantamount to a role reversal for the RCP and LCP photons; in other words, an oscillating dipole emits an RCP photon along a given direction with precisely the same probability amplitude as it emits an LCP photon in the opposite direction.

Next, let us consider a circularly-polarized (RCP or LCP) photon that arrives at the origin of the coordinate system along the $(\theta_{\text{in}},\varphi_{\text{in}})$ direction. A point-particle sitting at the origin with the polarizability tensor $\tilde{\alpha}(\omega)$ responds to the incident $E$-field $E_{\text{in}}^{(\pm)}(\boldsymbol{\mathcal{E}}'\pm \mathrm{i}\boldsymbol{\mathcal{E}}'')$ by producing an electric dipole moment $\boldsymbol{p}$, as follows:

$$\begin{pmatrix}p_x\\ p_y\\ p_z\end{pmatrix}=\begin{pmatrix}\alpha_{xx}&\alpha_{xy}&\alpha_{xz}\\ \alpha_{yx}&\alpha_{yy}&\alpha_{yz}\\ \alpha_{zx}&\alpha_{zy}&\alpha_{zz}\end{pmatrix}\begin{pmatrix}-\cos\theta_{\text{in}}\cos\varphi_{\text{in}}\pm \mathrm{i}\sin\varphi_{\text{in}}\\ -\cos\theta_{\text{in}}\sin\varphi_{\text{in}}\mp \mathrm{i}\cos\varphi_{\text{in}}\\ \sin\theta_{\text{in}}\end{pmatrix}E_{\text{in}}^{(\pm)}. \quad (6.7)$$

Unless the particle acts as a sink (by absorbing the photon and converting its energy to some other form, or emitting one or more photons at different frequencies), the incoming photon will be scattered, either as an RCP or an LCP photon of the same frequency $\omega$, along the $(\theta_{\text{out}},\varphi_{\text{out}})$ direction with a probability amplitude given by Eq.(6.6). Thus, the probability amplitude $E_{\text{out}}^{(\pm)}$ of the scattered photon is obtained by multiplying Eq.(6.7) on the left with the following row vector:

$$\frac{k_0^2}{8\pi\varepsilon_0}[(-\cos\theta_{\text{out}}\cos\varphi_{\text{out}}\mp \mathrm{i}\sin\varphi_{\text{out}})\quad(-\cos\theta_{\text{out}}\sin\varphi_{\text{out}}\pm \mathrm{i}\cos\varphi_{\text{out}})\quad \sin\theta_{\text{out}}]. \quad (6.8)$$

Note that the product of Eqs.(6.7) and (6.8) is really four equations in one, as the $\pm$ signs associated with the incident photon in Eq.(6.7) are independent of the $\pm$ signs associated with the scattered photon in Eq.(6.8). Stated differently, the incident photon can be RCP or LCP, and the scattered photon can also be RCP or LCP, completely independently of the incident photon.

Suppose now that both the incident photon and the scattered photon are made to retrace their propagation paths in reverse *without* changing their polarization states; that is, the scattered RCP photon propagating in the reverse direction now becomes the incident RCP photon, while the incident LCP photon propagating in reverse becomes a scattered LCP photon, and so on. This means that the $\pm$ and $\mp$ signs appearing in Eqs.(6.7) and (6.8) for each photon must flip. However, since $(\theta_{\text{in}},\varphi_{\text{in}})$ and $(\theta_{\text{out}},\varphi_{\text{out}})$ have switched places, for the product of these equations to remain precisely the same, it is also necessary that the polarizability tensor $\tilde{\alpha}(\omega)$ be symmetric; that is, $\tilde{\alpha}(\omega)=\tilde{\alpha}^T(\omega)$. It is thus seen that a symmetrically polarizable point-particle scatters an RCP photon arriving along $(\theta_{\text{in}},\varphi_{\text{in}})$ into an LCP photon departing along $(\theta_{\text{out}},\varphi_{\text{out}})$ with exactly the same probability amplitude as it scatters an LCP photon arriving along $(\theta_{\text{out}},\varphi_{\text{out}})$ into an RCP photon departing along $(\theta_{\text{in}},\varphi_{\text{in}})$. Similar statements, of course, can be made about all other combinations of the polarization states of the incident and scattered photons.

‡ These complex-valued probability amplitudes for RCP and LCP photons have the dimensions of volt in our SI system of units. The scattering probabilities $|E^{(\pm)}|^2$ must, of course, be normalized by the cross-sectional area of the photon wavepacket, which has the dimensions of meter$^2$, hence the apparent discrepancy between the SI units of the $E$-field (volt/meter) and those of the scattering probability amplitudes $E^{(\pm)}$ as given by Eq.(6.6).

Consider a source dipole $\boldsymbol{p}_s e^{-\mathrm{i}\omega t}$ located at the origin and radiating into the surrounding free space, producing the $E$-field $\boldsymbol{E}_d e^{\mathrm{i}(\boldsymbol{k}\cdot\boldsymbol{r}-\omega t)}$ at the destination point $\boldsymbol{r}$, which is a distance $r$ away from the origin in the $(\theta,\varphi)$ direction. If we place a dipole $\boldsymbol{p}_d e^{-\mathrm{i}\omega t}$ at the destination point $\boldsymbol{r}$ and proceed to monitor the radiated $E$-field $\boldsymbol{E}_s e^{\mathrm{i}(\boldsymbol{k}\cdot\boldsymbol{r}-\omega t)}$ at the original source location (i.e., at the origin), we will find that $\boldsymbol{E}_s\cdot\boldsymbol{p}_s=\boldsymbol{E}_d\cdot\boldsymbol{p}_d$, simply because the $3\times 3$ matrix appearing on the right-hand side of Eq.(6.4) remains intact when $(\theta,\varphi)$ is replaced by $(\pi-\theta,\varphi+\pi)$ and, moreover, because the matrix is symmetric. More generally, when an RCP or LCP photon of amplitude $E_{\mathrm{in}}^{(\pm)}$ arrives at the destination point $\boldsymbol{r}=(r,\theta,\varphi)$ along *any* direction $(\theta_{\mathrm{in}},\varphi_{\mathrm{in}})$, the dot-product of the incident $E$-field and the dipole moment $\boldsymbol{p}_d$ (sitting at $\boldsymbol{r}$) will be

$$\boldsymbol{E}_d\cdot\boldsymbol{p}_d=E_{\mathrm{in}}^{(\pm)}[(-\cos\theta_{\mathrm{in}}\cos\varphi_{\mathrm{in}}\pm\mathrm{i}\sin\varphi_{\mathrm{in}})\hat{\boldsymbol{x}}+(-\cos\theta_{\mathrm{in}}\sin\varphi_{\mathrm{in}}\mp\mathrm{i}\cos\varphi_{\mathrm{in}})\hat{\boldsymbol{y}}+\sin\theta_{\mathrm{in}}\hat{\boldsymbol{z}}]\cdot\boldsymbol{p}_d. \quad (6.9)$$

This equation resembles Eq.(6.6) in many ways, and their resemblance would be even closer if $\boldsymbol{p}_d$ were to become an emitter (rather than a receiver), radiating the same RCP or LCP photon in the reverse direction, i.e., in the direction of $(\theta_{\mathrm{out}},\varphi_{\mathrm{out}})=(\pi-\theta_{\mathrm{in}},\varphi_{\mathrm{in}}+\pi)$. The remaining difference is that the leading coefficient on the right-hand side of Eq.(6.6) is $k_0^2/(8\pi\varepsilon_0)$, which is needed to yield the probability amplitude $E_{\mathrm{out}}^{(\pm)}$ of a photon emitted by the source dipole $\boldsymbol{p}_s$ in the $(\theta_{\mathrm{out}},\varphi_{\mathrm{out}})$ direction, whereas the leading coefficient on the right-hand side of Eq.(6.9) is the probability amplitude $E_{\mathrm{in}}^{(\pm)}$ of an incoming photon received by the destination dipole $\boldsymbol{p}_d$. These coefficients will also switch places when the roles of $\boldsymbol{p}_s$ and $\boldsymbol{p}_d$ are reversed.

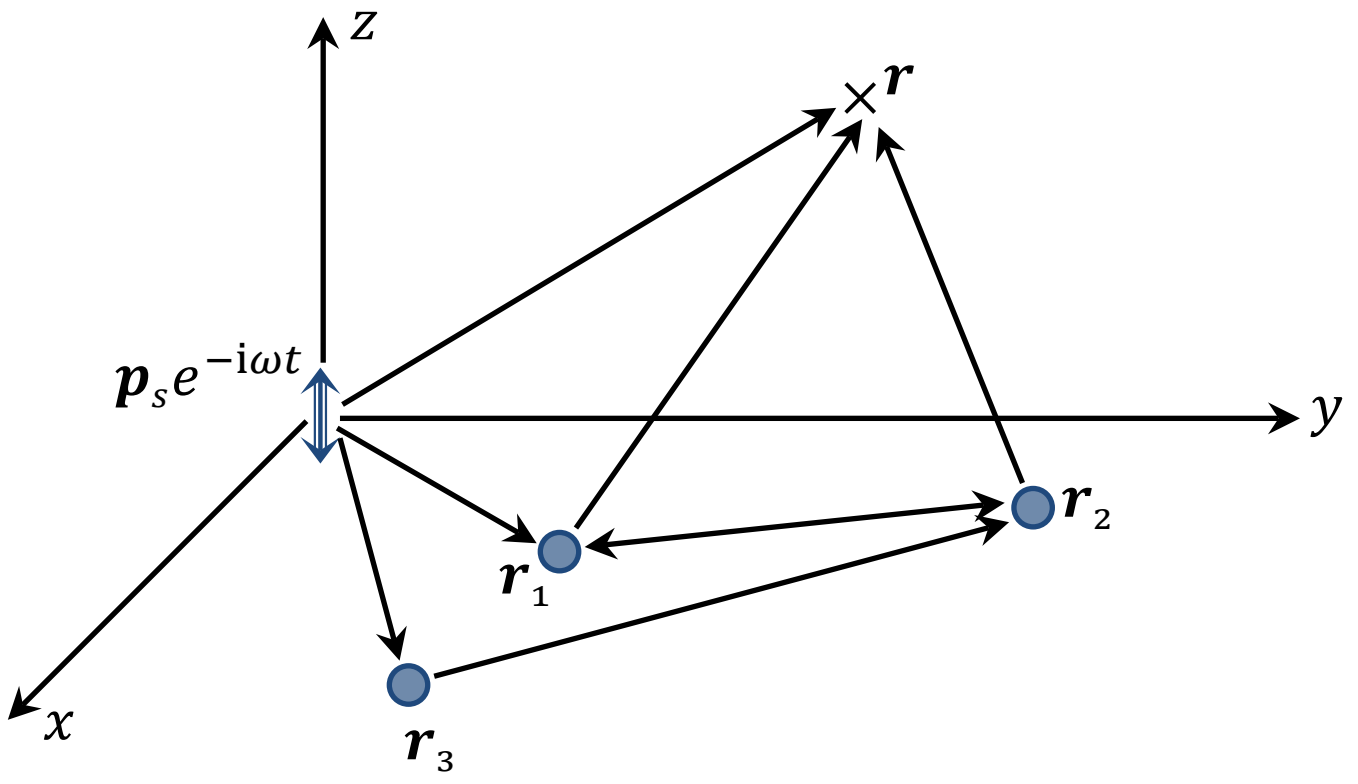

**Fig.7**. An RCP or LCP photon emitted by the oscillating dipole $\boldsymbol{p}_s e^{-\mathrm{i}\omega t}$ at the origin of coordinates may reach the destination point $\boldsymbol{r}=(r,\theta,\varphi)$ in many different ways. The photon may go directly to $\boldsymbol{r}$, or it may get there after a single scattering by one of the point-particles located at $\boldsymbol{r}_1,\boldsymbol{r}_2,\boldsymbol{r}_3$, etc. Also possible are multiple scatterings at any number of point-particles in any arbitrary order. During each scattering event, an incident photon may retain its polarization state, or change from RCP to LCP and vice-versa.

If, as depicted in Fig.7, a point-particle located at $\boldsymbol{r}_1$ happens to scatter a photon emitted by $\boldsymbol{p}_s$ toward the receiver $\boldsymbol{p}_d$ (sitting at $\boldsymbol{r}$) there will be an amplitude for the photon to be scattered at $\boldsymbol{r}_1$ and arrive at $\boldsymbol{p}_d$. This, of course, is in addition to the amplitude for the photon to go directly from the source to the destination point. Each of these possibilities has its own probability amplitude, and they both contribute to the observed dot-product $\boldsymbol{E}_d\cdot\boldsymbol{p}_d$. Similarly, another particle, this one sitting at $\boldsymbol{r}_2$, may also contribute to the received signal by providing additional scattering paths such as $\boldsymbol{p}_s\to\boldsymbol{p}_2\to\boldsymbol{p}_d$ and $\boldsymbol{p}_s\to\boldsymbol{p}_1\to\boldsymbol{p}_2\to\boldsymbol{p}_d$ and $\boldsymbol{p}_s\to\boldsymbol{p}_2\to\boldsymbol{p}_1\to\boldsymbol{p}_d$ and $\boldsymbol{p}_s\to\boldsymbol{p}_1\to\boldsymbol{p}_2\to\boldsymbol{p}_1\to\boldsymbol{p}_d$, and so forth. Thus, the received signal will be the sum of all $\boldsymbol{E}_d\cdot\boldsymbol{p}_d$ contributed by all the various paths that the emitted photon could possibly take from the source to the destination point. It is clear that the system may contain any number of scatterers located at various points in the vicinity of $\boldsymbol{p}_s$ and $\boldsymbol{p}_d$, and that computing the dot-product $\boldsymbol{E}_d \cdot \boldsymbol{p}_d$ at the destination point $\boldsymbol{r}$ requires that all possible paths for the photon (as well as all possible states of its polarization) be taken into account.

One can now appreciate the essence of the reciprocity theorem of classical electrodynamics.[15-18] If the roles of $\boldsymbol{p}_s$ and $\boldsymbol{p}_d$ are switched, so that $\boldsymbol{p}_d$ becomes the source and $\boldsymbol{p}_s$ the receiver, then every path that an emitted photon might have taken will be reversed. However, all scattering amplitudes in the reverse direction will be the same as those in the forward direction and, consequently, $\boldsymbol{E}_s \cdot \boldsymbol{p}_s$ will turn out to be identical to $\boldsymbol{E}_d \cdot \boldsymbol{p}_d$.

**7. Magnetic dipole radiator**. The results of the preceding section can be readily extended to cover radiation and scattering by magnetic dipole oscillators. Let a magnetic point-particle having magnetizability tensor $\tilde{\beta}(\omega)$ be excited by an incoming RCP or LCP photon along the direction of $(\theta_{\text{in}}, \varphi_{\text{in}})$, where the incident magnetic field is now given by

$$\boldsymbol{H}_{\text{in}}^{(\pm)} = Z_0^{-1} E_{\text{in}}^{(\pm)} [(\sin\varphi_{\text{in}} \pm \mathrm{i}\cos\theta_{\text{in}}\cos\varphi_{\text{in}})\hat{\boldsymbol{x}} + (-\cos\varphi_{\text{in}} \pm \mathrm{i}\cos\theta_{\text{in}}\sin\varphi_{\text{in}})\hat{\boldsymbol{y}} \mp \mathrm{i}\sin\theta_{\text{in}}\,\hat{\boldsymbol{z}}]. \quad (7.1)$$

The Cartesian components of the induced magnetic dipole moment $\boldsymbol{m}e^{-\mathrm{i}\omega t}$ are [§]

$$\begin{pmatrix} m_x \\ m_y \\ m_z \end{pmatrix} = \begin{pmatrix} \beta_{xx} & \beta_{xy} & \beta_{xz} \\ \beta_{yx} & \beta_{yy} & \beta_{yz} \\ \beta_{zx} & \beta_{zy} & \beta_{zz} \end{pmatrix} \begin{pmatrix} H_x \\ H_y \\ H_z \end{pmatrix} = Z_0^{-1} E_{\text{in}}^{(\pm)} \begin{pmatrix} \beta_{xx} & \beta_{xy} & \beta_{xz} \\ \beta_{yx} & \beta_{yy} & \beta_{yz} \\ \beta_{zx} & \beta_{zy} & \beta_{zz} \end{pmatrix} \begin{pmatrix} \sin\varphi_{\text{in}} \pm \mathrm{i}\cos\theta_{\text{in}}\cos\varphi_{\text{in}} \\ -\cos\varphi_{\text{in}} \pm \mathrm{i}\cos\theta_{\text{in}}\sin\varphi_{\text{in}} \\ \mp\mathrm{i}\sin\theta_{\text{in}} \end{pmatrix}. \quad (7.2)$$

In the far field (i.e., $r \gg \lambda_0$), the radiated $E$-field of this dipole is known to be[11,12]

$$\begin{pmatrix} E_x \\ E_y \\ E_z \end{pmatrix} = \frac{ck_0^2\, e^{\mathrm{i}k_0 r}}{4\pi r} \begin{pmatrix} 0 & \cos\theta_{\text{out}} & -\sin\theta_{\text{out}}\sin\varphi_{\text{out}} \\ -\cos\theta_{\text{out}} & 0 & \sin\theta_{\text{out}}\cos\varphi_{\text{out}} \\ \sin\theta_{\text{out}}\sin\varphi_{\text{out}} & -\sin\theta_{\text{out}}\cos\varphi_{\text{out}} & 0 \end{pmatrix} \begin{pmatrix} m_x \\ m_y \\ m_z \end{pmatrix}. \quad (7.3)$$

This radiated $E$-field can be expressed as the superposition of a pair of co-propagating RCP and LCP photons, whose respective probability amplitudes $E^{(+)}$ and $E^{(-)}$ are given by

$$E_{\text{out}}^{(\pm)} = \frac{ck_0^2}{8\pi} [(\sin\varphi_{\text{out}} \mp \mathrm{i}\cos\theta_{\text{out}}\cos\varphi_{\text{out}})m_x + (-\cos\varphi_{\text{out}} \mp \mathrm{i}\cos\theta_{\text{out}}\sin\varphi_{\text{out}})m_y \pm \mathrm{i}\sin\theta_{\text{out}}\, m_z]. \quad (7.4)$$

Thus, when an RCP or LCP photon arrives at the point-particle in the $(\theta_{\text{in}}, \varphi_{\text{in}})$ direction and leaves (either as an RCP or LCP photon) along the $(\theta_{\text{out}}, \varphi_{\text{out}})$ direction, the scattering amplitude can be found by multiplying Eq.(7.2) on the left-hand side by the following row vector:

$$\frac{ck_0^2}{8\pi} [(\sin\varphi_{\text{out}} \mp \mathrm{i}\cos\theta_{\text{out}}\cos\varphi_{\text{out}}) \quad (-\cos\varphi_{\text{out}} \mp \mathrm{i}\cos\theta_{\text{out}}\sin\varphi_{\text{out}}) \quad (\pm\mathrm{i}\sin\theta_{\text{out}})]. \quad (7.5)$$

As before, the product of Eqs.(7.2) and (7.5) is actually four equations in one, considering that the incident and scattered photons can be in different combinations of RCP and LCP states. When the propagation directions of both the incident and scattered photons are reversed (*without* changing their respective polarization states), the product of Eqs.(7.1) and (7.5) remains the same provided that the magnetizability tensor $\tilde{\beta}(\omega)$ is symmetric; that is, $\tilde{\beta}(\omega) = \tilde{\beta}^T(\omega)$. Thus, the assembly of point-particles surrounding the source dipole and the destination dipole shown in Fig.7 may comprise an arbitrary combination of symmetrically polarizable and symmetrically magnetizable particles, whose contributions to the overall scattering amplitude in the forward path and in the reverse path ensure the satisfaction of the reciprocity theorem of classical electrodynamics.

---

§ In our SI system of units, the magnetization $\boldsymbol{M}$ has the same units (weber/m$^2$) as the magnetic induction $\boldsymbol{B}$. Thus, $\boldsymbol{B} = \mu_0\boldsymbol{H} + \boldsymbol{M}$, where the permeability $\mu_0$ of free space has units of henry/m, and the magnetic field $\boldsymbol{H}$ has units of ampere/m. A magnetic dipole $\boldsymbol{m}$ is thus a small loop of circulating current $I$ around an area $A$, with the magnitude of its dipole moment defined as $m = \mu_0 IA$. Consequently, $\boldsymbol{m} \cdot \boldsymbol{H}$ and $\boldsymbol{p} \cdot \boldsymbol{E}$ have the same units of volt · coulomb.

If the goal is to monitor the magnetic field $\boldsymbol{H}_d$ (rather than the electric field $\boldsymbol{E}_d$) at the destination point $\boldsymbol{r}$, then a magnetic dipole $\boldsymbol{m}_d$ placed at $\boldsymbol{r}$ will, in accordance with Eq.(7.1), yield the following dot-product with the $H$-field of an incoming RCP or LCP photon:

$$\boldsymbol{H}_d \cdot \boldsymbol{m}_d = Z_0^{-1} E_{\text{in}}^{(\pm)} [(\sin\varphi_{\text{in}} \pm \mathrm{i}\cos\theta_{\text{in}}\cos\varphi_{\text{in}})\hat{\boldsymbol{x}} + (-\cos\varphi_{\text{in}} \pm \mathrm{i}\cos\theta_{\text{in}}\sin\varphi_{\text{in}})\hat{\boldsymbol{y}} \mp \mathrm{i}\sin\theta_{\text{in}}\,\hat{\boldsymbol{z}}] \cdot \boldsymbol{m}_d. \tag{7.6}$$

In the reverse path, this $\boldsymbol{m}_d$ will be the radiator and the propagation direction becomes $(\theta_{\text{out}}, \varphi_{\text{out}}) = (\pi - \theta_{\text{in}}, \varphi_{\text{in}} + \pi)$. The emitted RCP and LCP amplitudes will then be given by Eq.(7.4), which is similar to Eq.(7.6) except for an overall minus sign, and aside from the leading coefficients $ck_0^2/8\pi$ and $Z_0^{-1}E_{\text{in}}^{(\pm)}$. The leading coefficients manage to switch places between the forward path and the reverse path, without modifying the relevant dot-products at the source and at the destination point. As for the minus sign, it is cancelled with a similar minus sign when the source and the receiver are both magnetic dipoles, in which case $\boldsymbol{H}_s \cdot \boldsymbol{m}_s = \boldsymbol{H}_d \cdot \boldsymbol{m}_d$. However, when the source and the destination dipoles are of different types, retaining the minus sign yields the final result of the reciprocity theorem as $\boldsymbol{E}_s \cdot \boldsymbol{p}_s = -\boldsymbol{H}_d \cdot \boldsymbol{m}_d$ (or $\boldsymbol{H}_s \cdot \boldsymbol{m}_s = -\boldsymbol{E}_d \cdot \boldsymbol{p}_d$).

**8. Manifestation of reciprocity in the case of diffraction from gratings**. A typical example of application of the reciprocity theorem is depicted in Fig.8, where a photon leaving the source at $S$ is reflected at a perfectly electrically conducting diffraction grating,[7,8] then arrives at the destination $D$, where it is picked up by a photodetector. The probability of detection at $D$ is the squared magnitude of the sum of all the probability amplitudes associated with different paths that the photon could possibly take from $S$ to $D$. If the locations of the source and the detector are switched, that is, if the source is moved to $D$ and the detector is placed at $S$, all the allowed paths will be reversed. However, the probability amplitudes remain the same; hence, the probability of the photon leaving the source at its new location and getting detected at the new location of the detector will remain the same as before. Note that this argument requires that the scattering amplitudes at each point on the grating surface be the same in opposite directions. This is tantamount to requiring that the scattering amplitude of a single photon from a given atom (or scatterer) be the same whether the photon arrives in the $|\omega, \boldsymbol{k}_1, \hat{\boldsymbol{e}}_1\rangle$ state and scatters off in the $|\omega, \boldsymbol{k}_2, \hat{\boldsymbol{e}}_2\rangle$ state, or arrives in the $|\omega, -\boldsymbol{k}_2, \hat{\boldsymbol{e}}_2\rangle$ state and leaves in the $|\omega, -\boldsymbol{k}_1, \hat{\boldsymbol{e}}_1\rangle$ state. (Here, the complex unit-vector $\hat{\boldsymbol{e}} = \boldsymbol{e}' + \mathrm{i}\boldsymbol{e}''$, where $|\boldsymbol{e}'|^2 + |\boldsymbol{e}''|^2 = 1$, is used to represent the photon's state of polarization.)

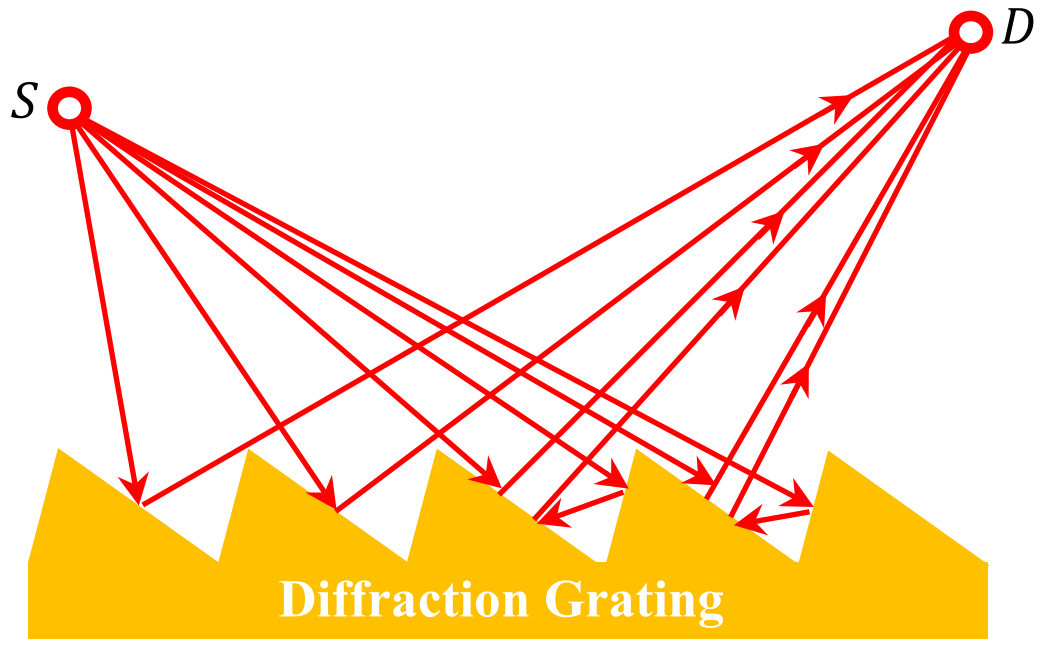

**Fig.8**. A photon emitted by the source $S$ arrives at the detector $D$ after propagating to a grating and being scattered at the grating's surface. All the probability amplitudes associated with the various paths that the photon could possibly take in its journey from $S$ to $D$ are added together to yield the overall probability amplitude for a detection event at $D$.

**9. Slabs, multiple reflections, and reciprocity**. With reference to Fig.9, let the Fresnel reflection and transmission coefficients of a semi-infinite medium of refractive index $n$ be $\rho$ and $\tau$, while those of a finite-thickness slab of the same material (thickness $= d$, refractive index $= n$) are $r$ and $t$. In the following analysis, the gap separating the slab from the substrate is assumed to shrink to zero, so that no phase shifts need be associated with propagations inside the gap. The incident plane-wave has amplitude $\boldsymbol{E}_{\text{in}}$, while that of the total $E$-field entering the substrate through the gap is $\tau e^{\mathrm{i}\varphi}\boldsymbol{E}_{\text{in}} = (1+\rho)e^{ink_0d}\boldsymbol{E}_{\text{in}}$. To account for the multiple reflections within the gap, we use the Fresnel coefficients $\rho$ and $\tau = 1+\rho$ at the front facet of the semi-infinite substrate, and $r$ and $t$ at the backside of the finite-thickness slab. With the gap-width between the two parts approaching zero, we will have

$$\rho = r + t\rho t + t\rho r\rho t + \cdots = r + \frac{\rho t^2}{1-\rho r}. \tag{9.1}$$

$$\tau e^{\mathrm{i}\varphi} = t\tau + t\rho r\tau + t\rho^2 r^2\tau + \cdots = \frac{t\tau}{1-\rho r} \quad \rightarrow \quad t = (1-\rho r)e^{\mathrm{i}\varphi}. \tag{9.2}$$

Substituting for $t$ from Eq.(9.2) into Eq.(9.1), we find

$$\rho = r + \rho(1-\rho r)e^{2\mathrm{i}\varphi} = r + \rho e^{2\mathrm{i}\varphi} - r\rho^2 e^{2\mathrm{i}\varphi} \quad \rightarrow \quad r = \frac{(1-e^{2\mathrm{i}\varphi})\rho}{1-\rho^2 e^{2\mathrm{i}\varphi}}. \tag{9.3}$$

This $r$ may now be substituted into Eq.(9.2) to yield the expression for $t$, as follows:

$$t = \left[1 - \frac{(1-e^{2\mathrm{i}\varphi})\rho^2}{1-\rho^2 e^{2\mathrm{i}\varphi}}\right]e^{\mathrm{i}\varphi} = \frac{(1-\rho^2)e^{\mathrm{i}\varphi}}{1-\rho^2 e^{2\mathrm{i}\varphi}}. \tag{9.4}$$

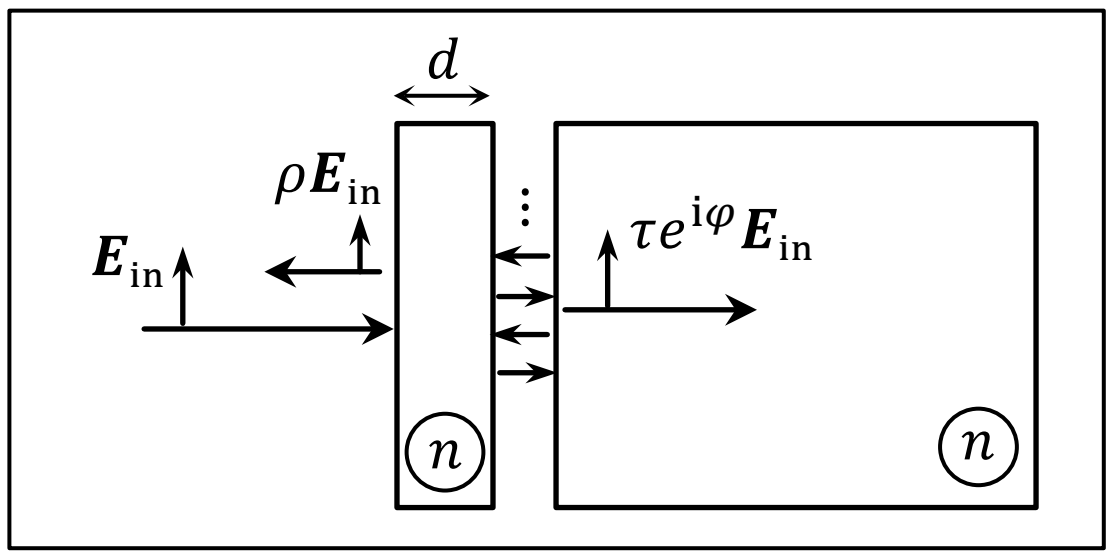

**Fig.9**. A normally-incident plane-wave of amplitude $\boldsymbol{E}_{\text{in}}$ enters a semi-infinite medium of refractive index $n$. The Fresnel reflection and transmission coefficients at the front facet of the semi-infinite medium are denoted by $\rho$ and $\tau$, respectively. To determine the Fresnel coefficients $r$ and $t$ of a finite-thickness slab of the same material, we introduce an infinitesimal gap between the slab and the semi-infinite medium, and proceed to express $\rho$ and $\tau$ in terms of $r$ and $t$ by accounting for multiple reflections within the gap. In the limit when the gap-width approaches zero, a single path through the slab introduces the phase shift $\varphi = nk_0 d$.

It is well known that $\rho = (1-n)/(1+n)$ and $\tau = 1+\rho = 2/(1+n)$.[7,11,12] Therefore, Eqs.(9.3) and (9.4) can be used to compute the Fresnel coefficients of the slab in terms of $n$ and $d$. Note that $r/t = -\mathrm{i}2\rho\sin(\varphi)/(1-\rho^2)$, which indicates that the phase difference between $r$ and $t$ is 90° when the slab's refractive index $n$ is real-valued. It is also not difficult to show that the magnitude of $r^2 - t^2$ equals 1.0 when $n$ is real.

**Proof**: $r^2 - t^2 = \frac{\rho^2(1-e^{2\mathrm{i}\varphi})^2 - (1-\rho^2)^2 e^{2\mathrm{i}\varphi}}{(1-\rho^2 e^{2\mathrm{i}\varphi})^2} = \frac{\rho^2(1-2e^{2\mathrm{i}\varphi}+e^{4\mathrm{i}\varphi}) - (1-2\rho^2+\rho^4)e^{2\mathrm{i}\varphi}}{(1-\rho^2 e^{2\mathrm{i}\varphi})^2} = \frac{\rho^2(1+e^{4\mathrm{i}\varphi}) - (1+\rho^4)e^{2\mathrm{i}\varphi}}{(1-\rho^2 e^{2\mathrm{i}\varphi})^2}$

$$= \frac{(\rho^2 - e^{2\mathrm{i}\varphi})(1-\rho^2 e^{2\mathrm{i}\varphi})}{(1-\rho^2 e^{2\mathrm{i}\varphi})^2} = \frac{\rho^2 - e^{2\mathrm{i}\varphi}}{1-\rho^2 e^{2\mathrm{i}\varphi}} = -\left(\frac{1-\rho^2 e^{-2\mathrm{i}\varphi}}{1-\rho^2 e^{2\mathrm{i}\varphi}}\right)e^{2\mathrm{i}\varphi}. \tag{9.5}$$

When $\rho$ is real, the final expression on the right-hand side of the above equation will have a magnitude of 1.

As another example of application of the method, consider the bilayer slab depicted in Fig.10. Here, the individual slabs have Fresnel reflection and transmission coefficients $(r_1, t_1)$ and $(r_2, t_2)$, which we assume to be symmetric — that is, each layer has the same Fresnel coefficients whether the light shines from the left- or from the right-hand side. By introducing a gap $\Delta$ between the two layers, accounting for multiple bounces of the photon within the gap, and eventually allowing $\Delta$ to shrink to zero, we compute the overall Fresnel coefficients $(\rho, \tau)$ of the bilayer, as follows:

$$\rho = lim_{\Delta\to 0}\left(r_1 + t_1 r_2 e^{2\mathrm{i}k_z\Delta} t_1 + t_1 r_2 r_1 r_2 e^{4\mathrm{i}k_z\Delta} t_1 + \cdots\right) = r_1 + lim_{\Delta\to 0}\left(\frac{t_1^2 r_2 e^{2\mathrm{i}k_z\Delta}}{1 - r_1 r_2 e^{2\mathrm{i}k_z\Delta}}\right)$$
$$= \frac{r_1 - (r_1^2 - t_1^2) r_2}{1 - r_1 r_2}. \tag{9.6}$$

$$\tau = lim_{\Delta\to 0}\left(t_1 e^{\mathrm{i}k_z\Delta} t_2 + t_1 r_2 r_1 e^{3\mathrm{i}k_z\Delta} t_2 + \cdots\right) = lim_{\Delta\to 0}\left(\frac{t_1 t_2 e^{\mathrm{i}k_z\Delta}}{1 - r_1 r_2 e^{2\mathrm{i}k_z\Delta}}\right) = \frac{t_1 t_2}{1 - r_1 r_2}. \tag{9.7}$$

The expression for $\tau$ in Eq.(9.7) is symmetric, meaning that the transmissivity of the bilayer is the same whether the light shines from the left- or from the right-hand side. This is true for transparent as well as partially absorptive layers. One may also consider adding a third symmetric layer $(r_3, t_3)$ after the second layer, then use the same argument as above to show that the transmissivity of the trilayer stack is the same from both sides. In fact, any number of (symmetric) layers may be added to the stack, without affecting the equality of its transmissivity from the opposite sides.[8] This general property of multilayers comprised of symmetric layers is a manifestation of the reciprocity theorem discussed in Secs.6 and 7. Note that the reflectivities from the opposite sides of a bilayer (or, in general, a multilayer) are not necessarily identical. For instance, a bilayer consisting of a highly reflective layer and a highly absorptive layer will reflect strongly from one side and absorb strongly from the opposite side, even though its transmissivity will be the same from both sides.

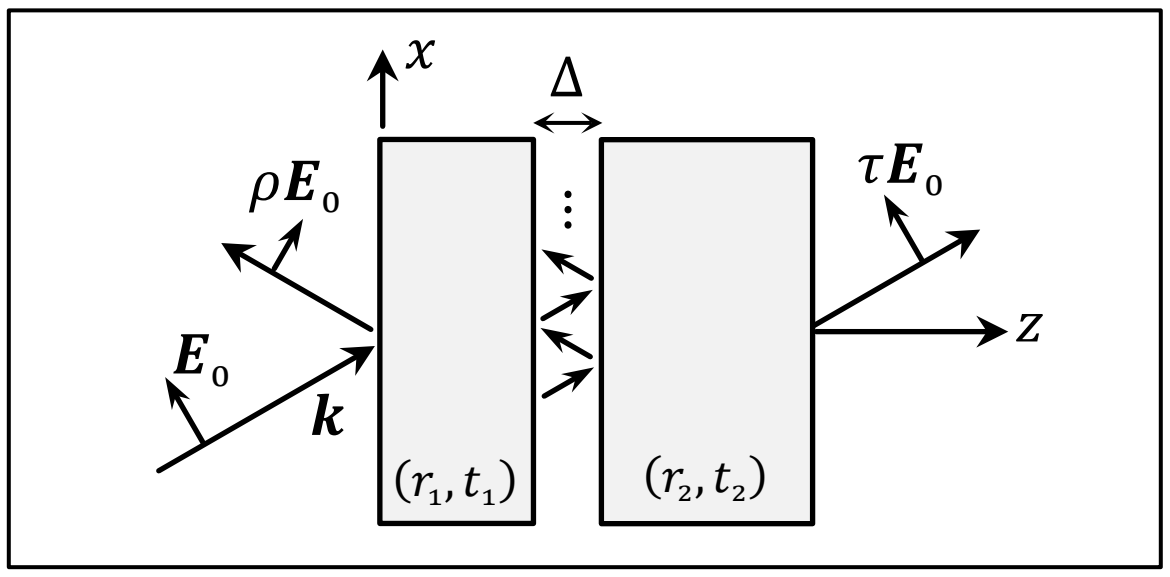

**Fig.10**. A slab having Fresnel reflection and transmission coefficients $(r_1, t_1)$ at the angle of incidence under consideration is separated by a small gap $\Delta$ from a second slab whose reflection and transmission coefficients are $(r_2, t_2)$. Both slabs are symmetric, meaning that their Fresnel coefficients are the same whether the light is incident from the right- or from the left-hand side — albeit at the same angle of incidence and with the same polarization state. By accounting for multiple reflections within the gap, then allowing the gap-width $\Delta$ to shrink to zero, we find expressions for the bilayer's reflection and transmission coefficients $(\rho, \tau)$ in terms of $(r_1, t_1)$ and $(r_2, t_2)$.

**10. Will there be interference when a photon scatters from a pair of small particles?** Shown in Fig.11 is a pair of small spherical particles separated from each other by a fixed distance $d$ along the $x$-axis. Also shown is a single RCP photon that propagates along the $z$-axis and gets scattered by the pair of particles. The particles are homogeneous and isotropic, having polarizability $\alpha$, and the scattering amplitudes for RCP and LCP photons arriving at a point $x$ within an observation plane located a distance $z_0$ from the particles are given by

$$E_{\text{out}}^{(\pm)} = \frac{\alpha k_0^2(\sin\theta \pm 1)}{8\pi\varepsilon_0}\left(\frac{e^{\mathrm{i}k_0 r_1}}{r_1} + \frac{e^{\mathrm{i}k_0 r_2}}{r_2}\right) E_{\text{in}}^{(+)}. \tag{10.1}$$

The distances $r_1$ and $r_2$ from the particles to the observation point differ by $\sim d\cos\theta$, which is small enough to be neglected where $r_1$ and $r_2$ appearing in the denominators in Eq.(10.1) are concerned, but needs to be taken into account in the phase-factors $e^{\mathrm{i}k_0 r_1}$ and $e^{\mathrm{i}k_0 r_2}$. Thus, the phase difference associated with the two paths to the observation point $(x, z_0)$ is $\sim e^{\mathrm{i}k_0 d\cos\theta}$. We conclude that, for an incident RCP photon, the probability amplitude for an RCP or LCP photon to arrive at the observation point should be

$$E_{\text{out}}^{(\pm)}/E_{\text{in}}^{(+)} \cong \frac{\alpha k_0^2 e^{\mathrm{i}k_0 r_0}}{4\pi\varepsilon_0 r_0}(\sin\theta \pm 1)\cos(\pi d x/\lambda_0 r_0). \tag{10.2}$$

At the center of the observation plane (i.e., in the vicinity of the $z$-axis), $\theta$ is close to 90°, which means that there is scant probability of receiving scattered LCP photons in this region. However, as one moves up or down along the $x$-axis, $\sin\theta$ is reduced, giving LCP photons a reasonable chance of being detected in regions where $|x|$ is large. Note that the conversion of an incident RCP photon into a scattered LCP photon entails a change of angular momentum, which should be accompanied by a transfer of an equal but opposite angular momentum to the scatterer.

An interesting question arises if the angular momenta of the small spherical particles could be monitored before and after the scattering. There does not appear to be any relevant uncertainty about the initial angular momenta of the particles that could prevent their accurate measurement following a scattering event. Consequently, the "which way" information gained by monitoring the angular momenta of the scatterers should, in principle, prevent the possibility of interference fringes forming at the observation plane.[19-23]

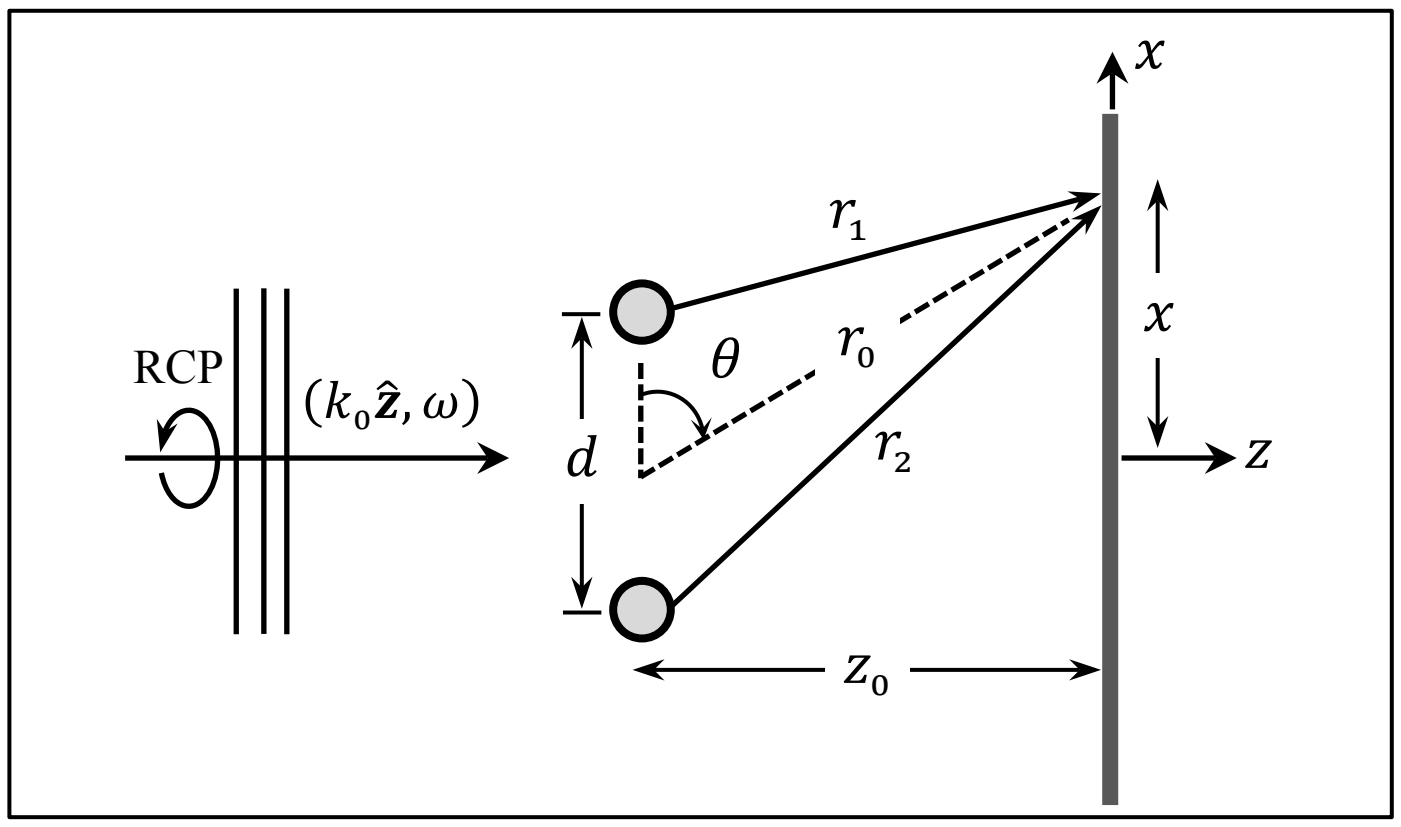

**Fig. 11**. Two small spherical particles at a fixed separation distance $d$ are suspended, say, in outer space, where gravitational as well as other forces are absent. A single RCP photon propagating along the $z$-axis is scattered from the particles and subsequently detected at a point $x$ in a distant observation plane. The uncertainties $\delta x, \delta y, \delta z$ about the positions of the particles are sufficiently small, and the corresponding momentum uncertainties $\delta\boldsymbol{p}$ sufficiently large, so that any measurement of the particles' linear momenta after the photon is scattered will not be able to identify the particle that scattered the photon. However, the angular momentum imparted by the scattered photon to either particle may be measureable, in which case the "which way" information acquired by measuring the particles' angular momenta should prevent the formation of interference fringes at the observation plane.

**11. Time-reversal symmetry**.[11] Let a thin sheet of a linear, homogeneous, isotropic material of thickness $d \ll \lambda_0$ and polarizability $\varepsilon_0\zeta(\omega) = \varepsilon_0|\zeta|e^{\mathrm{i}\varphi_\zeta}$ reside in the $xy$-plane of a Cartesian coordinate system, as shown in Fig.12. A circularly-polarized, normally incident plane-wave $E_{\text{in}}^{(\pm)} e^{\mathrm{i}(k_0 z - \omega t)}(\hat{\boldsymbol{x}} \pm \mathrm{i}\hat{\boldsymbol{y}})$ arrives from the left-hand side, producing the following polarization (i.e., electric dipole moment per unit volume) within the sheet:

$$\boldsymbol{P}(x, y, z = 0, t) = \varepsilon_0 \zeta(\omega) E_{\text{in}}^{(\pm)} e^{-\mathrm{i}\omega t}(\hat{\boldsymbol{x}} \pm \mathrm{i}\hat{\boldsymbol{y}}). \quad (11.1)$$

The excited sheet radiates two identical plane-waves, one propagating to the right along the $\hat{\boldsymbol{z}}$ direction, with the same sense of circular polarization as the incident wave, and a second one, propagating to the left (i.e., along $-\hat{\boldsymbol{z}}$) and having the opposite sense of polarization, as follows:[12]

$$\boldsymbol{E}_t^{(\pm)}(\boldsymbol{r}, t) = (\mathrm{i}\pi d/\lambda_0)\zeta(\omega) E_{\text{in}}^{(\pm)} e^{\mathrm{i}(k_0 z - \omega t)}(\hat{\boldsymbol{x}} \pm \mathrm{i}\hat{\boldsymbol{y}}), \quad (11.2)$$

$$\boldsymbol{E}_r^{(\mp)}(\boldsymbol{r}, t) = (\mathrm{i}\pi d/\lambda_0)\zeta(\omega) E_{\text{in}}^{(\pm)} e^{-\mathrm{i}(k_0 z + \omega t)}(\hat{\boldsymbol{x}} \pm \mathrm{i}\hat{\boldsymbol{y}}). \quad (11.3)$$

The subscripts $r$ and $t$ are used here to indicate the directions of reflection from and transmission through the sheet. The corresponding magnetic fields of the radiated plane-waves are readily found to be

$$\boldsymbol{H}_t^{(\pm)}(\boldsymbol{r}, t) = (\mathrm{i}\pi d/\lambda_0 Z_0)\zeta(\omega) E_{\text{in}}^{(\pm)} e^{\mathrm{i}(k_0 z - \omega t)}(\mp\mathrm{i}\hat{\boldsymbol{x}} + \hat{\boldsymbol{y}}), \quad (11.4)$$

$$\boldsymbol{H}_r^{(\mp)}(\boldsymbol{r}, t) = (\mathrm{i}\pi d/\lambda_0 Z_0)\zeta(\omega) E_{\text{in}}^{(\pm)} e^{-\mathrm{i}(k_0 z + \omega t)}(\pm\mathrm{i}\hat{\boldsymbol{x}} - \hat{\boldsymbol{y}}). \quad (11.5)$$

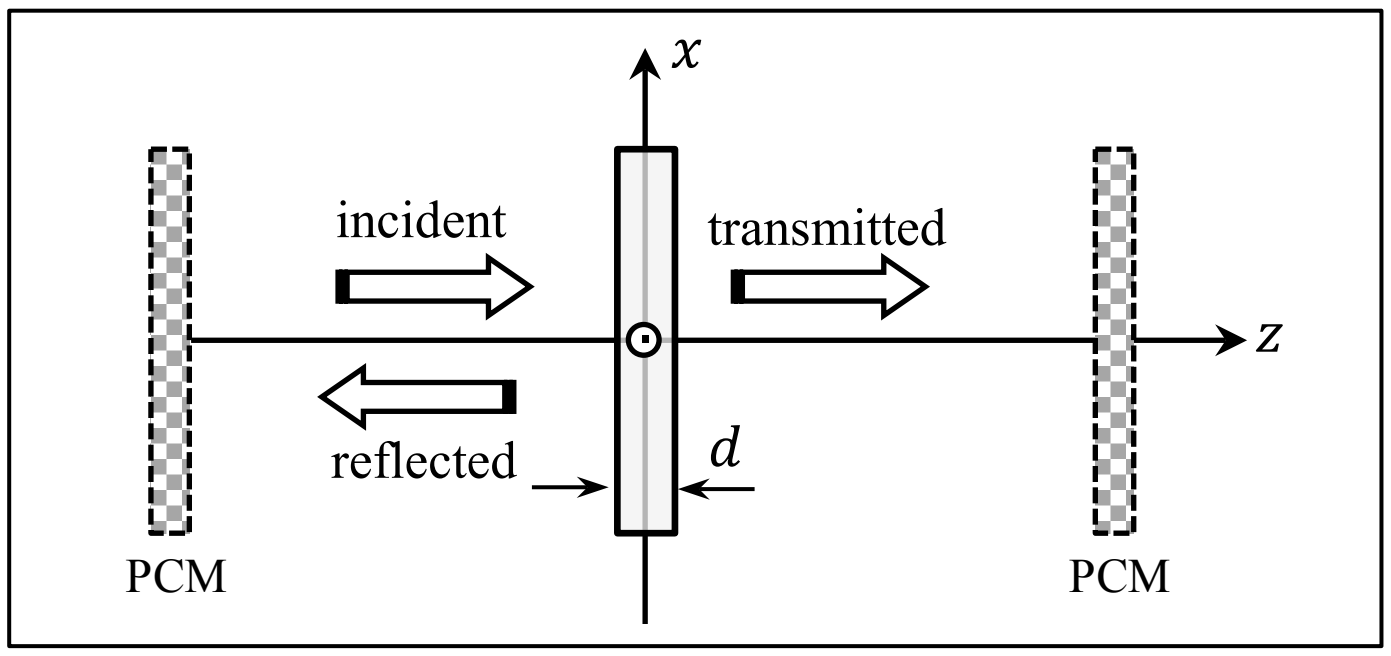

**Fig.12**. An electromagnetic plane-wave of frequency $\omega$ and vacuum wavelength $\lambda_0 = 2\pi c/\omega$ is normally incident on a thin sheet of linear, homogeneous, isotropic material of electrical polarizability $\varepsilon_0\zeta(\omega) = \varepsilon_0|\zeta|e^{\mathrm{i}\varphi_\zeta}$ and thickness $d \ll \lambda_0$. The incident beam may be either right- or left-circularly polarized (RCP or LCP). When the material medium of the sheet is non-absorptive, its electric susceptibility will be given by $\chi_e(\omega) = (\pi d/\lambda_0)^{-1}\tan(\varphi_\zeta)$. To describe the consequences of time-reversal symmetry for this system, a pair of phase-conjugate mirrors (PCMs)[24] are deployed that cause the reflected and transmitted beams to retrace their paths in the reverse direction.

Note that the radiated $E$-field is continuous at $z = 0$, whereas the radiated $H$-field has a discontinuity that matches the surface current density $\boldsymbol{J}_s(x, y, z = 0, t) = (\partial\boldsymbol{P}/\partial t)d$ of the sheet. (This matching of the boundary conditions at the sheet's surface is, in fact, all that is needed to justify the adopted expressions for the $E$-fields in Eqs.(11.2) and (11.3).) The time-averaged Poynting vectors on the opposite sides of the sheet are thus given by[11,12]

$$\langle\boldsymbol{S}_t(\boldsymbol{r}, t)\rangle = ½Re\left[\boldsymbol{E}_t^{(\pm)} \times \boldsymbol{H}_t^{(\pm)*}\right] = Z_0^{-1}(\pi d/\lambda_0)^2|\zeta(\omega)|^2|E_{\text{in}}^{(\pm)}|^2\hat{\boldsymbol{z}}, \quad (11.6)$$

$$\langle\boldsymbol{S}_r(\boldsymbol{r}, t)\rangle = ½Re\left[\boldsymbol{E}_r^{(\mp)} \times \boldsymbol{H}_r^{(\mp)*}\right] = -Z_0^{-1}(\pi d/\lambda_0)^2|\zeta(\omega)|^2|E_{\text{in}}^{(\pm)}|^2\hat{\boldsymbol{z}}. \quad (11.7)$$

The self $E$-field is given by either Eq.(11.2) or Eq.(11.3) at $z = 0$, and the rate of extraction of energy from the sheet (per unit area) is evaluated as follows:[12]

$$\tfrac{1}{2}Re(\boldsymbol{E}^*_{\text{self}} \cdot \partial\boldsymbol{P}/\partial t)d = \tfrac{1}{2}Re\big[(-\mathrm{i}\pi d/\lambda_0)\zeta^*(\omega)E_{\text{in}}^{(\pm)*}(\hat{\boldsymbol{x}} \mp \mathrm{i}\hat{\boldsymbol{y}}) \cdot (-\mathrm{i}\omega\varepsilon_0 d)\zeta(\omega)E_{\text{in}}^{(\pm)}(\hat{\boldsymbol{x}} \pm \mathrm{i}\hat{\boldsymbol{y}})\big]$$
$$= -2Z_0^{-1}(\pi d/\lambda_0)^2|\zeta(\omega)|^2|E_{\text{in}}^{(\pm)}|^2. \qquad (11.8)$$

The minus sign on the right-hand side of Eq.(11.8) merely indicates that the energy is being taken out of the dipoles — as opposed to going into the dipoles. The rate of extraction of energy from the sheet given by Eq.(11.8) matches the radiated energy flux on both sides of the sheet, as it must, in accordance with Eqs.(11.6) and (11.7). Assuming the sheet itself does not absorb any of the incident optical energy, the above rate of energy radiation must be the same as that taken from the incident plane-wave; that is,

$$\tfrac{1}{2}Re\big(\boldsymbol{E}_{\text{in}}^{(\pm)*} \cdot \partial\boldsymbol{P}/\partial t\big)d = \tfrac{1}{2}Re\big[E_{\text{in}}^{(\pm)*}(\hat{\boldsymbol{x}} \mp \mathrm{i}\hat{\boldsymbol{y}}) \cdot (-\mathrm{i}\omega\varepsilon_0 d)\zeta(\omega)E_{\text{in}}^{(\pm)}(\hat{\boldsymbol{x}} \pm \mathrm{i}\hat{\boldsymbol{y}})\big]$$
$$= (\varepsilon_0\omega d)|E_{\text{in}}^{(\pm)}|^2 Im[\zeta(\omega)] = 2Z_0^{-1}(\pi d/\lambda_0)|E_{\text{in}}^{(\pm)}|^2|\zeta(\omega)|\sin(\varphi_\zeta). \quad (11.9)$$

It is thus seen that $(\pi d/\lambda_0)|\zeta(\omega)| = \sin(\varphi_\zeta)$ is a necessary constraint on $\zeta(\omega)$ in cases where the sheet material itself is non-absorptive. The total $E$-field within the sheet is the sum of the incident plane-wave and the self $E$-field (responsible for radiation resistance), namely,

$$\boldsymbol{E}_{\text{total}}^{(\pm)}(x,y,z=0,t) = [1 + (\mathrm{i}\pi d/\lambda_0)\zeta(\omega)]E_{\text{in}}^{(\pm)}e^{-\mathrm{i}\omega t}(\hat{\boldsymbol{x}} \pm \mathrm{i}\hat{\boldsymbol{y}})$$
$$= \big[e^{\mathrm{i}\varphi_\zeta}e^{-\mathrm{i}\varphi_\zeta} + \mathrm{i}\sin(\varphi_\zeta)\,e^{\mathrm{i}\varphi_\zeta}\big]E_{\text{in}}^{(\pm)}e^{-\mathrm{i}\omega t}(\hat{\boldsymbol{x}} \pm \mathrm{i}\hat{\boldsymbol{y}})$$
$$= \cos(\varphi_\zeta)\,E_{\text{in}}^{(\pm)}e^{\mathrm{i}(\varphi_\zeta - \omega t)}(\hat{\boldsymbol{x}} \pm \mathrm{i}\hat{\boldsymbol{y}}). \qquad (11.10)$$

This total $E$-field is in-phase with $\boldsymbol{P}(x,y,z=0,t)$ given by Eq.(11.1), indicating that the material medium's susceptibility $\varepsilon_0\chi_e(\omega)$ is real-valued; that is, $\chi_e(\omega) = |\zeta(\omega)|/\cos(\varphi_\zeta)$. Given that $\varphi_\zeta$ is restricted to the $(0,\pi)$ interval, its cosine can be positive or negative, which indicates that $\chi_e$ can be positive or negative as well. In other words, our non-absorptive sheet could be made of a transparent dielectric or of a lossless metal.

The transmitted field on the right-hand side of the sheet is the sum of the incident plane-wave (which continues to propagate unhindered beyond the sheet) and the radiated wave; that is,

$$\boldsymbol{E}_{\text{trans}}^{(\pm)}(\boldsymbol{r},t) = [1 + (\mathrm{i}\pi d/\lambda_0)\zeta(\omega)]E_{\text{in}}^{(\pm)}e^{\mathrm{i}(k_0 z - \omega t)}(\hat{\boldsymbol{x}} \pm \mathrm{i}\hat{\boldsymbol{y}}) = \cos(\varphi_\zeta)\,E_{\text{in}}^{(\pm)}e^{\mathrm{i}(k_0 z - \omega t + \varphi_\zeta)}(\hat{\boldsymbol{x}} \pm \mathrm{i}\hat{\boldsymbol{y}}). \qquad (11.11)$$

The reflected $E$-field is given by Eq.(11.3), which may equivalently be written as

$$\boldsymbol{E}_{\text{ref}}^{(\mp)}(\boldsymbol{r},t) = \sin(\varphi_\zeta)\,E_{\text{in}}^{(\pm)}e^{-\mathrm{i}(k_0 z + \omega t - \varphi_\zeta - \frac{1}{2}\pi)}(\hat{\boldsymbol{x}} \pm \mathrm{i}\hat{\boldsymbol{y}}). \qquad (11.12)$$

The sheet's reflectance and transmittance are now seen to be given by $\sin^2(\varphi_\zeta)$ and $\cos^2(\varphi_\zeta)$, confirming that no fraction of the incident optical power is absorbed within the host medium; this is consistent with the fact that the susceptibility of the medium is assumed at the outset to be real-valued (positive or negative). An alternative way of expressing the susceptibility $\chi_e$ is

$$\chi_e(\omega) = (\pi d/\lambda_0)^{-1}\tan(\varphi_\zeta). \qquad (11.13)$$

As before, $\chi_e$ is seen to be positive if $\varphi_\zeta$ belongs to $(0, ½\pi)$, and negative if it belongs to $(½\pi, \pi)$. We may now demonstrate the time-reversal symmetry of the system by arranging for both the reflected and transmitted beams to bounce off a pair of distant phase-conjugate mirrors[24] (as shown in Fig.12) and return to the sheet located at $z = 0$. The returning plane-waves, whose $E$-fields are denoted by $\tilde{\boldsymbol{E}}_{\text{trans}}$ and $\tilde{\boldsymbol{E}}_{\text{ref}}$, are obtained from Eqs.(11.11) and (11.12) by conjugating every one of their terms except for $e^{-\mathrm{i}\omega t}$, as follows:[8]

$$\tilde{\boldsymbol{E}}_{\text{trans}}^{(\pm)}(\boldsymbol{r}, t) = \cos(\varphi_\zeta)\, E_{\text{in}}^{(\pm)*} e^{-\mathrm{i}(k_0 z + \varphi_\zeta)} e^{-\mathrm{i}\omega t} (\hat{\boldsymbol{x}} \mp \mathrm{i}\hat{\boldsymbol{y}}), \tag{11.14}$$

$$\tilde{\boldsymbol{E}}_{\text{ref}}^{(\mp)}(\boldsymbol{r}, t) = \sin(\varphi_\zeta)\, E_{\text{in}}^{(\pm)*} e^{\mathrm{i}(k_0 z - \varphi_\zeta - ½\pi)} e^{-\mathrm{i}\omega t} (\hat{\boldsymbol{x}} \mp \mathrm{i}\hat{\boldsymbol{y}}). \tag{11.15}$$

Upon arriving at the sheet, $\tilde{\boldsymbol{E}}_{\text{trans}}$ will get multiplied by $\cos(\varphi_\zeta)\, e^{\mathrm{i}\varphi_\zeta}$ and pass through to the other side, while $\tilde{\boldsymbol{E}}_{\text{ref}}$ gets multiplied by $\sin(\varphi_\zeta)\, e^{\mathrm{i}(\varphi_\zeta + ½\pi)}$ and bounces back (with its $k$-vector reversed). These two beams combine to yield $E_{\text{in}}^{(\pm)*} e^{-\mathrm{i}(k_0 z + \omega t)} (\hat{\boldsymbol{x}} \mp \mathrm{i}\hat{\boldsymbol{y}})$, which is the phase-conjugated version of the original incident beam (i.e., $\tilde{\boldsymbol{E}}_{\text{in}}^{(\pm)}$) that now reverse-propagates in the half-space to the left of the sheet. On the right-hand side, $\tilde{\boldsymbol{E}}_{\text{trans}}$ bounces back with its $k$-vector reversed and its amplitude multiplied by $\sin(\varphi_\zeta)\, e^{\mathrm{i}(\varphi_\zeta + ½\pi)}$, while $\tilde{\boldsymbol{E}}_{\text{ref}}$ passes through the sheet and gets multiplied by $\cos(\varphi_\zeta)\, e^{\mathrm{i}\varphi_\zeta}$; these two beams then exactly cancel out.

The end result is that, upon returning to the sheet, the phase-conjugated beams ($\tilde{\boldsymbol{E}}_{\text{trans}}$ and $\tilde{\boldsymbol{E}}_{\text{ref}}$) reconstitute the original incident beam, albeit phase-conjugated and propagating backward, on the left-hand side. The half-space to the right of the sheet remains empty, as was the case prior to the arrival of the original incident beam. This has been an elementary demonstration of time-reversal symmetry for the system of Fig.12, a well-known general feature of EM waves in the presence of material media whose electric and magnetic susceptibility tensors $\tilde{\chi}_e$ and $\tilde{\chi}_m$ are real-valued.[11]

Returning to the analysis of the sheet material's electric susceptibility in the more general case when the medium is partially absorptive, it is necessary to have $(\pi d/\lambda_0)|\zeta(\omega)| < \sin(\varphi_\zeta)$ to ensure that the dipoles absorb more energy than they radiate. The susceptibility, obtained by equating $\varepsilon_0 \chi_e(\omega) \boldsymbol{E}_{\text{total}}^{(\pm)}$ with $\varepsilon_0 \zeta(\omega) \boldsymbol{E}_{\text{in}}^{(\pm)}$, now acquires a positive imaginary part, as seen below.

$$\chi_e(\omega) = \frac{\zeta(\omega)}{1 + (\mathrm{i}\pi d/\lambda_0)\zeta(\omega)} = \frac{[1 - (\mathrm{i}\pi d/\lambda_0)\zeta^*]\zeta}{[1 + (\mathrm{i}\pi d/\lambda_0)\zeta][1 - (\mathrm{i}\pi d/\lambda_0)\zeta^*]} = \frac{|\zeta|\{\cos(\varphi_\zeta) + \mathrm{i}[\sin(\varphi_\zeta) - (\pi d/\lambda_0)|\zeta|]\}}{1 + (\pi d/\lambda_0)^2|\zeta|^2 - (2\pi d/\lambda_0)|\zeta| \sin(\varphi_\zeta)}. \tag{11.16}$$

While the presence of an imaginary part in $\chi_e$ breaks down the argument for the time-reversal symmetry of the system, it does *not* violate the requirements of the reciprocity theorem. Reciprocity continues to hold for the system of Fig.12, irrespective of whether $\chi_e(\omega)$ is real or complex, although its prediction (namely, that the transmissivity of the sheet is the same whether the incident beam arrives from the left- or from the right-hand side) is rather trivial in this case.

**12. The Ewald-Oseen Extinction Theorem.**[7,8] We saw in the preceding section that the single-photon approach to analyzing certain optical systems closely parallels their classical Maxwellian treatment based on the system's response to EM plane-waves. The present section provides yet another example where the standard classical analysis of a problem could stand for its single-photon treatment. Figure 13 shows a plane-wave of amplitude $\boldsymbol{E}_{\text{in}}$ arriving at normal incidence and being partially reflected at the front facet of a semi-infinite medium of refractive index $n$. The incidence medium being free space, the Fresnel reflection and transmission coefficients at the front facet are known to be $\rho = (1 - n)/(1 + n)$ and $\tau = 2/(1 + n)$, respectively.[7,11,12] Thus, the $E$-field inside a thin sheet (thickness $= d$) parallel to the $xy$-plane inside the medium at some distance $z$ from the surface is $\tau \boldsymbol{E}_{\text{in}} e^{\mathrm{i}(nk_0 z-\omega t)}$. According to Eqs.(11.2), (11.3) and (11.10), the ratio of the forward as well as backward radiated $E$-fields by the sheet to the total $E$-field inside the sheet is $(\mathrm{i}\pi d/\lambda_0)\zeta(\omega)/[1+(\mathrm{i}\pi d/\lambda_0)\zeta(\omega)]$, which equals $(\mathrm{i}\pi d/\lambda_0)\chi_e(\omega)$ in accordance with Eq.(11.16). Considering that $\chi_e(\omega)=n^2-1$ and writing $\mathrm{d}z$ for the sheet thickness $d$, we may integrate the backward radiated $E$-fields that return to the front facet of the semi-infinite medium at $z=0$ to find

$$\int_{z=0}^{\infty} \tau \boldsymbol{E}_{\text{in}} e^{\mathrm{i}(nk_0 z-\omega t)}(\mathrm{i}\pi/\lambda_0)\chi_e(\omega) e^{\mathrm{i}k_0 z}\mathrm{d}z = \tfrac{1}{2}\mathrm{i}k_0\chi_e(\omega)\tau \boldsymbol{E}_{\text{in}} e^{-\mathrm{i}\omega t}\int_{z=0}^{\infty} e^{\mathrm{i}(n+1)k_0 z}\mathrm{d}z$$

$$= -\frac{\tfrac{1}{2}\mathrm{i}k_0\chi_e(\omega)\tau}{\mathrm{i}(n+1)k_0}\boldsymbol{E}_{\text{in}} e^{-\mathrm{i}\omega t} = -\frac{n^2-1}{(n+1)^2}\boldsymbol{E}_{\text{in}} e^{-\mathrm{i}\omega t} = \left(\frac{1-n}{1+n}\right)\boldsymbol{E}_{\text{in}} e^{-\mathrm{i}\omega t}. \qquad (12.1)$$

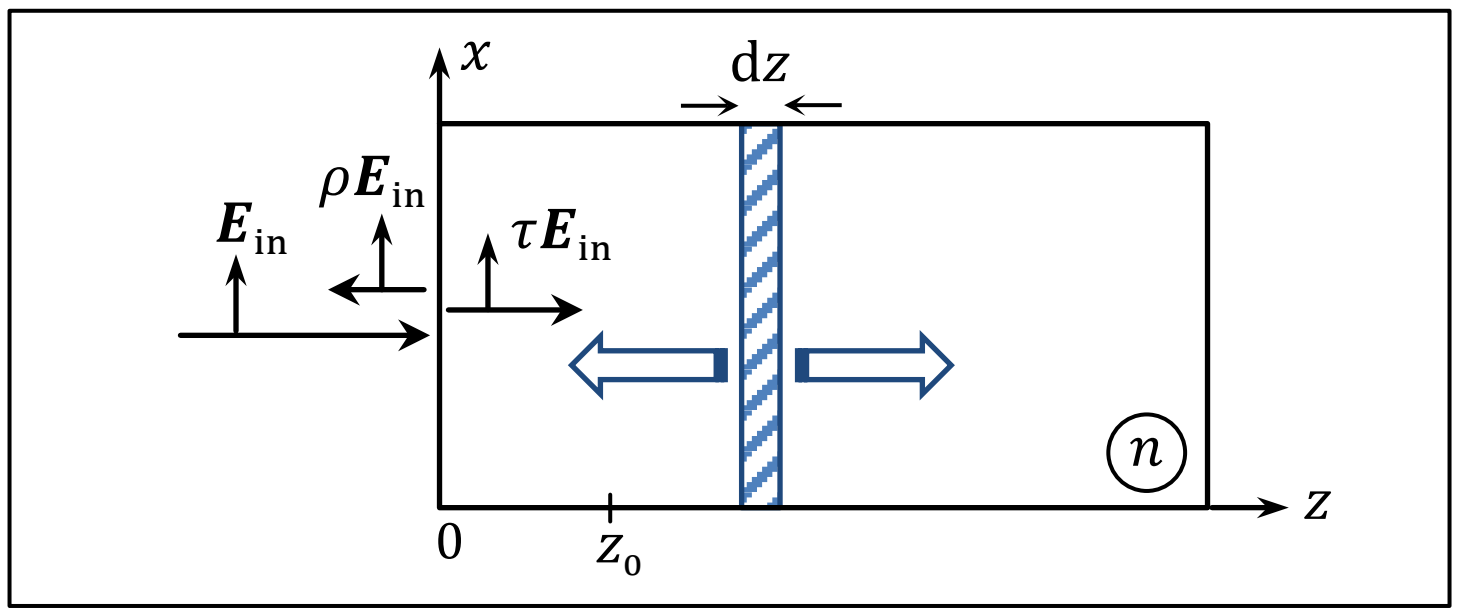

**Fig.13**. A plane-wave having $E$-field amplitude $\boldsymbol{E}_{\text{in}}$ arrives at the surface of a semi-infinite medium of refractive index $n$ at normal incidence. The transmitted wave $\tau \boldsymbol{E}_{\text{in}} e^{\mathrm{i}(nk_0 z-\omega t)}$ excites the various slices of the material medium, which proceed to radiate in both forward and backward directions. The totality of backward radiated waves constitutes the reflected beam at the front facet of the medium. Similarly, the superposition of forward radiated waves from $z=0$ to $z_0$, backward radiated waves from $z=z_0$ to $\infty$, and the incident beam (which continues to propagate unhindered through the medium) constitutes the wave that is the transmitted beam at $z=z_0$.

It is seen that the Fresnel reflection coefficient $\rho$ emerges from the above argument by adding up all the backward radiated fields that originate throughout the semi-infinite medium and return to the front facet at $z=0$ upon propagating in free space. A similar argument for the $E$-field inside the medium at $z=z_0$ yields

$$\int_{z=0}^{z_0} \tau \boldsymbol{E}_{\text{in}} e^{\mathrm{i}(nk_0 z-\omega t)}(\mathrm{i}\pi/\lambda_0)\chi_e(\omega) e^{\mathrm{i}k_0(z_0-z)}\mathrm{d}z + \int_{z=z_0}^{\infty} \tau \boldsymbol{E}_{\text{in}} e^{\mathrm{i}(nk_0 z-\omega t)}(\mathrm{i}\pi/\lambda_0)\chi_e(\omega) e^{\mathrm{i}k_0(z-z_0)}\mathrm{d}z$$

$$= \tfrac{1}{2}\mathrm{i}k_0\chi_e(\omega)\tau\left[e^{\mathrm{i}k_0 z_0}\int_0^{z_0} e^{\mathrm{i}(n-1)k_0 z}\mathrm{d}z + e^{-\mathrm{i}k_0 z_0}\int_{z_0}^{\infty} e^{\mathrm{i}(n+1)k_0 z}\mathrm{d}z\right]\boldsymbol{E}_{\text{in}} e^{-\mathrm{i}\omega t}$$

$$= \frac{\mathrm{i}k_0(n^2-1)}{n+1}\left\{\frac{e^{\mathrm{i}k_0 z_0}}{\mathrm{i}(n-1)k_0}\left[e^{\mathrm{i}(n-1)k_0 z_0}-1\right] - \frac{e^{-\mathrm{i}k_0 z_0}}{\mathrm{i}(n+1)k_0} e^{\mathrm{i}(n+1)k_0 z_0}\right\}\boldsymbol{E}_{\text{in}} e^{-\mathrm{i}\omega t}$$

$$= \left[\left(1-\frac{n-1}{n+1}\right)e^{\mathrm{i}nk_0 z_0} - e^{\mathrm{i}k_0 z}\right]\boldsymbol{E}_{\text{in}} e^{-\mathrm{i}\omega t} = \left[\left(\frac{2}{n+1}\right)e^{\mathrm{i}(nk_0 z_0-\omega t)} - e^{\mathrm{i}(k_0 z-\omega t)}\right]\boldsymbol{E}_{\text{in}}. \qquad (12.2)$$

On the right-hand side of the above equation, the first term is the expected $E$-field at $z=z_0$, whereas the second term must be cancelled by the incident $E$-field. In these calculations, we see the essence of the Ewald-Oseen extinction theorem,[7,8] since the reflected beam is seen to have been produced by the radiated waves (propagating backward in free space) from all those thin sheets within the medium, while the transmitted beam is the superposition of the incident beam and the radiated waves (propagating forward from $z=0$ to $z_0$, and backward from $z=z_0$ to $\infty$) from all the excited thin sheets.

**13. The Optical Theorem**. The scattering amplitudes of single photons from electric and magnetic dipoles were derived in Secs.6 and 7. We now use the results of these earlier sections to relate the scattering cross-section of a collection of such dipoles (e.g., in the form of a material object) to their collective scattering amplitude along the direction of the incident light — also known as the forward scattering amplitude.[7,11] Let an RCP or LCP plane-wave of frequency $\omega$ propagating along the $z$-axis illuminate an electric point-dipole $\boldsymbol{p}e^{-\mathrm{i}\omega t}$ located at $(x_d, y_d, z_d)$, as shown in Fig.14. The time-averaged rate of delivery of EM energy to the dipole by either an RCP or an LCP wave is given by

$$\begin{aligned}\frac{\partial\mathcal{E}}{\partial t} &= \tfrac{1}{2}Re\left\{\left[E_{\text{in}}^{(\pm)}e^{\mathrm{i}(k_0 z_d-\omega t)}(\hat{\boldsymbol{x}}\pm\mathrm{i}\hat{\boldsymbol{y}})\right]^*\cdot\frac{\partial}{\partial t}(\boldsymbol{p}e^{-\mathrm{i}\omega t})\right\}\\ &= \tfrac{1}{2}Re\left[E_{\text{in}}^{(\pm)*}e^{-\mathrm{i}k_0 z_d}(\hat{\boldsymbol{x}}\mp\mathrm{i}\hat{\boldsymbol{y}})\cdot(-\mathrm{i}\omega\boldsymbol{p})\right] = \tfrac{1}{2}\omega\, Im\left[E_{\text{in}}^{(\pm)*}e^{-\mathrm{i}k_0 z_d}(p_x\mp\mathrm{i}p_y)\right]. \qquad (13.1)\end{aligned}$$

Thus, when the incident beam is a superposition of RCP and LCP plane-waves, the energy delivery rate may be written as

$$\frac{\partial\mathcal{E}}{\partial t} = \tfrac{1}{4}\omega\, Im\left\{e^{-\mathrm{i}k_0 z_d}\left[E_{\text{in}}^{(+)}(\hat{\boldsymbol{x}}+\mathrm{i}\hat{\boldsymbol{y}})+E_{\text{in}}^{(-)}(\hat{\boldsymbol{x}}-\mathrm{i}\hat{\boldsymbol{y}})\right]^*\cdot\left[(p_x-\mathrm{i}p_y)(\hat{\boldsymbol{x}}+\mathrm{i}\hat{\boldsymbol{y}})+(p_x+\mathrm{i}p_y)(\hat{\boldsymbol{x}}-\mathrm{i}\hat{\boldsymbol{y}})\right]\right\}. \qquad (13.2)$$

As for the incident beam's energy flux per unit cross-sectional area, its time-averaged Poynting vector is given by[7,11,12]

$$\begin{aligned}\langle\boldsymbol{S}_{\text{in}}(\boldsymbol{r},t)\rangle &= \tfrac{1}{2}Re\left[E_{\text{in}}^{(+)}(\hat{\boldsymbol{x}}+\mathrm{i}\hat{\boldsymbol{y}})+E_{\text{in}}^{(-)}(\hat{\boldsymbol{x}}-\mathrm{i}\hat{\boldsymbol{y}})\right]\times\overbrace{Z_0^{-1}\left[E_{\text{in}}^{(+)}(-\mathrm{i}\hat{\boldsymbol{x}}+\hat{\boldsymbol{y}})+E_{\text{in}}^{(-)}(\mathrm{i}\hat{\boldsymbol{x}}+\hat{\boldsymbol{y}})\right]^*}^{\boldsymbol{H}_{\text{in}}^*}\\ &= \tfrac{1}{2}Z_0^{-1}Re\Big[\overbrace{\left(E_{\text{in}}^{(+)}+E_{\text{in}}^{(-)}\right)}^{E_x}\hat{\boldsymbol{x}}+\mathrm{i}\overbrace{\left(E_{\text{in}}^{(+)}-E_{\text{in}}^{(-)}\right)}^{E_y}\hat{\boldsymbol{y}}\Big]\times\left[\mathrm{i}\left(E_{\text{in}}^{(+)*}-E_{\text{in}}^{(-)*}\right)\hat{\boldsymbol{x}}+\left(E_{\text{in}}^{(+)*}+E_{\text{in}}^{(-)*}\right)\hat{\boldsymbol{y}}\right]\\ &= \tfrac{1}{2}Z_0^{-1}Re\left[\left(E_{\text{in}}^{(+)}+E_{\text{in}}^{(-)}\right)\left(E_{\text{in}}^{(+)*}+E_{\text{in}}^{(-)*}\right)+\left(E_{\text{in}}^{(+)}-E_{\text{in}}^{(-)}\right)\left(E_{\text{in}}^{(+)*}-E_{\text{in}}^{(-)*}\right)\right]\hat{\boldsymbol{z}}\\ &= Z_0^{-1}\left(\left|E_{\text{in}}^{(+)}\right|^2+\left|E_{\text{in}}^{(-)}\right|^2\right)\hat{\boldsymbol{z}}. \qquad (13.3)^{**}\end{aligned}$$

Normalizing the energy delivery rate $\partial\mathcal{E}/\partial t$ of Eq.(13.2) by the incident energy flux per unit area $\langle S_{\text{in}}(\boldsymbol{r},t)\rangle$ given by Eq.(13.3) now yields the dipole's scattering cross-section $\mathcal{A}$, as follows:[7,11]

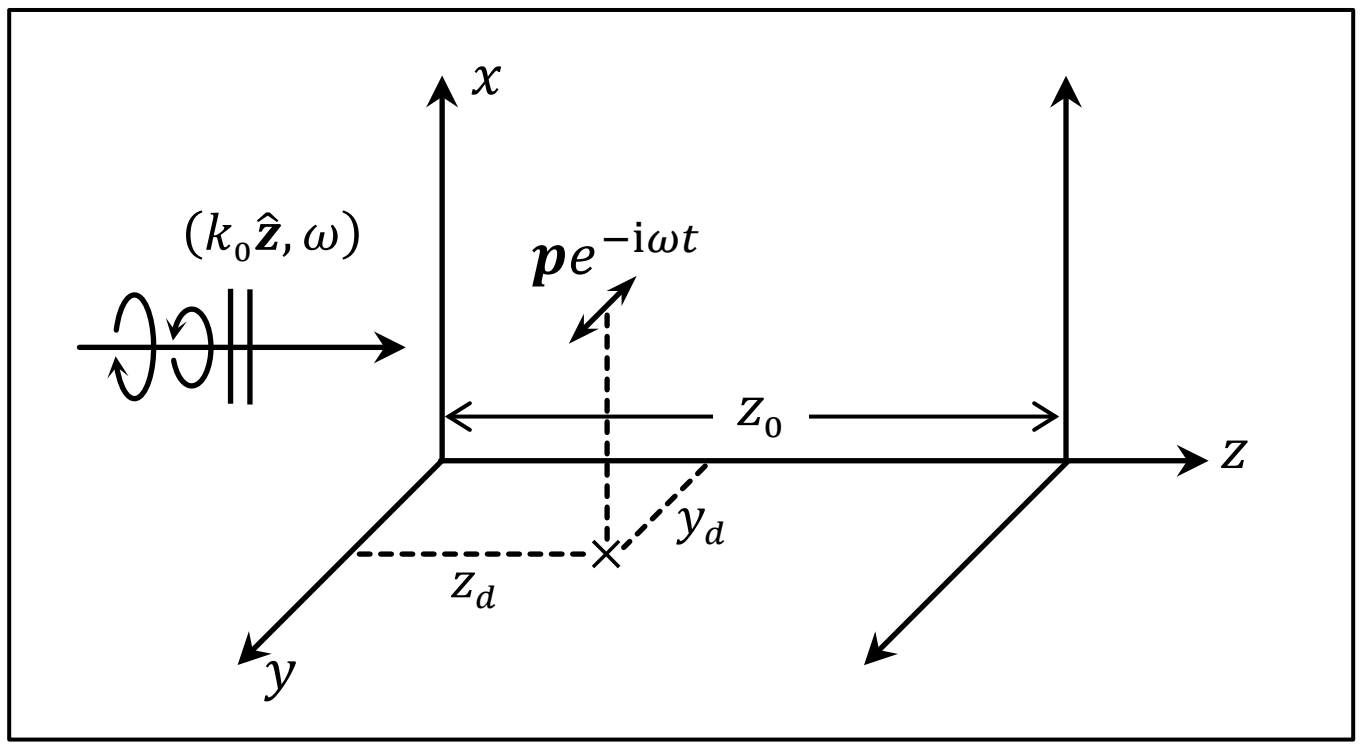

**Fig.14**. A plane-wave of frequency $\omega$, whose polarization state is a superposition of RCP and LCP, propagates along the $z$-axis. A point dipole $\boldsymbol{p}e^{-\mathrm{i}\omega t}$ located at $(x_d, y_d, z_d)$ extracts energy from the incident wave, which energy is either absorbed locally or re-emitted as scattered light.

---

** Given that $E_x = E_{\text{in}}^{(+)} + E_{\text{in}}^{(-)}$ and $E_y = \mathrm{i}E_{\text{in}}^{(+)} - \mathrm{i}E_{\text{in}}^{(-)}$, it is readily seen that $|E_{\text{in}}^{(+)}|^2 + |E_{\text{in}}^{(-)}|^2 = \tfrac{1}{2}(|E_x|^2 + |E_y|^2)$.

$$\mathcal{A} = \tfrac{1}{4} Z_0 \omega\, Im\Big\{\big[E_{\text{in}}^{(+)}(\hat{\boldsymbol{x}} + \mathrm{i}\hat{\boldsymbol{y}}) + E_{\text{in}}^{(-)}(\hat{\boldsymbol{x}} - \mathrm{i}\hat{\boldsymbol{y}})\big]^*$$

$$\cdot\, e^{-\mathrm{i}k_0 z_d}\big[(p_x - \mathrm{i}p_y)(\hat{\boldsymbol{x}} + \mathrm{i}\hat{\boldsymbol{y}}) + (p_x + \mathrm{i}p_y)(\hat{\boldsymbol{x}} - \mathrm{i}\hat{\boldsymbol{y}})\big]\Big\} \Big/ \Big(\big|E_{\text{in}}^{(+)}\big|^2 + \big|E_{\text{in}}^{(-)}\big|^2\Big). \quad (13.4)$$

Now, the amplitudes of the RCP and LCP photons emitted by the oscillating dipole $\boldsymbol{p}e^{-\mathrm{i}\omega t}$ in the forward direction (i.e., $\theta_{\text{out}} = \tfrac{1}{2}\pi, \varphi_{\text{out}} = \tfrac{1}{2}\pi$) are given by Eq.(6.6) as

**Note**: The axes in Fig.6 are labeled differently than those in Fig.14.

$$E_{\text{out}}^{(\pm)} = \frac{k_0^2}{8\pi\varepsilon_0}(p_x \mp \mathrm{i}p_y). \quad (13.5)$$

Thus, upon propagating the distance $z_0 - z_d$ along $\hat{\boldsymbol{z}}$, the forward-scattered $E$-field at the destination plane (located at $z = z_0$) will be

$$\boldsymbol{E}_s(x, y, z_0)e^{-\mathrm{i}\omega t} = \frac{k_0^2}{8\pi\varepsilon_0(z_0 - z_d)} e^{\mathrm{i}[k_0(z_0 - z_d) - \omega t]}\big[(p_x - \mathrm{i}p_y)(\hat{\boldsymbol{x}} + \mathrm{i}\hat{\boldsymbol{y}}) + (p_x + \mathrm{i}p_y)(\hat{\boldsymbol{x}} - \mathrm{i}\hat{\boldsymbol{y}})\big]. \quad (13.6)$$

Substitution into Eq.(13.4) now yields

$$\mathcal{A} = (2\pi/k_0)\, Im\big[\boldsymbol{E}_{\text{in}}^* \cdot (z_0 - z_d)e^{-\mathrm{i}k_0 z_0}\boldsymbol{E}_s(x, y, z_0)\big] \Big/ \Big(\big|E_{\text{in}}^{(+)}\big|^2 + \big|E_{\text{in}}^{(-)}\big|^2\Big). \quad (13.7)$$

In the case of a magnetic dipole $\boldsymbol{m}e^{-\mathrm{i}\omega t}$, the incident $H$-field of an RCP or LCP plane-wave, namely, $\boldsymbol{H}_{\text{in}}^{(\pm)} = Z_0^{-1}E_{\text{in}}^{(\pm)}e^{\mathrm{i}(k_0 z_d - \omega t)}(\mp\mathrm{i}\hat{\boldsymbol{x}} + \hat{\boldsymbol{y}})$, delivers energy to the dipole at the following rate:

$$\frac{\partial \mathcal{E}}{\partial t} = \tfrac{1}{2} Re\Big[Z_0^{-1}E_{\text{in}}^{(\pm)*}e^{-\mathrm{i}(k_0 z_d - \omega t)}(\pm\mathrm{i}\hat{\boldsymbol{x}} + \hat{\boldsymbol{y}}) \cdot \frac{\partial}{\partial t}(\boldsymbol{m}e^{-\mathrm{i}\omega t})\Big] = \tfrac{1}{2} Z_0^{-1}\omega\, Im\big[E_{\text{in}}^{(\pm)*}e^{-\mathrm{i}k_0 z_d}(\pm\mathrm{i}m_x + m_y)\big]. \quad (13.8)$$

According to Eq.(7.4), the radiated $E$-field by this oscillating dipole is

**Note**: The axes in Fig.6 are labeled differently than those in Fig.14.

$$E_{\text{out}}^{(\pm)} = \frac{ck_0^2}{8\pi}(\pm\mathrm{i}m_x + m_y). \quad (13.9)$$

Consequently, the preceding arguments for an electric dipole $\boldsymbol{p}$ apply to a magnetic dipole $\boldsymbol{m}$ as well, and the formula relating the scattering cross-section $\mathcal{A}$ to the scattered $E$-field $\boldsymbol{E}_s$ given in Eq.(13.7) becomes equally applicable to both types of dipole. We emphasize that the EM energy delivered by the incident beam to either dipole (in accordance with Eq.(13.1) or Eq.(13.8)) may be absorbed within the dipole or radiated out in the form of scattered EM energy. Thus, the scattering cross-section $\mathcal{A}$ of Eq.(13.7) is a measure of energy loss from the incident beam, be it in the form of absorption, scattering, or any combination thereof.

If there happens to be a number of dipoles in the system, irrespective of how many are electric and how many magnetic, and regardless of their individual coordinates $(x_d, y_d, z_d)$, Eq.(13.7) will continue to hold. The overall scattering cross-section then becomes the sum of the cross-sections of all the individual dipoles, while the forward scattered $E$-fields $\boldsymbol{E}_s$ produced by the various dipoles are simply added together. The only caveat is that the coefficient $(z_0 - z_d)$ appearing in Eq.(13.7) must be replaced by some sort of average distance between the various dipoles and the observation plane located at $z = z_0$. When the dipoles collectively constitute an object sitting in the vicinity of the origin at $(x, y, z) = (0,0,0)$ while the observation plane is in the far field of the object, one may use the approximation $z_0 - z_d \cong z_0$ and proceed to define the object's overall scattering amplitude $\boldsymbol{F}(\boldsymbol{k}_{\text{out}} = k_0\hat{\boldsymbol{z}}; \boldsymbol{k}_{\text{in}} = k_0\hat{\boldsymbol{z}})$ as follows:

$$\boldsymbol{E}_s(x, y, z_0) = \frac{\exp(\mathrm{i}k_0 z_0)}{z_0}\boldsymbol{F}(\boldsymbol{k}_{\text{out}} = k_0\hat{\boldsymbol{z}}; \boldsymbol{k}_{\text{in}} = k_0\hat{\boldsymbol{z}}). \quad (13.10)$$

Substitution into Eq.(13.7) now yields

$$\mathcal{A} \cong (2\pi/k_0)\, Im[\boldsymbol{E}_{\text{in}}^* \cdot \boldsymbol{F}(\boldsymbol{k}_{\text{out}} = k_0\hat{\boldsymbol{z}}; \boldsymbol{k}_{\text{in}} = k_0\hat{\boldsymbol{z}})]/\left(\left|E_{\text{in}}^{(+)}\right|^2 + \left|E_{\text{in}}^{(-)}\right|^2\right). \tag{13.11}$$

This is the well-known optical theorem of classical electrodynamics,[25] according to which the scattering cross-section $\mathcal{A}$ of an object is directly relatable to the far field forward scattering amplitude $\boldsymbol{F}$ along the direction of an incident plane-wave.[7,11,25,26] Note that, in the above arguments, the radiated $E$-field emanating from the various dipoles in the system does not need to be taken into account when calculating the absorbed/scattered energy by other dipoles, since the energy that they deliver to or extract from a dipole has already been accounted for at the dipole that is the source of that radiation.

**14. Scalar and vector diffraction**. Let a single photon arrive at normal incidence at an aperture (or a partially transmissive region) within an otherwise opaque screen, as depicted in Fig.15(a). Upon scattering at the aperture, the various modes of the EM field in the free space region beyond the aperture will each have an amplitude for being occupied by a single photon. The superposition of all these modes should, of course, yield the probability amplitude for finding the photon in the $xy$-plane immediately after the aperture at $z = 0^+$. In the far field, the probability amplitude profile is related to the Fourier transform of the distribution in the immediate vicinity of the aperture, as will be seen below.[7,8,11] We treat this problem by ignoring the polarization states of the incident and diffracted photons at first, focusing our effort on relating the scalar probability amplitudes before and after diffraction. This will be followed by an analysis of the effects of diffraction (i.e., scattering at the aperture) on the polarization state of the photon.

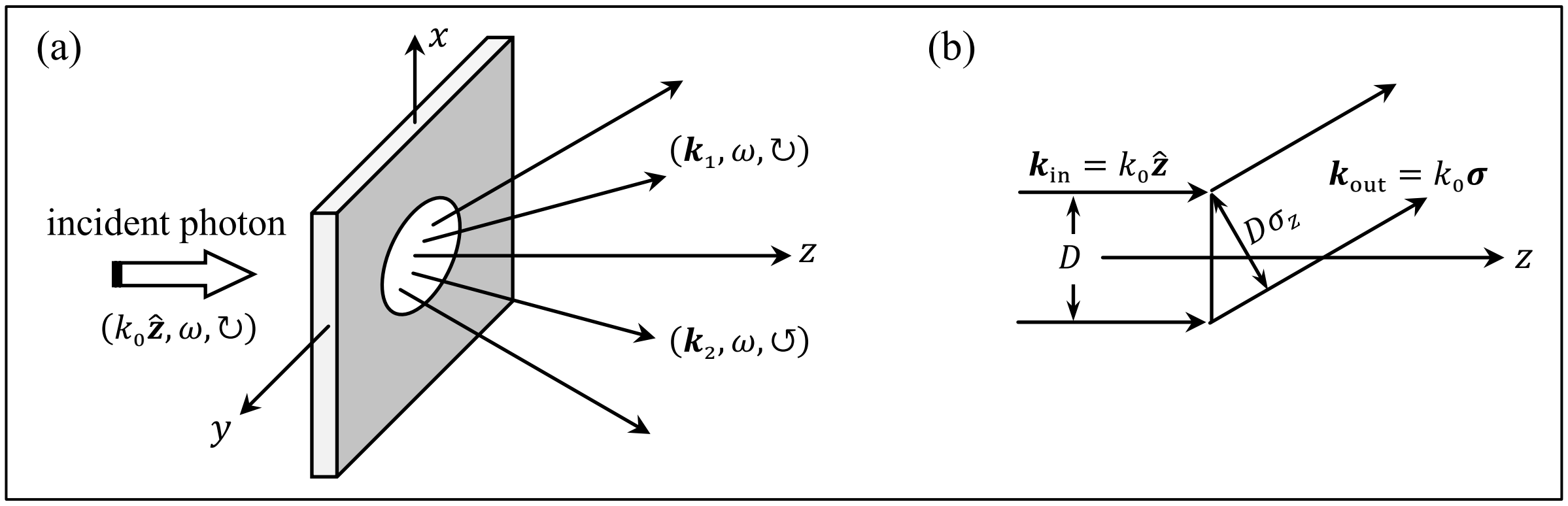

**Fig.15**. (a) A single photon of frequency $\omega$ propagating along the $z$-axis arrives at an aperture (or a partially transmissive region) in an otherwise opaque screen located in the $xy$-plane at $z = 0$. The various diffracted plane-waves (also of frequency $\omega$) are specified by their $k$-vectors $\boldsymbol{k}$ and RCP (↺) or LCP (↻) polarization states. (b) Given that a deflection of the incident $k$-vector from $k_0\hat{\boldsymbol{z}}$ to $k_0\boldsymbol{\sigma}$ (in consequence of diffraction at the aperture) entails a reduction of the beam's cross-sectional area by a factor of $\sigma_z$, the corresponding plane-wave's amplitude must be multiplied by $\sigma_z^{-1/2}$ to conserve the energy content of the incident beam.

We begin by assuming that the (scalar) field amplitude in the $xy$-plane at $z = 0^+$ is known to be $a_0(x, y)e^{-\mathrm{i}\omega t}$. The spatial Fourier transform of this initial amplitude profile is given by

$$\tilde{a}_0(k_x, k_y) = \iint_{-\infty}^{\infty} a_0(x, y)e^{-\mathrm{i}(k_x x + k_y y)}\mathrm{d}x\mathrm{d}y. \tag{14.1}$$

An inverse Fourier transformation then reconstitutes the initial distribution, as follows:

$$a_0(x, y) = a(x, y, z = 0^+) = (2\pi)^{-2} \iint_{-\infty}^{\infty} \tilde{a}_0(k_x, k_y)e^{\mathrm{i}(k_x x + k_y y)}\mathrm{d}k_x\mathrm{d}k_y. \tag{14.2}$$

The exponential term in the integrand of Eq.(14.2) is the reduced phase-factor $e^{\mathrm{i}(\boldsymbol{k}\cdot\boldsymbol{r}-\omega t)}$ of an EM plane-wave propagating in free space along the $k$-vector $\boldsymbol{k} = (k_x, k_y, k_z)$, evaluated at $\boldsymbol{r} = (x, y, z = 0^+)$, with its temporal factor $e^{-\mathrm{i}\omega t}$ stripped off. Maxwell's equations impose on each plane-wave the dispersion relation $\boldsymbol{k}\cdot\boldsymbol{k} = (\omega/c)^2$, which can be written as $k_x^2 + k_y^2 + k_z^2 = (2\pi/\lambda_0)^2$, where $\lambda_0 = 2\pi c/\omega$ is the photon's vacuum wavelength. One may thus define a unit-vector $\boldsymbol{\sigma}$ along the direction of the $k$-vector for each diffracted mode and proceed to write

$$\boldsymbol{k} = k_x\hat{\boldsymbol{x}} + k_y\hat{\boldsymbol{y}} + k_z\hat{\boldsymbol{z}} = (2\pi/\lambda_0)(\sigma_x\hat{\boldsymbol{x}} + \sigma_y\hat{\boldsymbol{y}} + \sigma_z\hat{\boldsymbol{z}}) = k_0\boldsymbol{\sigma}. \tag{14.3}$$

Here, $\sigma_z = (1 - \sigma_x^2 - \sigma_y^2)^{1/2}$. To compute the light amplitude distribution in the far field at $\boldsymbol{r} = (x, y, z = z_0)$, the expression of the initial distribution in Eq.(14.2) must be modified in several ways. First and foremost, we allow for the propagation of individual modes along their respective $k$-vectors by adding $\mathrm{i}k_z z_0$ to the exponent of the integrand — this is standard treatment for plane-wave propagation in free space.[7,8,11] Next, recognizing that evanescent waves do not survive propagation by more than a few wavelengths beyond the initial $xy$-plane at $z = 0^+$, we confine the spatial frequency domain of integration in Eq.(14.2) to the unit-circle $\sigma_x^2 + \sigma_y^2 < 1$, thereby limiting $\sigma_z$ to real values in the $[0,1)$ interval. And finally, considering that the cross-sectional area of a plane-wave incident along the $z$-axis is reduced by a factor of $\sigma_z$ when it is deflected to propagate along $\boldsymbol{\sigma}$, as shown in Fig.15(b), we scale the corresponding plane-wave's amplitude by $\sigma_z^{-1/2}$, to ensure the conservation of energy upon refraction. All in all, we will have

$$a(x, y, z = z_0) = \lambda_0^{-2} \iint_{\sigma_x^2+\sigma_y^2<1} \sigma_z^{-1/2}\tilde{a}_0(k_x, k_y) \exp[\mathrm{i}(2\pi/\lambda_0)(\sigma_x x + \sigma_y y + \sigma_z z_0)]\, \mathrm{d}\sigma_x \mathrm{d}\sigma_y. \tag{14.4}$$

The integral in Eq.(14.4) can be expressed as an integral over the hemispherical surface of radius 1 shown in Fig.16.[27] Since the projection of the differential surface element $\mathrm{d}s$ onto the $xy$-plane must equal $\mathrm{d}\sigma_x\mathrm{d}\sigma_y$, we replace $\mathrm{d}\sigma_x\mathrm{d}\sigma_y$ by $\sigma_z\mathrm{d}s$. Over the narrow ring of radius $\sin\psi$ (ring area $= 2\pi\sin\psi\,\mathrm{d}\psi$) shown in Fig.16(b), the exponential term in the integrand remains constant and can be written as $\exp(\mathrm{i}k_0\boldsymbol{\sigma}\cdot\boldsymbol{r}) = \exp(\mathrm{i}2\pi r\cos\psi/\lambda_0)$. The remainder of the integrand can be averaged over the area of the ring and written as $f(\psi) = \langle\sigma_z^{1/2}\tilde{a}_0(k_x, k_y)\rangle$. We will have

$$a(x, y, z_0) = \lambda_0^{-2} \int_{\psi=0}^{\pi} f(\psi) \exp(\mathrm{i}2\pi r\cos\psi/\lambda_0)\, 2\pi\sin\psi\, \mathrm{d}\psi. \tag{14.5}$$

Integrating by parts, we find

$$a(x, y, z_0) = (\mathrm{i}/\lambda_0 r)\left[f(\psi)e^{\mathrm{i}2\pi r\cos\psi/\lambda_0}\Big|_{\psi=0}^{\pi} - \int_0^{\pi} f'(\psi)e^{\mathrm{i}2\pi r\cos\psi/\lambda_0}\mathrm{d}\psi\right]. \tag{14.6}$$

The remaining integral, evaluated in the far field via stationary-phase approximation,[7] can be shown to be proportional to $(k_0 r)^{-1/2}$, which is negligible. Given that $f(\pi) = 0$, and that $f(0) = \sigma_z^{1/2}\tilde{a}_0(k_0\sigma_x, k_0\sigma_y)$ where $\boldsymbol{\sigma}|_{\psi=0} = (x/r, y/r, z_0/r)$, we finally arrive at

$$a(x, y, z_0) \cong -(\mathrm{i}/\lambda_0 r)f(0)e^{\mathrm{i}k_0 r} = -(\mathrm{i}/\lambda_0 r)(z_0/r)^{1/2}\tilde{a}_0(k_0 x/r, k_0 y/r)e^{\mathrm{i}k_0 r}. \tag{14.7}$$

The diffracted field amplitude $a(x, y, z_0)$ in the aperture's far field at $\boldsymbol{r} = (x, y, z_0)$ is thus seen to be intimately related to the Fourier transform $\tilde{a}_0(k_x, k_y)$ of the initial distribution $a(x, y, z = 0^+)$ evaluated at $(k_x, k_y) = (k_0 x/r, k_0 y/r)$.[7,8]

Extending the above result to account for the vectorial nature of the EM field requires that we compute the Fourier transforms of the $x$ and $y$ components of the $E$-field at the initial $xy$-plane at $z = 0^+$; that is,

$$\tilde{E}_{x,y}(k_x,k_y) = \iint_{-\infty}^{\infty} E_{x,y}(x,y,z=0^+)e^{-\mathrm{i}(k_x x+k_y y)}\mathrm{d}x\mathrm{d}y. \quad (14.8)$$

The RCP and LCP components of the refracted plane-wave in the direction of a $k$-vector are readily computed from the above $\tilde{E}_x$ and $\tilde{E}_y$, as follows:††

$$\tilde{E}_{\mathrm{out}}^{(\pm)}(k_x,k_y) = \left[(\sigma_z \mp \mathrm{i}\sigma_x\sigma_y)\tilde{E}_x \mp \mathrm{i}(1-\sigma_x^2)\tilde{E}_y\right]/2\sigma_z(1-\sigma_x^2)^{½}. \quad (14.9)$$

As before, $(\sigma_x,\sigma_y,\sigma_z) = (x/r\,,y/r\,,z_0/r)$. The expression for $\tilde{E}_{\mathrm{out}}^{(\pm)}$ in Eq.(14.9) now replaces $\tilde{a}_0(k_0x/r\,,k_0y/r)$ in Eq.(14.7) to yield the diffracted RCP and LCP amplitudes at the observation point in the far field. Note that the knowledge of the initial $E$-field amplitudes $E_{x,y}(x,y,z=0^+)$ is essential in these diffraction calculations starting with Eq.(14.8). Unless an exact solution of Maxwell's equations is available to account for the interaction between the incident light and the aperture, one must resort to judicious approximations[7,11] to estimate the emergent $E$-field profile immediately after the aperture in the $xy$-plane at $z=0^+$.

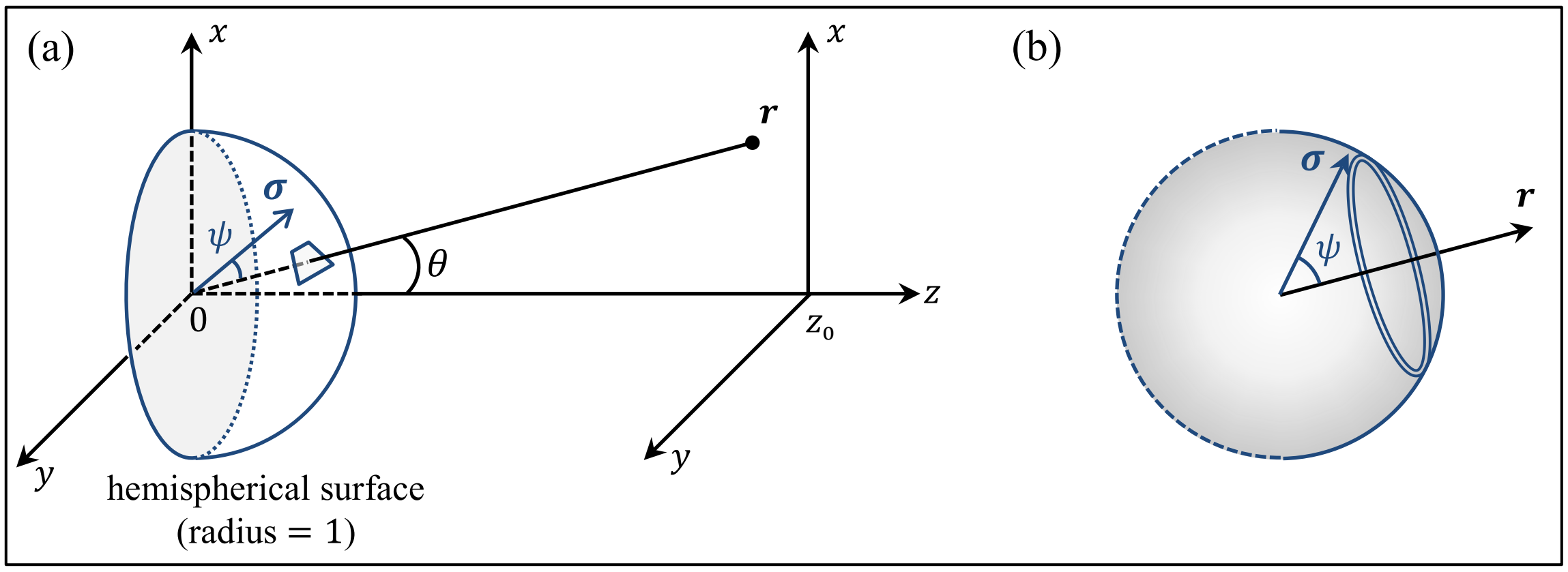

**Fig.16**. (a) An initial light amplitude distribution $a_0(x,y)$ in the $xy$-plane at $z=0^+$ propagates in free space to arrive at a parallel $xy$-plane at $z=z_0$. A hemispherical surface of radius 1.0 centered at the origin is the locus of the tips of the unit-vectors $\boldsymbol{\sigma} = \sigma_x\hat{\boldsymbol{x}} + \sigma_y\hat{\boldsymbol{y}} + \sigma_z\hat{\boldsymbol{z}}$ that define the propagation directions of the various plane-waves that constitute the initial distribution. The observation point $\boldsymbol{r} = r\hat{\boldsymbol{r}} = x\hat{\boldsymbol{x}} + y\hat{\boldsymbol{y}} + z_0\hat{\boldsymbol{z}}$ in the aperture's far field is where we desire to compute the amplitude of the diffracted light. In this far field, where $k_0r \gg 1$, the only plane-waves that significantly contribute to the observed light amplitude at $\boldsymbol{r}$ are those whose $\boldsymbol{\sigma}$-vectors fall within a small patch surrounding the point where the vector $\boldsymbol{r}$ pierces the hemisphere's surface. Within this patch, the angle $\psi$ between $\boldsymbol{r}$ and $\boldsymbol{\sigma}$ is close to 0°. (b) For any given angle $\psi$, the unit-vector $\boldsymbol{\sigma}$ describes a ring of radius $\sin\psi$ circling $\boldsymbol{r}$ on the hemisphere's surface.

Questions arise as to the distribution of photons among the various diffracted modes if two or more photons happen to be present in the incident beam. In particular, what if the incident beam is taken to be in a coherent state? The latter question may be answered by imagining that the diffracted modes are produced by a sequence of beam-splitters, as shown in Fig.17(a). The well-known formulas relating the number-state distributions in the input and output ports of ideal beam-splitters can then be used to determine the states of the various emergent beams (see Sec.16; also Sec.20).[1-4] Specifically, for a coherent incident beam, the input and output states of all the beam-splitters depicted in Fig.17(a) turn out to be coherent as well;[4] this will be shown in Sec.21. The RCP and LCP states may also be treated separately but in a similar way, since these orthogonal polarization states can be physically combined or split apart as suggested by the optical system of Fig.17(b).‡‡ All in all, a coherent beam illuminating the aperture (or the partially transmissive region) within the opaque screen of Fig.15(a) will be diffracted in such a way that the RCP and LCP light rays arriving at the various $(x, y, z_0)$ locations in the aperture's far field will maintain their coherence.

†† In terms of the polar and azimuthal angles $(\theta,\varphi)$ of Fig.6, the Cartesian components of the unit-vector $\boldsymbol{\sigma}$ are $\sigma_x = \cos\theta$, $\sigma_y = \sin\theta\cos\varphi$, and $\sigma_z = \sin\theta\sin\varphi$. Invoking Eqs.(6.1)-(6.3), one arrives at Eq.(14.9) by relating the RCP and LCP amplitudes $\tilde{E}_{\mathrm{out}}^{(\pm)}$ of the diffracted wave along $\boldsymbol{k} = k_0\boldsymbol{\sigma}$ to its $E$-field amplitudes $(\tilde{E}_x,\tilde{E}_y)$; see Appendix A.

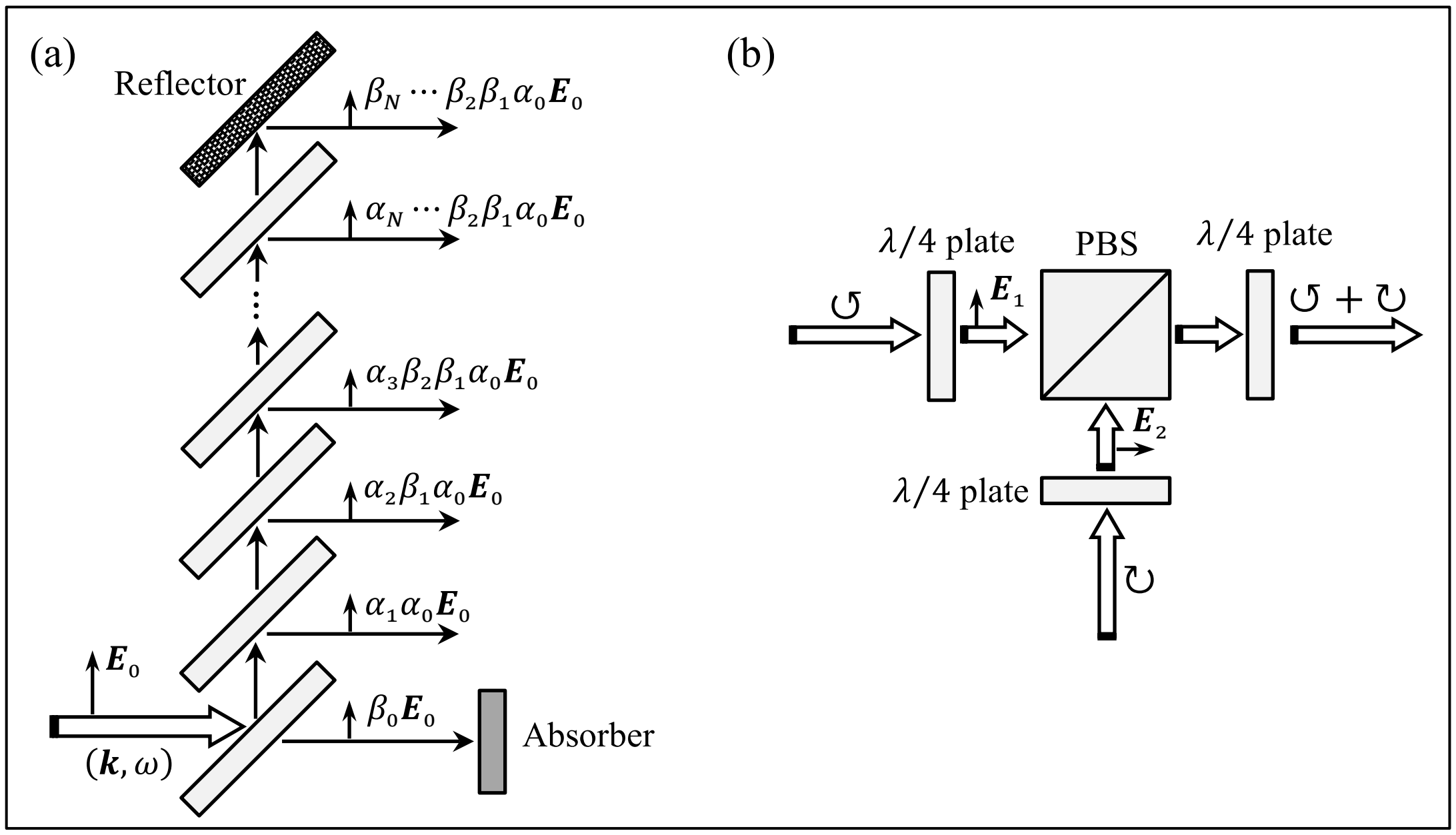

**Fig.17**. (a) A coherent beam is divided into a number of weaker beams by means of beam-splitters whose reflection and transmission coefficients $(\alpha_n, \beta_n)$ can have arbitrary values so long as $|\alpha_n|^2 + |\beta_n|^2 = 1$ and $\varphi_{\alpha_n} - \varphi_{\beta_n} = \pm 90°$. Each splitter receives a coherent incident beam at one of its input ports and produces a pair of coherent beams at its output ports. The various emergent beams on the right-hand side of the diagram are, therefore, coherent. (b) An RCP and an LCP beam pass through a pair of quarter-wave plates and become linearly polarized in orthogonal directions. The two beams are then combined at a polarizing beam-splitter (PBS) and subsequently converted back to RCP and LCP via a third quarter-wave plate. The same system operated in reverse can be used to separate the RCP and LCP contents of a diffracted beam.

**15. The Hong-Ou-Mandel effect**.[6] This is a two-photon interference phenomenon that occurs when a pair of identical photons arrives at a symmetric 50/50 beam-splitter, one photon coming through port 1, the other through port 2. At each port, the reflection amplitude is $e^{\mathrm{i}\varphi}/\sqrt{2}$, while the transmission amplitude is $\mathrm{i}e^{\mathrm{i}\varphi}/\sqrt{2}$. (The phase angle $\varphi$ is a characteristic of the beam-splitter, which could be anywhere between zero and $2\pi$, although, in the context of the present argument, its exact value is inconsequential.) Thus, the probability amplitude that both photons emerge at port 3 is $\mathrm{i}e^{\mathrm{i}2\varphi}/2$, since this requires one photon to be reflected and the other to be transmitted at the beam-splitter. Similarly, the amplitude that both photons emerge at port 4 is $\mathrm{i}e^{\mathrm{i}2\varphi}/2$. Now, for a single photon to appear at each of the output ports, there exist two possibilities: either both photons are reflected (amplitude $= e^{\mathrm{i}2\varphi}/2$), or both are transmitted (amplitude $= \mathrm{i}^2 e^{\mathrm{i}2\varphi}/2$). Considering that these two events are indistinguishable, their amplitudes must be added together, yielding a zero amplitude for detecting a coincidence event. In other words, there is zero chance of detecting a single photon at each of the output ports.

‡‡ With reference to Fig.6, a quarter-wave plate whose fast and slow axes are aligned with $\boldsymbol{\mathcal{E}}'$ and $\boldsymbol{\mathcal{E}}''$ in the far field, modifies the polarization state of a diffracted plane-wave to $[E^{(+)} + E^{(-)}]\boldsymbol{\mathcal{E}}' + [E^{(+)} - E^{(-)}]\boldsymbol{\mathcal{E}}''$. Subsequently, a polarizing beam-splitter oriented at 45° relative to $\boldsymbol{\mathcal{E}}'$ and $\boldsymbol{\mathcal{E}}''$ resolves the beam into its $E^{(+)}$ and $E^{(-)}$ components.

A lingering question concerning the above analysis is why the probability of detecting both photons together at port 3 (or together at port 4) turns out to be ¼ rather than ½. The answer is that the correct probability amplitude for two identical Bose particles — i.e., indistinguishable photons in the present example — to be scattered into a single state at port 3 (or a single state at port 4) is actually $\sqrt{2}$ times the product of the amplitudes that either particle alone is scattered into that state;[5] this is further elaborated in the next section. With this additional $\sqrt{2}$ factor taken into account, the probability of detecting both photons together at port 3 (or together at port 4) rises to ½, as it must.

**16. Wavepackets in number states $|n_1\rangle$ and $|n_2\rangle$ simultaneously arriving at a beam-splitter**. An insightful argument in *The Feynman Lectures on Physics* (Vol. III, Chap. 4, Secs. 2,3)[5] concerns the scattering of two or more indistinguishable Bose particles (e.g., identical photons) into a given quantum state. Here Feynman explains why the probability of $n$ identical bosons piling up within a single number state $|n\rangle$ is greater than the corresponding probability for distinguishable particles by a factor of $n!$. In the present section, we extend Feynman's argument to the problem of single-mode wavepackets $(\omega, \boldsymbol{k}_1, \hat{\boldsymbol{e}})$ and $(\omega, \boldsymbol{k}_2, \hat{\boldsymbol{e}})$ occupied by respective number-states $|n_1\rangle$ and $|n_2\rangle$ that enter the input ports of a lossless beam-splitter. The simultaneously arriving beams proceed to produce wavepackets in a superposition of the number-states $|m\rangle_3|n_1 + n_2 - m\rangle_4$ at the splitter's output ports.

With reference to Fig.18, consider a pair of single-mode light pulses (i.e., wavepackets) in the number states $|n_1\rangle$ and $|n_2\rangle$ entering ports 1 and 2 of a conventional beam-splitter. The amplitude reflection and transmission coefficients of the splitter are $\rho$ and $\tau$, respectively, with $|\rho|^2 + |\tau|^2 = 1$ and $\phi_\rho - \phi_\tau = \pm 90°$. The splitter is assumed to be symmetric, with $\rho$ and $\tau$ representing the Fresnel coefficients at both ports 1 and 2.[§§] We take the detectors in ports 3 and 4 to be finely divided photodetector arrays in the beam's cross-sectional plane, and also in the time domain, so that a single-photon detection event can be attributed to a small area $\Delta x \Delta y$ in the cross-sectional plane of one detector, and also to a short time interval $\Delta t$. The total number of cells associated with each detector in space-time is thus $N = N_x N_y N_t$. Detection of all $n_1 + n_2$ photons within a specific set of $n_1 + n_2$ cells from the two arrays constitutes a complete detection event.

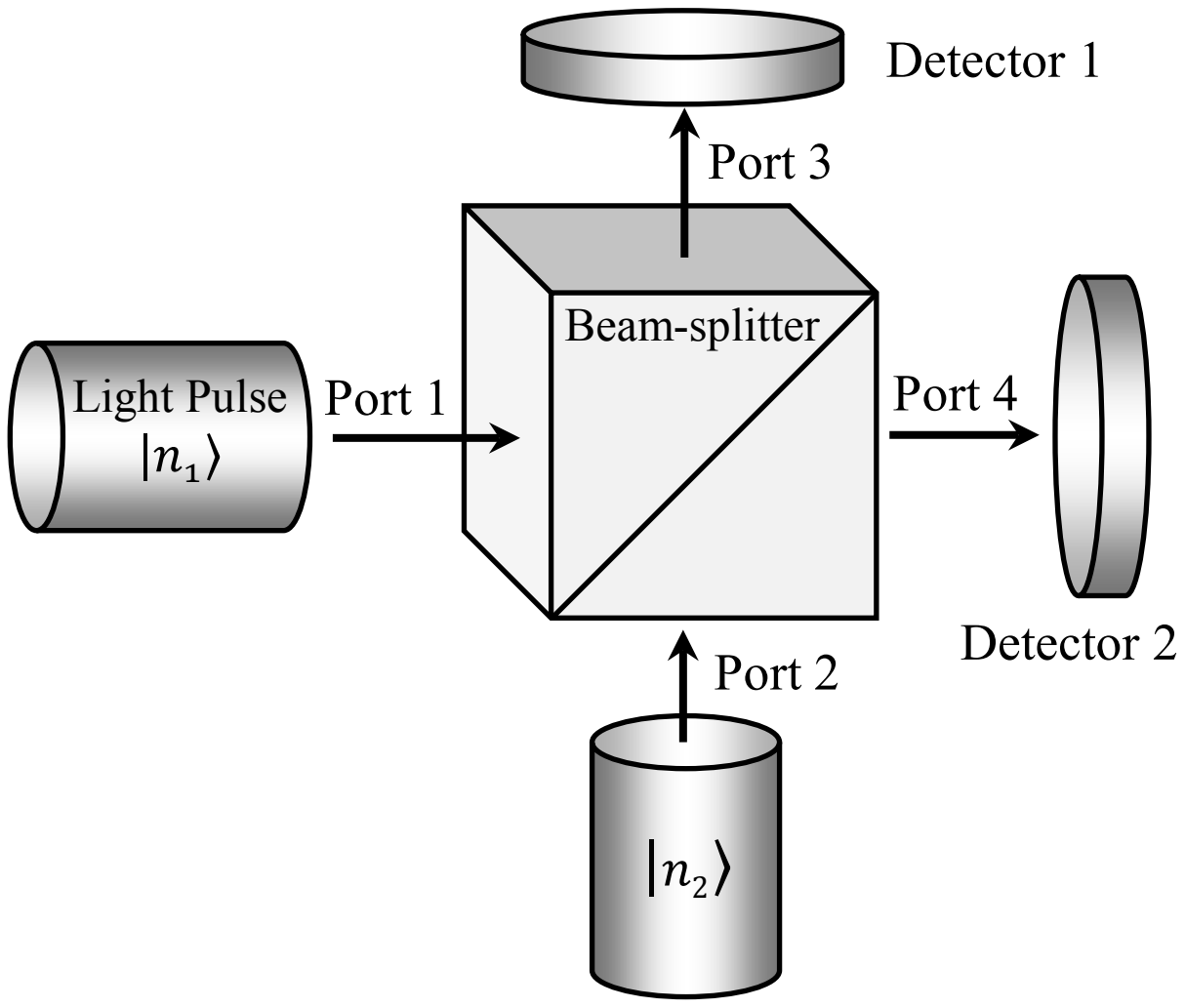

**Fig. 18**. A pair of single-mode light pulses of frequency $\omega$, wave-vectors $\boldsymbol{k}_1$ and $\boldsymbol{k}_2$, and a given polarization state (i.e., both $p$- or both $s$-polarized) arrive in the number states $|n_1\rangle$ and $|n_2\rangle$ at ports 1 and 2 of a lossless beam-splitter whose Fresnel reflection and transmission coefficients are $\rho$ and $\tau$. Each detector is a finely divided array in the beam's cross-sectional plane, and also in the time domain, so that a single-photon detection event can be attributed to a small area $\Delta x \Delta y$ in the cross-sectional $xy$ plane of a single detector, and also to a short time interval $\Delta t$. The total number $N = N_x N_y N_t$ of detection cells in each port is far greater than the number $n_1 + n_2$ of the incoming photons.

We assume at first that all $n_1 + n_2$ incoming photons are distinguishable — as would be the case, for instance, with small balls of differing colors — then impose the indistinguishability constraint at a later stage of the analysis. The calculation proceeds by identifying the photodetection events that would be physically distinct (i.e., clicks heard from different detector cells, or at different times from the same cells) and those that would, in principle, be indistinguishable if all the photons were identical. We then add up the probability *amplitudes* of all the indistinguishable events and the *ordinary* probabilities of all the distinct events that would result in $m$ clicks at detector 1 and $n_1 + n_2 - m$ clicks at detector 2. Details of this calculation appear in Ref.[28]. The end result is that the probability amplitude of detecting $m$ photons in port 3 and the remaining $n_1 + n_2 - m$ photons in port 4 is given by

$$\sum_{\substack{m_1=0 \\ m_1+m_2=m}}^{n_1} \sum_{m_2=0}^{n_2} \frac{\sqrt{n_1!\, n_2!\, (m_1+m_2)!\, (n_1+n_2-m_1-m_2)!}\; \rho^{n_2+m_1-m_2} \tau^{n_1-m_1+m_2}}{m_1!\, m_2!\, (n_1-m_1)!\, (n_2-m_2)!}. \qquad (16.1)$$

[§§] Strictly speaking, the symmetry of the splitter is merely convenient but not necessary for the validity of the argument.

The above expression is precisely the same as that obtained by applying the operator algebra to the system depicted in Fig.18, when light pulses in the number states $|n_1\rangle$ and $|n_2\rangle$ simultaneously arrive at ports 1 and 2.[4] An analysis of this problem with the aid of operator algebra is the subject of Sec.20. Before getting there, however, we need to review some of the basic principles of quantum optics and describe the characteristic features of the EM field in its various modes in free space; this will be done in the next three sections.

**17. Modes of the electromagnetic field in free space**. A single mode of the classical EM field in vacuum has a time-dependence $\exp(-\mathrm{i}\omega_m t)$ at a given angular frequency $\omega_m$, spatial dependence $u_m(\boldsymbol{r})$, where $u_m(\cdot)$ is a scalar function of the position $\boldsymbol{r} = x\hat{\boldsymbol{x}} + y\hat{\boldsymbol{y}} + z\hat{\boldsymbol{z}}$ in three-dimensional Euclidean space, and polarization state $\tilde{\boldsymbol{e}}_m$, where the complex unit-vector $\hat{\boldsymbol{e}}_m = \boldsymbol{e}'_m + \mathrm{i}\boldsymbol{e}''_m$ represents either of the two orthogonal states of polarization of the mode.[1,2] Note that $\hat{\boldsymbol{e}}_m \cdot \hat{\boldsymbol{e}}_m^* = |\boldsymbol{e}'_m|^2 + |\boldsymbol{e}''_m|^2 = 1$. In general, it is always possible to choose $\hat{\boldsymbol{e}}_m$ such that $\boldsymbol{e}'_m \cdot \boldsymbol{e}''_m = 0$, although, strictly speaking, this constraint is not necessary. (See Appendix B for a discussion of the polarization states of plane-waves and the conditions under which they can be considered mutually orthogonal.)

In a cavity, the function $u_m(\boldsymbol{r})\exp(-\mathrm{i}\omega_m t)\,\hat{\boldsymbol{e}}_m$ would represent the trapped harmonic mode $m$ in one of its allowed polarization states. In open (i.e., unbounded) free space, the spatial distribution of the mode could be that of a monochromatic plane-wave propagating along a specific (real-valued) $k$-vector satisfying $|\boldsymbol{k}_m| = \omega_m/c$, namely, $u_m(\boldsymbol{r}) = \exp(\mathrm{i}\boldsymbol{k}_m \cdot \boldsymbol{r})$. Maxwell's divergence equations $\boldsymbol{\nabla} \cdot \boldsymbol{E}(\boldsymbol{r}, t) = 0$ and $\boldsymbol{\nabla} \cdot \boldsymbol{B}(\boldsymbol{r}, t) = 0$ in this case require that $\boldsymbol{k}_m \cdot \hat{\boldsymbol{e}}_m = 0$.

In quantum optics, each allowed mode may be occupied by a superposition of number states, say, $|\psi\rangle = \sum_n c_n |n\rangle$, where the complex amplitudes $c_n$ are normalized to satisfy $\sum_n |c_n|^2 = 1$. States such as $|\psi\rangle$ which are superpositions (with complex probability amplitudes) of certain eigenstates of the system are generally referred to as pure states.[1,2] Alternatively, some modes may be occupied by a mixture of pure states, say, $|\psi_1\rangle, |\psi_2\rangle, \cdots, |\psi_K\rangle$, with ordinary (i.e., real-valued) probabilities $p_1, p_2, \cdots, p_K$, respectively.[1,2,4] These probabilities, of course, must be non-negative and add up to 1.

**18. Annihilation and creation operators**. The annihilation (or lowering) operator acting on the number state $|n\rangle$ yields $\hat{a}|n\rangle = \sqrt{n}|n-1\rangle$. Similarly, the creation (or raising) operator acting on $|n\rangle$ yields $\hat{a}^\dagger|n\rangle = \sqrt{n+1}|n+1\rangle$. When acted upon by $\hat{a}$, the vacuum state (i.e., the state with zero photons) is eliminated, that is, $\hat{a}|0\rangle = 0$, whereas under the action of $\hat{a}^\dagger$ the vacuum state acquires a single photon, namely, $\hat{a}^\dagger|0\rangle = |1\rangle$. By repeating the action of $\hat{a}^\dagger$ on the vacuum state, one can generate any number state, as follows: $(\hat{a}^\dagger)^n|0\rangle = \sqrt{n!}\,|n\rangle$.[1,2,4]

Neither $\hat{a}$ nor $\hat{a}^\dagger$ are Hermitian operators. When represented by matrices, the conjugate transpose of $\hat{a}$ will be $\hat{a}^\dagger$, and vice-versa. In general, the norm of $\hat{a}|\psi\rangle$ is written as $\langle\psi|\hat{a}^\dagger\hat{a}|\psi\rangle$ while the norm of $\hat{a}^\dagger|\chi\rangle$ is $\langle\chi|\hat{a}\hat{a}^\dagger|\chi\rangle$.

Glauber's coherent states $|\gamma\rangle$ are eigenstates of the annihilation operator $\hat{a}$ —the operator being non-Hermitian, its corresponding eigenvalues may or may not be real-valued.[3,4] We have

$$\hat{a}|\gamma\rangle = \exp(-½|\gamma|^2)\sum_{n=0}^{\infty}(\gamma^n/\sqrt{n!})\hat{a}|n\rangle = \exp(-½|\gamma|^2)\sum_{n=1}^{\infty}(\gamma^n/\sqrt{n!})\sqrt{n}|n-1\rangle$$
$$= \gamma\exp(-½|\gamma|^2)\sum_{n=1}^{\infty}\left[\gamma^{n-1}/\sqrt{(n-1)!}\right]|n-1\rangle = \gamma|\gamma\rangle. \quad (18.1)$$

The coherent state $|\gamma\rangle$ is *not* an eigenstate of the creation operator $\hat{a}^\dagger$, as revealed by the following argument:

$$\hat{a}^\dagger|\gamma\rangle = \exp(-½|\gamma|^2)\sum_{n=0}^{\infty}(\gamma^n/\sqrt{n!})\hat{a}^\dagger|n\rangle = \exp(-½|\gamma|^2)\sum_{n=0}^{\infty}(\gamma^n/\sqrt{n!})\sqrt{n+1}|n+1\rangle$$
$$= \gamma^{-1}\exp(-½|\gamma|^2)\sum_{n=0}^{\infty}n(\gamma^n/\sqrt{n!})|n\rangle. \quad (18.2)$$

Clearly, $\hat{a}^\dagger|\gamma\rangle$ cannot be expressed as a constant (i.e., an eigenvalue) times $|\gamma\rangle$.

The photon-number operator $\hat{a}^\dagger\hat{a}$ is clearly Hermitian, having the number states $|n\rangle$ as its eigenstates, that is, $\hat{a}^\dagger\hat{a}|n\rangle = \sqrt{n}\hat{a}^\dagger|n-1\rangle = \sqrt{n}\sqrt{n-1+1}|n-1+1\rangle = n|n\rangle$. The operators $\hat{a}$ and $\hat{a}^\dagger$ do *not* commute. One can readily verify that $\hat{a}\hat{a}^\dagger|n\rangle = \sqrt{n+1}\hat{a}|n+1\rangle = (n+1)|n\rangle$. Consequently, $[\hat{a},\hat{a}^\dagger] = \hat{a}\hat{a}^\dagger - \hat{a}^\dagger\hat{a} = \hat{1}$, where $\hat{1}$ is the identity operator.

---

**Example 1**. The average number of photons in the coherent state $|\gamma\rangle$ is evaluated as follows:

$$\langle n\rangle = \langle\gamma|\hat{a}^\dagger\hat{a}|\gamma\rangle = \langle\gamma|\gamma^*\gamma|\gamma\rangle = |\gamma|^2\langle\gamma|\gamma\rangle = |\gamma|^2. \quad (18.3)$$

The average photon-number squared $\langle n^2\rangle$ is similarly found as the expectation value of the operator $\hat{a}^\dagger\hat{a}\hat{a}^\dagger\hat{a}$, namely,

$$\langle n^2\rangle = \langle\gamma|\hat{a}^\dagger\hat{a}\hat{a}^\dagger\hat{a}|\gamma\rangle = \exp(-|\gamma|^2)\sum_{n=0}^{\infty}n(\gamma^{*n}/\sqrt{n!})\langle n|\sum_{m=0}^{\infty}m(\gamma^m/\sqrt{m!})|m\rangle$$
$$= \exp(-|\gamma|^2)\sum_{n=0}^{\infty}n^2|\gamma|^{2n}/n! = \exp(-|\gamma|^2)\sum_{n=1}^{\infty}n|\gamma|^{2n}/(n-1)!$$
$$= |\gamma|^2\exp(-|\gamma|^2)\sum_{n=1}^{\infty}(n-1+1)|\gamma|^{2(n-1)}/(n-1)!$$
$$= |\gamma|^2\exp(-|\gamma|^2)\left[\sum_{n=1}^{\infty}(n-1)|\gamma|^{2(n-1)}/(n-1)! + \sum_{n=1}^{\infty}|\gamma|^{2(n-1)}/(n-1)!\right]$$
$$= |\gamma|^2\exp(-|\gamma|^2)\left[\sum_{n=2}^{\infty}|\gamma|^{2(n-1)}/(n-2)! + \sum_{n=0}^{\infty}|\gamma|^{2n}/n!\right]$$
$$= |\gamma|^2\exp(-|\gamma|^2)\left[|\gamma|^2\sum_{n=0}^{\infty}|\gamma|^{2n}/n! + \exp(|\gamma|^2)\right]$$
$$= |\gamma|^4 + |\gamma|^2. \quad (18.4)$$

← A somewhat simpler proof starts by writing $\hat{a}^\dagger\hat{a}\hat{a}^\dagger\hat{a}$ as $\hat{a}^\dagger(\hat{a}^\dagger\hat{a}+1)\hat{a}$.

The variance of the photon number is thus seen to be $\langle n^2\rangle - \langle n\rangle^2 = |\gamma|^2$. The equality of the average and the variance of the photon number in a coherent state $|\gamma\rangle$ is a characteristic feature of its Poissonian probability distribution over the photon number $n$, namely, $p_n = e^{-|\gamma|^2}|\gamma|^{2n}/n!$.

---

**Example 2**. In general, the superposition of two coherent states $|\gamma_1\rangle$ and $|\gamma_2\rangle$ is *not* a coherent state; specifically, $c_1|\gamma_1\rangle + c_2|\gamma_2\rangle \neq |c_1\gamma_1 + c_2\gamma_2\rangle$. Writing

$$c_1|\gamma_1\rangle + c_2|\gamma_2\rangle = c_1 \exp\left(-\tfrac{1}{2}|\gamma_1|^2\right) \textstyle\sum_{n=0}^{\infty}\left(\gamma_1^n/\sqrt{n!}\right)|n\rangle + c_2 \exp\left(-\tfrac{1}{2}|\gamma_2|^2\right) \sum_{n=0}^{\infty}\left(\gamma_2^n/\sqrt{n!}\right)|n\rangle, \quad (18.5)$$

we note that the above expression cannot be further simplified and that, therefore, the superposition state is not expressible as a coherent state. However, it will be shown in Sec.21 that combining two coherent states at a beam-splitter yields coherent states at both output ports of the splitter.

---

**Example 3**. The operator $\hat{\Gamma}(\gamma) = \exp(-\tfrac{1}{2}|\gamma|^2)\sum_{n=0}^{\infty}\gamma^n(\hat{a}^\dagger)^n/n!$ acts on the vacuum state $|0\rangle$ to generate the coherent state $|\gamma\rangle$. One can show that $\hat{\Gamma}(\gamma_1)\hat{\Gamma}(\gamma_2) = \exp[\mathrm{Re}(\gamma_1\gamma_2^*)]\,\hat{\Gamma}(\gamma_1+\gamma_2)$, as follows:

$$\begin{aligned}\hat{\Gamma}(\gamma_1)\hat{\Gamma}(\gamma_2) &= \exp[-\tfrac{1}{2}(|\gamma_1|^2 + |\gamma_2|^2)]\,[\textstyle\sum_{n=0}^{\infty}\gamma_1^n(\hat{a}^\dagger)^n/n!][\sum_{m=0}^{\infty}\gamma_2^m(\hat{a}^\dagger)^m/m!]\\ &= \exp[-\tfrac{1}{2}(\gamma_1\gamma_1^* + \gamma_2\gamma_2^*)]\textstyle\sum_{n=0}^{\infty}\sum_{m=0}^{\infty}\gamma_1^n\gamma_2^m(\hat{a}^\dagger)^{n+m}/(n!\,m!)\\ &= \exp[-\tfrac{1}{2}|\gamma_1+\gamma_2|^2 + \mathrm{Re}(\gamma_1\gamma_2^*)]\textstyle\sum_{k=0}^{\infty}\{\sum_{n=0}^{k}\gamma_1^n\gamma_2^{k-n}/[n!\,(k-n)!]\}\,(\hat{a}^\dagger)^k\\ &= \exp[\mathrm{Re}(\gamma_1\gamma_2^*)]\exp(-\tfrac{1}{2}|\gamma_1+\gamma_2|^2)\textstyle\sum_{k=0}^{\infty}(\gamma_1+\gamma_2)^k(\hat{a}^\dagger)^k/k!\\ &= \exp[\mathrm{Re}(\gamma_1\gamma_2^*)]\,\hat{\Gamma}(\gamma_1+\gamma_2). \qquad (18.6)\end{aligned}$$

---

**19. Electromagnetic field operators (single mode)**. Other useful operators expressible in terms of $\hat{a}$ and $\hat{a}^\dagger$ are the vector potential operator $\hat{\boldsymbol{A}}(\boldsymbol{r},t)$, the electric field operator $\hat{\boldsymbol{E}}(\boldsymbol{r},t)$, the magnetic field operator $\hat{\boldsymbol{B}}(\boldsymbol{r},t)$, and the energy operator $\hat{\mathcal{H}}$ for a traveling plane-wave in free space. Denoting by $V$ the free space volume occupied by the $(\omega,\boldsymbol{k},\hat{\boldsymbol{e}})$ mode, we will have[1,2]

$$\hat{\boldsymbol{A}}(\boldsymbol{r},t) = \sqrt{\hbar/(2\varepsilon_0 V\omega)}\,\{\hat{\boldsymbol{e}}\exp[\mathrm{i}(\boldsymbol{k}\cdot\boldsymbol{r}-\omega t)]\,\hat{a} + \hat{\boldsymbol{e}}^*\exp[-\mathrm{i}(\boldsymbol{k}\cdot\boldsymbol{r}-\omega t)]\,\hat{a}^\dagger\}. \qquad (19.1)$$

$$\hat{\boldsymbol{E}}(\boldsymbol{r},t) = \mathrm{i}\sqrt{\hbar\omega/(2\varepsilon_0 V)}\,\{\hat{\boldsymbol{e}}\exp[\mathrm{i}(\boldsymbol{k}\cdot\boldsymbol{r}-\omega t)]\,\hat{a} - \hat{\boldsymbol{e}}^*\exp[-\mathrm{i}(\boldsymbol{k}\cdot\boldsymbol{r}-\omega t)]\,\hat{a}^\dagger\}. \qquad (19.2)$$

$$\hat{\boldsymbol{B}}(\boldsymbol{r},t) = \mathrm{i}\sqrt{\hbar/(2\varepsilon_0 V\omega)}\,\{(\boldsymbol{k}\times\hat{\boldsymbol{e}})\exp[\mathrm{i}(\boldsymbol{k}\cdot\boldsymbol{r}-\omega t)]\,\hat{a} - (\boldsymbol{k}\times\hat{\boldsymbol{e}}^*)\exp[-\mathrm{i}(\boldsymbol{k}\cdot\boldsymbol{r}-\omega t)]\,\hat{a}^\dagger\}. \qquad (19.3)$$

$$\hat{\mathcal{H}} = \hbar\omega(\hat{a}^\dagger\hat{a} + \tfrac{1}{2}). \qquad (19.4)$$

In the above equations, we have (i) used the Heisenberg picture in which the spacetime dependence is embedded in the operators; (ii) allowed an arbitrary (i.e., elliptical) polarization state by setting $\hat{\boldsymbol{e}} = \boldsymbol{e}' + \mathrm{i}\boldsymbol{e}''$, where $\hat{\boldsymbol{e}}\cdot\hat{\boldsymbol{e}}^* = |\boldsymbol{e}'|^2 + |\boldsymbol{e}''|^2 = 1$; (iii) ensured the satisfaction of the classical identities $\boldsymbol{E}(\boldsymbol{r},t) = -\partial\boldsymbol{A}(\boldsymbol{r},t)/\partial t$ and $\boldsymbol{B}(\boldsymbol{r},t) = \boldsymbol{\nabla}\times\boldsymbol{A}(\boldsymbol{r},t)$; and (iv) incorporated the ground-state (or vacuum) energy $\tfrac{1}{2}\hbar\omega$ associated with each harmonic oscillator. Note that the operators $\hat{\boldsymbol{A}},\hat{\boldsymbol{E}},\hat{\boldsymbol{B}},\hat{\mathcal{H}}$ defined in Eqs.(19.1)-(19.4) are Hermitian, and that the $\boldsymbol{E}$ and $\boldsymbol{B}$ fields satisfy Maxwell's equations, $\boldsymbol{\nabla}\times\boldsymbol{E}(\boldsymbol{r},t) = -\partial\boldsymbol{B}(\boldsymbol{r},t)/\partial t$ and $\boldsymbol{\nabla}\times\boldsymbol{B}(\boldsymbol{r},t) = c^{-2}\,\partial\boldsymbol{E}(\boldsymbol{r},t)/\partial t$. The $E$- and $B$-field energy-density operators are readily found from Eqs.(19.2) and (19.3), as follows:

$$\begin{aligned}\tfrac{1}{2}\varepsilon_0\hat{\boldsymbol{E}}(\boldsymbol{r},t)\cdot\hat{\boldsymbol{E}}(\boldsymbol{r},t) = -\tfrac{1}{4}(\hbar\omega/V)&\{\hat{\boldsymbol{e}}\exp[\mathrm{i}(\boldsymbol{k}\cdot\boldsymbol{r}-\omega t)]\,\hat{a} - \hat{\boldsymbol{e}}^*\exp[-\mathrm{i}(\boldsymbol{k}\cdot\boldsymbol{r}-\omega t)]\,\hat{a}^\dagger\}\\ &\cdot\{\hat{\boldsymbol{e}}\exp[\mathrm{i}(\boldsymbol{k}\cdot\boldsymbol{r}-\omega t)]\,\hat{a} - \hat{\boldsymbol{e}}^*\exp[-\mathrm{i}(\boldsymbol{k}\cdot\boldsymbol{r}-\omega t)]\,\hat{a}^\dagger\}\\ = -\tfrac{1}{4}(\hbar\omega/V)&\{-(\hat{\boldsymbol{e}}\cdot\hat{\boldsymbol{e}}^*)\hat{a}\hat{a}^\dagger - (\hat{\boldsymbol{e}}^*\cdot\hat{\boldsymbol{e}})\hat{a}^\dagger\hat{a}\\ &+(\hat{\boldsymbol{e}}\cdot\hat{\boldsymbol{e}})\exp[2\mathrm{i}(\boldsymbol{k}\cdot\boldsymbol{r}-\omega t)]\,\hat{a}\hat{a} + (\hat{\boldsymbol{e}}^*\cdot\hat{\boldsymbol{e}}^*)\exp[-2\mathrm{i}(\boldsymbol{k}\cdot\boldsymbol{r}-\omega t)]\,\hat{a}^\dagger\hat{a}^\dagger\}. \qquad (19.5)\end{aligned}$$

$$\tfrac{1}{2}\mu_0^{-1}\hat{\boldsymbol{B}}(\boldsymbol{r},t)\cdot\hat{\boldsymbol{B}}(\boldsymbol{r},t) = -\tfrac{1}{4}(\hbar c^2/V\omega)\{(\boldsymbol{k}\times\hat{\boldsymbol{e}})\exp[\mathrm{i}(\boldsymbol{k}\cdot\boldsymbol{r}-\omega t)]\,\hat{a} - (\boldsymbol{k}\times\hat{\boldsymbol{e}}^*)\exp[-\mathrm{i}(\boldsymbol{k}\cdot\boldsymbol{r}-\omega t)]\,\hat{a}^\dagger\}$$
$$\cdot\{(\boldsymbol{k}\times\hat{\boldsymbol{e}})\exp[\mathrm{i}(\boldsymbol{k}\cdot\boldsymbol{r}-\omega t)]\,\hat{a} - (\boldsymbol{k}\times\hat{\boldsymbol{e}}^*)\exp[-\mathrm{i}(\boldsymbol{k}\cdot\boldsymbol{r}-\omega t)]\,\hat{a}^\dagger\}$$
$$= -\tfrac{1}{4}(\hbar\omega/V)\{-(\hat{\boldsymbol{e}}\cdot\hat{\boldsymbol{e}}^*)\hat{a}\hat{a}^\dagger - (\hat{\boldsymbol{e}}^*\cdot\hat{\boldsymbol{e}})\hat{a}^\dagger\hat{a}$$
$$+(\hat{\boldsymbol{e}}\cdot\hat{\boldsymbol{e}})\exp[2\mathrm{i}(\boldsymbol{k}\cdot\boldsymbol{r}-\omega t)]\,\hat{a}\hat{a} + (\hat{\boldsymbol{e}}^*\cdot\hat{\boldsymbol{e}}^*)\exp[-2\mathrm{i}(\boldsymbol{k}\cdot\boldsymbol{r}-\omega t)]\,\hat{a}^\dagger\hat{a}^\dagger\}. \quad (19.6)^{***}$$

The last two terms on the right-hand sides of Eqs.(19.5) and (19.6) are complex conjugate pairs which produce a real, sinusoidally varying field in space and time; averaging over the volume $V$ of the field would eliminate these terms. The remaining terms yield

$$\int_V [\tfrac{1}{2}\varepsilon_0\hat{\boldsymbol{E}}(\boldsymbol{r},t)\cdot\hat{\boldsymbol{E}}(\boldsymbol{r},t) + \tfrac{1}{2}\mu_0^{-1}\hat{\boldsymbol{B}}(\boldsymbol{r},t)\cdot\hat{\boldsymbol{B}}(\boldsymbol{r},t)]\mathrm{d}v = \tfrac{1}{2}\hbar\omega(\hat{a}\hat{a}^\dagger + \hat{a}^\dagger\hat{a}) = \hbar\omega(\hat{a}^\dagger\hat{a} + \tfrac{1}{2}). \quad (19.7)$$

The above equation confirms the consistency of the energy operator, i.e., the Hamiltonian $\hat{\mathcal{H}}$ of Eq.(19.4), with the $E$- and $B$-field operators of Eqs.(19.2) and (19.3). We mention in passing that the time-evolution factor $e^{-\mathrm{i}\omega t}$ of the field operators in the Heisenberg picture is distinct from the corresponding factor $e^{-\mathrm{i}n\omega t}$ that might accompany a number state $|n\rangle$ in the Schrödinger picture.

---

**Example 4**. The number states $|n\rangle$ are peculiar in that the average values of their $E$- and $B$-fields vanish, but, as expected, their field variances remain nonzero, as shown below.

$$\langle n|\hat{\boldsymbol{E}}(\boldsymbol{r},t)|n\rangle = \mathrm{i}\sqrt{\hbar\omega/(2\varepsilon_0 V)}\{\hat{\boldsymbol{e}}\exp[\mathrm{i}(\boldsymbol{k}\cdot\boldsymbol{r}-\omega t)]\langle n|\hat{a}|n\rangle - \hat{\boldsymbol{e}}^*\exp[-\mathrm{i}(\boldsymbol{k}\cdot\boldsymbol{r}-\omega t)]\langle n|\hat{a}^\dagger|n\rangle\} = 0. \quad (19.8)$$

$$\langle n|\hat{\boldsymbol{E}}(\boldsymbol{r},t)\cdot\hat{\boldsymbol{E}}(\boldsymbol{r},t)|n\rangle = [\hbar\omega/(2\varepsilon_0 V)]\{(\hat{\boldsymbol{e}}\cdot\hat{\boldsymbol{e}}^*)\langle n|\hat{a}\hat{a}^\dagger|n\rangle + (\hat{\boldsymbol{e}}^*\cdot\hat{\boldsymbol{e}})\langle n|\hat{a}^\dagger\hat{a}|n\rangle$$
$$-(\hat{\boldsymbol{e}}\cdot\hat{\boldsymbol{e}})\exp[2\mathrm{i}(\boldsymbol{k}\cdot\boldsymbol{r}-\omega t)]\underbrace{\langle n|\hat{a}\hat{a}|n\rangle}_{0} - (\hat{\boldsymbol{e}}^*\cdot\hat{\boldsymbol{e}}^*)\exp[-2\mathrm{i}(\boldsymbol{k}\cdot\boldsymbol{r}-\omega t)]\underbrace{\langle n|\hat{a}^\dagger\hat{a}^\dagger|n\rangle}_{0}\}$$
$$= (n+\tfrac{1}{2})\hbar\omega/(\varepsilon_0 V). \quad (19.9)$$

Similarly, we find that $\langle n|\hat{\boldsymbol{B}}(\boldsymbol{r},t)|n\rangle = 0$ and $\langle n|\hat{\boldsymbol{B}}(\boldsymbol{r},t)\cdot\hat{\boldsymbol{B}}(\boldsymbol{r},t)|n\rangle = \mu_0(n+\tfrac{1}{2})\hbar\omega/V$. Thus, the total EM energy content of the number state $|n\rangle$ is $(n+\tfrac{1}{2})\hbar\omega$, which can be associated with the energy $\hbar\omega$ of each one of its $n$ photons, plus the $\tfrac{1}{2}\hbar\omega$ energy of the ever-present vacuum field.

---

**Example 5**. The average and variance of the $E$- and $B$-fields of the (quasi-classical) coherent state $|\gamma\rangle$ are evaluated as follows:

$$\langle\gamma|\hat{\boldsymbol{E}}(\boldsymbol{r},t)|\gamma\rangle = \mathrm{i}\sqrt{\hbar\omega/(2\varepsilon_0 V)}\{\hat{\boldsymbol{e}}\exp[\mathrm{i}(\boldsymbol{k}\cdot\boldsymbol{r}-\omega t)]\langle\gamma|\hat{a}|\gamma\rangle - \hat{\boldsymbol{e}}^*\exp[-\mathrm{i}(\boldsymbol{k}\cdot\boldsymbol{r}-\omega t)]\langle\gamma|\hat{a}^\dagger|\gamma\rangle\}$$
$$= \mathrm{i}\sqrt{\hbar\omega/(2\varepsilon_0 V)}\{\hat{\boldsymbol{e}}\gamma\exp[\mathrm{i}(\boldsymbol{k}\cdot\boldsymbol{r}-\omega t)] - \hat{\boldsymbol{e}}^*\gamma^*\exp[-\mathrm{i}(\boldsymbol{k}\cdot\boldsymbol{r}-\omega t)]\}$$
$$= \mathrm{i}\sqrt{\hbar\omega/(2\varepsilon_0 V)}\,|\gamma|\{(\boldsymbol{e}'+\mathrm{i}\boldsymbol{e}'')[\cos(\boldsymbol{k}\cdot\boldsymbol{r}-\omega t+\varphi_\gamma) + \mathrm{i}\sin(\boldsymbol{k}\cdot\boldsymbol{r}-\omega t+\varphi_\gamma)]$$
$$-(\boldsymbol{e}'-\mathrm{i}\boldsymbol{e}'')[\cos(\boldsymbol{k}\cdot\boldsymbol{r}-\omega t+\varphi_\gamma) - \mathrm{i}\sin(\boldsymbol{k}\cdot\boldsymbol{r}-\omega t+\varphi_\gamma)]\}$$
$$= -\sqrt{2\hbar\omega/(\varepsilon_0 V)}\,|\gamma|[\boldsymbol{e}'\sin(\boldsymbol{k}\cdot\boldsymbol{r}-\omega t+\varphi_\gamma) + \boldsymbol{e}''\cos(\boldsymbol{k}\cdot\boldsymbol{r}-\omega t+\varphi_\gamma)]. \quad (19.10)$$

$$\langle\gamma|\hat{\boldsymbol{E}}(\boldsymbol{r},t)\cdot\hat{\boldsymbol{E}}(\boldsymbol{r},t)|\gamma\rangle = [\hbar\omega/(2\varepsilon_0 V)]\{(\hat{\boldsymbol{e}}\cdot\hat{\boldsymbol{e}}^*)\langle\gamma|\hat{a}\hat{a}^\dagger|\gamma\rangle + (\hat{\boldsymbol{e}}^*\cdot\hat{\boldsymbol{e}})\langle\gamma|\hat{a}^\dagger\hat{a}|\gamma\rangle$$

set $\boldsymbol{e}'\cdot\boldsymbol{e}'' = 0$ →
$$-(\hat{\boldsymbol{e}}\cdot\hat{\boldsymbol{e}})\exp[2\mathrm{i}(\boldsymbol{k}\cdot\boldsymbol{r}-\omega t)]\langle\gamma|\hat{a}\hat{a}|\gamma\rangle - (\hat{\boldsymbol{e}}^*\cdot\hat{\boldsymbol{e}}^*)\exp[-2\mathrm{i}(\boldsymbol{k}\cdot\boldsymbol{r}-\omega t)]\langle\gamma|\hat{a}^\dagger\hat{a}^\dagger|\gamma\rangle\}$$

---

*** $(\boldsymbol{k}\times\hat{\boldsymbol{e}})\cdot(\boldsymbol{k}\times\hat{\boldsymbol{e}}^*) = [\hat{\boldsymbol{e}}\times(\boldsymbol{k}\times\hat{\boldsymbol{e}}^*)]\cdot\boldsymbol{k} = [(\hat{\boldsymbol{e}}\cdot\hat{\boldsymbol{e}}^*)\boldsymbol{k} - \underbrace{(\hat{\boldsymbol{e}}\cdot\boldsymbol{k})}_{0}\hat{\boldsymbol{e}}^*]\cdot\boldsymbol{k} = \underbrace{(\hat{\boldsymbol{e}}\cdot\hat{\boldsymbol{e}}^*)}_{1}k^2 = (\omega/c)^2.$

$(\boldsymbol{k}\times\hat{\boldsymbol{e}})\cdot(\boldsymbol{k}\times\hat{\boldsymbol{e}}) = [\hat{\boldsymbol{e}}\times(\boldsymbol{k}\times\hat{\boldsymbol{e}})]\cdot\boldsymbol{k} = [(\hat{\boldsymbol{e}}\cdot\hat{\boldsymbol{e}})\boldsymbol{k} - \underbrace{(\hat{\boldsymbol{e}}\cdot\boldsymbol{k})}_{0}\hat{\boldsymbol{e}}]\cdot\boldsymbol{k} = (\hat{\boldsymbol{e}}\cdot\hat{\boldsymbol{e}})k^2 = (\omega/c)^2(\hat{\boldsymbol{e}}\cdot\hat{\boldsymbol{e}}).$

$$= (\hbar\omega/\varepsilon_0 V)\{½ + |\gamma|^2 - |\gamma|^2(|\boldsymbol{e}'|^2 - |\boldsymbol{e}''|^2)\cos[2(\boldsymbol{k}\cdot\boldsymbol{r} - \omega t + \varphi_\gamma)]\}. \tag{19.11}$$

$$\boxed{\mathrm{Var}[\boldsymbol{E}(\boldsymbol{r},t)] = \langle\gamma|\hat{\boldsymbol{E}}(\boldsymbol{r},t)\cdot\hat{\boldsymbol{E}}(\boldsymbol{r},t)|\gamma\rangle - \langle\gamma|\hat{\boldsymbol{E}}(\boldsymbol{r},t)|\gamma\rangle\cdot\langle\gamma|\hat{\boldsymbol{E}}(\boldsymbol{r},t)|\gamma\rangle = \hbar\omega/(2\varepsilon_0 V).} \tag{19.12}$$

$$\langle\gamma|\hat{\boldsymbol{B}}(\boldsymbol{r},t)|\gamma\rangle = \mathrm{i}\sqrt{\hbar/(2\varepsilon_0 V\omega)}\,\{(\boldsymbol{k}\times\hat{\boldsymbol{e}})\exp[\mathrm{i}(\boldsymbol{k}\cdot\boldsymbol{r} - \omega t)]\,\langle\gamma|\hat{a}|\gamma\rangle - (\boldsymbol{k}\times\hat{\boldsymbol{e}}^*)\exp[-\mathrm{i}(\boldsymbol{k}\cdot\boldsymbol{r} - \omega t)]\,\langle\gamma|\hat{a}^\dagger|\gamma\rangle\}$$

$$= \mathrm{i}\sqrt{\hbar/(2\varepsilon_0 V\omega)}\,\{(\boldsymbol{k}\times\hat{\boldsymbol{e}})\gamma\exp[\mathrm{i}(\boldsymbol{k}\cdot\boldsymbol{r} - \omega t)] - (\boldsymbol{k}\times\hat{\boldsymbol{e}}^*)\gamma^*\exp[-\mathrm{i}(\boldsymbol{k}\cdot\boldsymbol{r} - \omega t)]\}$$

$$= \mathrm{i}\sqrt{\hbar/(2\varepsilon_0 V\omega)}\,|\gamma|\boldsymbol{k}\times\{(\boldsymbol{e}' + \mathrm{i}\boldsymbol{e}'')[\cos(\boldsymbol{k}\cdot\boldsymbol{r} - \omega t + \varphi_\gamma) + \mathrm{i}\sin(\boldsymbol{k}\cdot\boldsymbol{r} - \omega t + \varphi_\gamma)]$$
$$-(\boldsymbol{e}' - \mathrm{i}\boldsymbol{e}'')[\cos(\boldsymbol{k}\cdot\boldsymbol{r} - \omega t + \varphi_\gamma) - \mathrm{i}\sin(\boldsymbol{k}\cdot\boldsymbol{r} - \omega t + \varphi_\gamma)]\}$$

$$= -\sqrt{2\hbar/(\varepsilon_0 V\omega)}\,|\gamma|[(\boldsymbol{k}\times\boldsymbol{e}')\sin(\boldsymbol{k}\cdot\boldsymbol{r} - \omega t + \varphi_\gamma) + (\boldsymbol{k}\times\boldsymbol{e}'')\cos(\boldsymbol{k}\cdot\boldsymbol{r} - \omega t + \varphi_\gamma)]. \tag{19.13}$$

$$\langle\gamma|\hat{\boldsymbol{B}}(\boldsymbol{r},t)\cdot\hat{\boldsymbol{B}}(\boldsymbol{r},t)|\gamma\rangle = ½(\mu_0\hbar\omega/V)\{\langle\gamma|\hat{a}\hat{a}^\dagger|\gamma\rangle + \langle\gamma|\hat{a}^\dagger\hat{a}|\gamma\rangle$$

$$\boxed{\text{set } \boldsymbol{e}'\cdot\boldsymbol{e}'' = 0} \rightarrow -(\hat{\boldsymbol{e}}\cdot\hat{\boldsymbol{e}})\exp[2\mathrm{i}(\boldsymbol{k}\cdot\boldsymbol{r} - \omega t)]\,\langle\gamma|\hat{a}\hat{a}|\gamma\rangle - (\hat{\boldsymbol{e}}^*\cdot\hat{\boldsymbol{e}}^*)\exp[-2\mathrm{i}(\boldsymbol{k}\cdot\boldsymbol{r} - \omega t)]\,\langle\gamma|\hat{a}^\dagger\hat{a}^\dagger|\gamma\rangle\}$$

$$= (\mu_0\hbar\omega/V)\{½ + |\gamma|^2 - |\gamma|^2(|\boldsymbol{e}'|^2 - |\boldsymbol{e}''|^2)\cos[2(\boldsymbol{k}\cdot\boldsymbol{r} - \omega t + \varphi_\gamma)]\}. \tag{19.14}$$

$$\boxed{\mathrm{Var}[\boldsymbol{B}(\boldsymbol{r},t)] = \langle\gamma|\hat{\boldsymbol{B}}(\boldsymbol{r},t)\cdot\hat{\boldsymbol{B}}(\boldsymbol{r},t)|\gamma\rangle - \langle\gamma|\hat{\boldsymbol{B}}(\boldsymbol{r},t)|\gamma\rangle\cdot\langle\gamma|\hat{\boldsymbol{B}}(\boldsymbol{r},t)|\gamma\rangle = ½\mu_0\hbar\omega/V.} \tag{19.15}$$

It is seen that the average field amplitudes in Eqs.(19.10) and (19.13) behave as one would expect from a stable, monochromatic EM field, while the standard deviations deduced from Eqs.(19.12) and (19.15), being independent of the field amplitudes, are purely due to vacuum fluctuations. It must be emphasized that the shot noise in photodetection is an altogether different kind of fluctuation that is caused by the photon-counting mechanism—as opposed to the $E$- and $B$-field amplitude fluctuations.

---

**Example 6**. The Poynting vector for a single-mode coherent state $|\gamma\rangle$ is determined as follows:

$$\boldsymbol{S}(\boldsymbol{r},t) = \langle\gamma|\hat{\boldsymbol{E}}(\boldsymbol{r},t)\times\hat{\boldsymbol{H}}(\boldsymbol{r},t)|\gamma\rangle$$

$$= -[\hbar/(2\mu_0\varepsilon_0 V)]\langle\gamma|\{\hat{\boldsymbol{e}}\exp[\mathrm{i}(\boldsymbol{k}\cdot\boldsymbol{r} - \omega t)]\,\hat{a} - \hat{\boldsymbol{e}}^*\exp[-\mathrm{i}(\boldsymbol{k}\cdot\boldsymbol{r} - \omega t)]\,\hat{a}^\dagger\}$$
$$\times\{(\boldsymbol{k}\times\hat{\boldsymbol{e}})\exp[\mathrm{i}(\boldsymbol{k}\cdot\boldsymbol{r} - \omega t)]\,\hat{a} - (\boldsymbol{k}\times\hat{\boldsymbol{e}}^*)\exp[-\mathrm{i}(\boldsymbol{k}\cdot\boldsymbol{r} - \omega t)]\,\hat{a}^\dagger\}|\gamma\rangle$$

$$= ½(\hbar c^2/V)\{(\hat{\boldsymbol{e}}\cdot\hat{\boldsymbol{e}}^*)\langle\gamma|\hat{a}\hat{a}^\dagger|\gamma\rangle + (\hat{\boldsymbol{e}}^*\cdot\hat{\boldsymbol{e}})\langle\gamma|\hat{a}^\dagger\hat{a}|\gamma\rangle$$

$$\boxed{\text{set } \boldsymbol{e}'\cdot\boldsymbol{e}'' = 0} \rightarrow -(\hat{\boldsymbol{e}}\cdot\hat{\boldsymbol{e}})\exp[2\mathrm{i}(\boldsymbol{k}\cdot\boldsymbol{r} - \omega t)]\,\langle\gamma|\hat{a}\hat{a}|\gamma\rangle - (\hat{\boldsymbol{e}}^*\cdot\hat{\boldsymbol{e}}^*)\exp[-2\mathrm{i}(\boldsymbol{k}\cdot\boldsymbol{r} - \omega t)]\,\langle\gamma|\hat{a}^\dagger\hat{a}^\dagger|\gamma\rangle\}\boldsymbol{k}$$

$$= (\hbar\omega c/V)\{½ + |\gamma|^2 - (|\boldsymbol{e}'|^2 - |\boldsymbol{e}''|^2)\,|\gamma|^2\cos[2(\boldsymbol{k}\cdot\boldsymbol{r} - \omega t + \varphi_\gamma)]\}\hat{\boldsymbol{\kappa}}. \leftarrow \boxed{\boldsymbol{k} = (\omega/c)\hat{\boldsymbol{\kappa}}} \tag{19.16}$$

There exist strong similarities between the classical Poynting vector and its quantum counterpart given by Eq.(19.16). The main difference is the appearance of $½(\hbar\omega c/V)\hat{\boldsymbol{k}}$ in Eq.(19.16), which indicates that perhaps vacuum fluctuations make a net contribution to the energy flux along the direction of propagation. This is an artifact of choosing to work with a single-mode of the EM field. Needless to say, other modes of the vacuum field with arbitrary values of $\hat{\boldsymbol{k}}$ and $\omega$ always exist in the space occupied by the coherent state $|\gamma\rangle$ under consideration. The photodetection process is best described as a photon-counting process to which vacuum fluctuations and the cosine term in Eq.(19.16) make no contributions.[4]

---

**Example 7**. The $(\omega, \boldsymbol{k}, \hat{\boldsymbol{e}})$ mode of thermal radiation propagating in (unbounded) free space is a *mixed* mode that could be in number-states $|0\rangle, |1\rangle, |2\rangle, \cdots, |n\rangle, \cdots$ in accordance with Planck's probability distribution $p_n = [1 - \exp(-\hbar\omega/k_B T)] \exp(-n\hbar\omega/k_B T)$. Here $k_B$ is the Boltzmann constant and $T$ is the absolute temperature of the EM field. The average and the variance of the mode's photon number are thus given by

$$\langle n \rangle = \sum_{n=0}^{\infty} p_n \langle n|\hat{a}^\dagger \hat{a}|n\rangle = [1 - \exp(-\hbar\omega/k_B T)] \sum_{n=0}^{\infty} n \exp(-n\hbar\omega/k_B T) = \frac{1}{\exp(\hbar\omega/k_B T) - 1}. \tag{19.17}$$

$$\langle n^2 \rangle = \sum_{n=0}^{\infty} p_n \langle n|\hat{a}^\dagger \hat{a} \hat{a}^\dagger \hat{a}|n\rangle = [1 - \exp(-\hbar\omega/k_B T)] \sum_{n=0}^{\infty} n^2 \exp(-n\hbar\omega/k_B T) = \frac{\exp(\hbar\omega/k_B T) + 1}{[\exp(\hbar\omega/k_B T) - 1]^2}. \tag{19.18}$$

$$\text{Var}(n) = \langle n^2 \rangle - \langle n \rangle^2 = \frac{\exp(\hbar\omega/k_B T)}{[\exp(\hbar\omega/k_B T) - 1]^2} = \langle n \rangle + \langle n \rangle^2. \tag{19.19}$$

The variance of the photon number $n$ exceeding its average makes thermal radiation super Poissonian. As for the $E$- and $B$-fields, both fields have zero average because each number-state has zero field average. Moreover, it was shown in Example 4 that the squared $E$- and $B$-fields average to $(n + ½)\hbar\omega/(\varepsilon_0 V)$ and $(n + ½)\mu_0 \hbar\omega/V$, respectively. Thus, the expected values of the $E$- and $B$-field intensities for single-mode thermal radiation are $(\langle n \rangle + ½)\hbar\omega/(\varepsilon_0 V)$ and $(\langle n \rangle + ½)\mu_0 \hbar\omega/V$.

The Planck distribution may equivalently be written as $p_n = \zeta/(\zeta + 1)^{n+1}$, where $\zeta = \langle n \rangle^{-1} = \exp(\hbar\omega/k_B T) - 1$. (When put in this form, it is often referred to as the Bose-Einstein distribution.)

---

**20. Analysis of a beam-splitter using operator algebra**. Consider a beam-splitter having amplitude reflection and transmission coefficients $\rho$ and $\tau$, respectively, with $|\rho|^2 + |\tau|^2 = 1$ and $\phi_\rho - \phi_\tau = \pm 90°$. The splitter is assumed to be symmetric, with $\rho$ and $\tau$ representing the Fresnel coefficients for both input ports 1 and 2. Starting with the $|0\rangle_1 |0\rangle_2$ vacuum mode in the beam-splitter's input space (ports 1 and 2), we use creation operators to generate the number state $|n_1\rangle_1 |n_2\rangle_2$ in the input space, that is,

$$|n_1\rangle_1 |n_2\rangle_2 = \frac{(\hat{a}_1^\dagger)^{n_1} (\hat{a}_2^\dagger)^{n_2}}{\sqrt{n_1!\, n_2!}} |0\rangle_1 |0\rangle_2. \tag{20.1}$$

In the output space (ports 3 and 4), the annihilation and creation operators are given by $\hat{a}_3 = \rho \hat{a}_1 + \tau \hat{a}_2$, $\hat{a}_4 = \tau \hat{a}_1 + \rho \hat{a}_2$, $\hat{a}_3^\dagger = \rho^* \hat{a}_1^\dagger + \tau^* \hat{a}_2^\dagger$, and $\hat{a}_4^\dagger = \tau^* \hat{a}_1^\dagger + \rho^* \hat{a}_2^\dagger$. It is straightforward to show that $[\hat{a}_3, \hat{a}_3^\dagger] = 1$, $[\hat{a}_4, \hat{a}_4^\dagger] = 1$, $[\hat{a}_3, \hat{a}_4] = [\hat{a}_3^\dagger, \hat{a}_4^\dagger] = [\hat{a}_3, \hat{a}_4^\dagger] = [\hat{a}_3^\dagger, \hat{a}_4] = 0$. The above equations may be solved for the input operators $\hat{a}_1^\dagger$, $\hat{a}_2^\dagger$ in terms of the output operators $\hat{a}_3^\dagger$, $\hat{a}_4^\dagger$, as follows:

$$\hat{a}_1^\dagger = \rho \hat{a}_3^\dagger + \tau \hat{a}_4^\dagger, \tag{20.2}$$

$$\hat{a}_2^\dagger = \tau \hat{a}_3^\dagger + \rho \hat{a}_4^\dagger. \tag{20.3}$$

Substitution into the right-hand side of Eq.(20.1) then yields

$$\frac{(\hat{a}_1^\dagger)^{n_1} (\hat{a}_2^\dagger)^{n_2}}{\sqrt{n_1! n_2!}} |0\rangle_1 |0\rangle_2 = \frac{(\rho \hat{a}_3^\dagger + \tau \hat{a}_4^\dagger)^{n_1} (\tau \hat{a}_3^\dagger + \rho \hat{a}_4^\dagger)^{n_2}}{\sqrt{n_1! n_2!}} |0\rangle_1 |0\rangle_2$$

$$= \frac{\sum_{m_1=0}^{n_1} \binom{n_1}{m_1} \rho^{m_1} \tau^{(n_1 - m_1)} (\hat{a}_3^\dagger)^{m_1} (\hat{a}_4^\dagger)^{(n_1 - m_1)} \sum_{m_2=0}^{n_2} \binom{n_2}{m_2} \tau^{m_2} \rho^{(n_2 - m_2)} (\hat{a}_3^\dagger)^{m_2} (\hat{a}_4^\dagger)^{(n_2 - m_2)}}{\sqrt{n_1! n_2!}} |0\rangle_1 |0\rangle_2$$

$$= \frac{\sum_{m_1=0}^{n_1}\sum_{m_2=0}^{n_2}\binom{n_1}{m_1}\binom{n_2}{m_2}\rho^{(n_2+m_1-m_2)}\tau^{(n_1-m_1+m_2)}\left(\hat{a}_3^\dagger\right)^{m_1+m_2}\left(\hat{a}_4^\dagger\right)^{n_1+n_2-m_1-m_2}}{\sqrt{n_1!n_2!}}|0\rangle_1|0\rangle_2$$

$$= \sum_{m_1=0}^{n_1}\sum_{m_2=0}^{n_2}\frac{\sqrt{n_1!\,n_2!\,(m_1+m_2)!(n_1+n_2-m_1-m_2)!}}{m_1!\,m_2!(n_1-m_1)!\,(n_2-m_2)!}\,\rho^{(n_2+m_1-m_2)}\tau^{(n_1-m_1+m_2)}|m_1+m_2\rangle_3|n_1+n_2-m_1-m_2\rangle_4. \qquad (20.4)$$

The last expression contains the probability amplitudes for the various number states $|m_3\rangle_3|m_4 = n_1 + n_2 - m_3\rangle_4$ emerging at ports 3 and 4. Note that for each value of $m_3$ one must add up all the probability amplitudes associated with terms $m_1$ and $m_2$ whose sum equals $m_3$.

---

**Example 8**. Let $n_1 = n_2 = 1$. The probability amplitudes of the output states $|0\rangle_3|2\rangle_4$, $|1\rangle_3|1\rangle_4$, and $|2\rangle_3|0\rangle_4$ will then be $\sqrt{2}\rho\tau$, $\rho^2 + \tau^2$, and $\sqrt{2}\rho\tau$, respectively. Considering that $|\rho^2 + \tau^2| = |\rho|^2 - |\tau|^2$, it is clear that the three probabilities will add up to 1. Also, when $|\rho| = |\tau| = 1/\sqrt{2}$, the probability of the output being in the $|1\rangle_3|1\rangle_4$ state vanishes. The two input photons will then coalesce, appearing together either in port 3 or in port 4 with equal probabilities. This, of course, is the Hong-Ou-Mandel effect that was the subject of Sec.15.

---

**Example 9**: For a wavepacket in the number state $|n\rangle$ passing through a beam-splitter, we compute the expected value of the reflected $E$-field. Defining $c_m = \binom{n}{m}^{½}\rho^m\tau^{n-m}$, we write

$$\langle r|\hat{\boldsymbol{E}}(\boldsymbol{r},t)|r\rangle = \left(\sum c_m^*\langle m|\right)\mathrm{i}\sqrt{\hbar\omega/(2\varepsilon_0 V)}\,\{\hat{\boldsymbol{e}}\exp[\mathrm{i}(\boldsymbol{k}\cdot\boldsymbol{r}-\omega t)]\,\hat{a} - \hat{\boldsymbol{e}}^*\exp[-\mathrm{i}(\boldsymbol{k}\cdot\boldsymbol{r}-\omega t)]\,\hat{a}^\dagger\}\left(\sum c_m|m\rangle\right)$$

$$= \mathrm{i}\sqrt{\hbar\omega/(2\varepsilon_0 V)}\left(\sum_{m=0}^{n} c_m^*\langle m|\right)\Big\{\hat{\boldsymbol{e}}\exp[\mathrm{i}(\boldsymbol{k}\cdot\boldsymbol{r}-\omega t)]\left(\sum_{m=0}^{n} c_m\sqrt{m}|m-1\rangle\right)$$

$$-\hat{\boldsymbol{e}}^*\exp[-\mathrm{i}(\boldsymbol{k}\cdot\boldsymbol{r}-\omega t)]\left(\sum_{m=0}^{n} c_m\sqrt{m+1}|m+1\rangle\right)\Big\}$$

$$= \mathrm{i}\sqrt{\hbar\omega/(2\varepsilon_0 V)}\left(\sum_{m=0}^{n} c_m^*\langle m|\right)\Big\{\hat{\boldsymbol{e}}\exp[\mathrm{i}(\boldsymbol{k}\cdot\boldsymbol{r}-\omega t)]\left(\sum_{m=0}^{n-1} c_{m+1}\sqrt{m+1}|m\rangle\right)$$

$$-\hat{\boldsymbol{e}}^*\exp[-\mathrm{i}(\boldsymbol{k}\cdot\boldsymbol{r}-\omega t)]\left(\sum_{m=1}^{n+1} c_{m-1}\sqrt{m}|m\rangle\right)\Big\}$$

$$= \mathrm{i}\sqrt{\hbar\omega/(2\varepsilon_0 V)}\,\Big\{\hat{\boldsymbol{e}}\exp[\mathrm{i}(\boldsymbol{k}\cdot\boldsymbol{r}-\omega t)]\left(\sum_{m=1}^{n}\sqrt{m}\,c_{m-1}^*c_m\right)$$

$$-\hat{\boldsymbol{e}}^*\exp[-\mathrm{i}(\boldsymbol{k}\cdot\boldsymbol{r}-\omega t)]\left(\sum_{m=1}^{n}\sqrt{m}\,c_{m-1}c_m^*\right)\Big\}$$

$$= -\sqrt{2\hbar\omega/(\varepsilon_0 V)}\,\mathrm{Im}\left\{\hat{\boldsymbol{e}}\exp[\mathrm{i}(\boldsymbol{k}\cdot\boldsymbol{r}-\omega t)]\left[\sum_{m=1}^{n}\sqrt{m}\binom{n}{m-1}^{½}\binom{n}{m}^{½}(\rho^*)^{m-1}(\tau^*)^{n-m+1}\rho^m\tau^{n-m}\right]\right\}$$

$$= -\sqrt{2\hbar\omega/(\varepsilon_0 V)}\left[\sum_{m=0}^{n-1}\frac{n}{\sqrt{n-m}}\binom{n-1}{m}|\rho^2|^m\,|\tau^2|^{n-m-1}\right]\mathrm{Im}\{\hat{\boldsymbol{e}}\exp[\mathrm{i}(\boldsymbol{k}\cdot\boldsymbol{r}-\omega t)]\,\rho\tau^*\}. \qquad (20.5)$$

---

**21. Combining coherent states at a beam-splitter**. The creation operator for a single-mode coherent (i.e., quasi-classical or Glauber) state may be written as follows:

$$|\gamma\rangle = \exp(-½|\gamma|^2)\sum_{n=0}^{\infty}(\gamma^n/\sqrt{n!})|n\rangle = \exp(-½|\gamma|^2)\sum_{n=0}^{\infty}(\gamma^n/n!)(\hat{a}^\dagger)^n|0\rangle$$

$$= \exp(-½|\gamma|^2)\exp(\gamma\hat{a}^\dagger)\,|0\rangle. \qquad (21.1)$$

Suppose now that the single-mode beams entering the beam-splitter through ports 1 and 2 are described by the product state $|\gamma_1\rangle_1|\gamma_2\rangle_2$. Recalling that $[a_1^\dagger, a_2^\dagger] = 0$, we will have

$$|\gamma_1\rangle_1|\gamma_2\rangle_2 = \exp[-½(|\gamma_1|^2 + |\gamma_2|^2)]\exp(\gamma_1\hat{a}_1^\dagger + \gamma_2\hat{a}_2^\dagger)\,|0\rangle_1|0\rangle_2. \qquad (21.2)$$

Substitution for the input operators $\hat{a}_1^\dagger$, $\hat{a}_2^\dagger$ in terms of the output operators $\hat{a}_3^\dagger$, $\hat{a}_4^\dagger$ in accordance with Eqs.(20.2) and (20.3), followed by an invocation of the identity $[a_3^\dagger, a_4^\dagger] = 0$ now yields

$$\exp[-½(|\gamma_1|^2 + |\gamma_2|^2)] \exp\left[(\rho\gamma_1 + \tau\gamma_2)\hat{a}_3^\dagger + (\tau\gamma_1 + \rho\gamma_2)\hat{a}_4^\dagger\right] |0\rangle_1 |0\rangle_2$$

$$= |\rho\gamma_1 + \tau\gamma_2\rangle_3 |\tau\gamma_1 + \rho\gamma_2\rangle_4. \qquad (21.3)$$

In arriving at Eq.(21.3), we have used the fact that

$$|\rho\gamma_1 + \tau\gamma_2|^2 = |\rho\gamma_1|^2 + |\tau\gamma_2|^2 - 2|\rho\tau|\text{Im}(\gamma_1\gamma_2^*), \qquad (21.4)$$

$$|\tau\gamma_1 + \rho\gamma_2|^2 = |\tau\gamma_1|^2 + |\rho\gamma_2|^2 + 2|\rho\tau|\text{Im}(\gamma_1\gamma_2^*), \qquad (21.5)$$

$$|\rho\gamma_1 + \tau\gamma_2|^2 + |\tau\gamma_1 + \rho\gamma_2|^2 = (|\rho|^2 + |\tau|^2)(|\gamma_1|^2 + |\gamma_2|^2) = |\gamma_1|^2 + |\gamma_2|^2. \qquad (21.6)$$

The single-mode beams in ports 3 and 4 are thus seen to be coherent modes as well, with their (average) $E$-field amplitudes given by a proper combination of the $E$-field amplitudes of the beams in ports 1 and 2. This, of course, is what one expects from classical monochromatic plane-waves arriving at a beam-splitter.

**22. Thermal light passing through a beam-splitter**. When a light beam in the number state $|n\rangle$ arrives at a beam-splitter with Fresnel reflection and transmission coefficients $(\rho, \tau)$, the reflected beam emerges in a superposition state $|r\rangle = \sum_{m=0}^{n} \binom{n}{m}^{½} \rho^m \tau^{n-m} |m\rangle$; see Eq.(20.4) with $|n_2\rangle = 0$. The average number and the average of the squared number of reflected photons are readily found to be

$$\begin{aligned}\langle m\rangle = \langle r|\hat{a}^\dagger\hat{a}|r\rangle &= \textstyle\sum_{m=0}^{n} m \binom{n}{m} |\rho|^{2m}|\tau|^{2(n-m)} \\ &= n|\rho|^2 \textstyle\sum_{m=1}^{n} \binom{n-1}{m-1} |\rho|^{2(m-1)}|\tau|^{2(n-m)} \\ &= n|\rho|^2 \textstyle\sum_{m=0}^{n-1} \binom{n-1}{m} |\rho|^{2m}|\tau|^{2(n-1-m)} = n|\rho|^2(|\rho|^2 + |\tau|^2)^{n-1} = n|\rho|^2. \qquad (22.1)\end{aligned}$$

$$\begin{aligned}\langle m^2\rangle = \langle r|\hat{a}^\dagger\hat{a}\hat{a}^\dagger\hat{a}|r\rangle &= \textstyle\sum_{m=0}^{n} m^2 \binom{n}{m} |\rho|^{2m}|\tau|^{2(n-m)} \\ &= n|\rho|^2 \textstyle\sum_{m=1}^{n} m \binom{n-1}{m-1} |\rho|^{2(m-1)}|\tau|^{2(n-m)} \\ &= n|\rho|^2 + n(n-1)|\rho|^4 \textstyle\sum_{m=2}^{n} \binom{n-2}{m-2} |\rho|^{2(m-2)}|\tau|^{2(n-m)} \\ &= n|\rho|^2 + n(n-1)|\rho|^4 \textstyle\sum_{m=0}^{n-2} \binom{n-2}{m} |\rho|^{2m}|\tau|^{2(n-2-m)} \\ &= n|\rho|^2 + n(n-1)|\rho|^4(|\rho|^2 + |\tau|^2)^{n-2} = n^2|\rho|^4 + n|\rho|^2|\tau|^2. \qquad (22.2)\end{aligned}$$

If we now take the beam entering the splitter to be a single-mode $(\omega, \boldsymbol{k}, \hat{\boldsymbol{e}})$ of thermal radiation, upon invoking Eqs.(19.17) and (19.18), we find the average number and the average of the squared number of photons in the reflected beam to be $\langle n\rangle|\rho|^2$ and $\langle n^2\rangle|\rho|^4 + \langle n\rangle|\rho|^2|\tau|^2$. The variance of the number of reflected photons is, therefore, given by

$$\text{Var}(n) = (\langle n^2\rangle - \langle n\rangle^2)|\rho|^4 + \langle n\rangle|\rho|^2|\tau|^2 = (\langle n\rangle + \langle n\rangle^2)|\rho|^4 + \langle n\rangle|\rho|^2|\tau|^2 = |\rho|^2\langle n\rangle + (|\rho|^2\langle n\rangle)^2. \qquad (22.3)$$

The reflected beam thus appears to have the same photon-number distribution as the incident light, albeit scaled by the splitter's reflectance, $|\rho|^2$. Appendix C shows that this is indeed the case.

**23. The Sudarshan-Glauber P-representation**. In the complex $\gamma$-plane, each point $\gamma_0 = \gamma_0' + i\gamma_0''$ represents a coherent state $|\gamma_0\rangle$. The P-representation of this pure state uses the quasi-probability distribution $P(\gamma)=\delta(\gamma'-\gamma_0')\delta(\gamma''-\gamma_0'')$ defined over the entire $\gamma$-plane to arrive at the density-matrix $\hat{\rho}=|\gamma_0\rangle\langle\gamma_0|$ associated with the pure state $|\gamma_0\rangle$, as follows:

$$\hat{\rho}=\iint_{-\infty}^{\infty}P(\gamma)|\gamma\rangle\langle\gamma|\mathrm{d}\gamma'\mathrm{d}\gamma''. \tag{23.1}$$

As an elementary example of the type of calculations that involve the P-representation, we proceed from the standard definition of the coherent state, $|\gamma\rangle=\exp(-½|\gamma|^2)\sum_{n=0}^{\infty}(\gamma^n/\sqrt{n!})|n\rangle$, to demonstrate that $\pi^{-1}\iint_{-\infty}^{\infty}|\gamma\rangle\langle\gamma|\mathrm{d}\gamma'\mathrm{d}\gamma''$ equals the identity operator $\hat{I}$.

$$\begin{aligned}\iint_{-\infty}^{\infty}|\gamma\rangle\langle\gamma|\mathrm{d}\gamma'\mathrm{d}\gamma'' &= \sum_{n=0}^{\infty}\sum_{m=0}^{\infty}(|n\rangle\langle m|/\sqrt{n!\,m!})\iint_{-\infty}^{\infty}e^{-|\gamma|^2}\gamma^n\gamma^{*m}\,\mathrm{d}\gamma'\mathrm{d}\gamma''\\ &= \sum_{n=0}^{\infty}\sum_{m=0}^{\infty}(|n\rangle\langle m|/\sqrt{n!\,m!})\int_{|\gamma|=0}^{\infty}|\gamma|^{n+m+1}e^{-|\gamma|^2}\mathrm{d}|\gamma|\underbrace{\int_{\varphi=0}^{2\pi}e^{\mathrm{i}(n-m)\varphi}\mathrm{d}\varphi}_{2\pi\delta_{mn}}\\ &= \sum_{n=0}^{\infty}(|n\rangle\langle n|/n!)\int_0^{\infty}2\pi r^{2n+1}e^{-r^2}\mathrm{d}r\\ &= \pi\sum_{n=0}^{\infty}(|n\rangle\langle n|/n!)\underbrace{\int_0^{\infty}x^n e^{-x}\mathrm{d}x}_{n!\text{ (repeated use of integration by parts)}}\\ &= \pi\sum_{n=0}^{\infty}|n\rangle\langle n| = \pi\hat{I}.\end{aligned} \tag{23.2}$$

Next, we examine the quasi-probability distribution $P(\gamma)=(\zeta/\pi)\exp(-\zeta|\gamma|^2)$ defined over the entire $\gamma$-plane, where $\zeta$ is an arbitrary, positive, real-valued constant. Note that $P(\gamma)$ is properly normalized, so that $\iint_{-\infty}^{\infty}P(\gamma)\mathrm{d}\gamma'\mathrm{d}\gamma''=1$. The associated density operator $\hat{\rho}$ is calculated along the same lines as those followed in Eq.(23.2), yielding

$$\begin{aligned}\hat{\rho} &= \iint_{-\infty}^{\infty}(\zeta/\pi)e^{-\zeta|\gamma|^2}|\gamma\rangle\langle\gamma|\mathrm{d}\gamma'\mathrm{d}\gamma'' = (\zeta/\pi)\sum_{n=0}^{\infty}(|n\rangle\langle n|/n!)\iint_{-\infty}^{\infty}|\gamma|^{2n}e^{-(\zeta+1)|\gamma|^2}\mathrm{d}\gamma'\mathrm{d}\gamma''\\ &= (\zeta/\pi)\sum_{n=0}^{\infty}(|n\rangle\langle n|/n!)\int_0^{\infty}2\pi r^{2n+1}e^{-(\zeta+1)r^2}\mathrm{d}r\\ &= \sum_{n=0}^{\infty}(|n\rangle\langle n|/n!)[\zeta/(\zeta+1)^{n+1}]\underbrace{\int_0^{\infty}x^n e^{-x}\mathrm{d}x}_{n!} = \sum_{n=0}^{\infty}[\zeta/(\zeta+1)^{n+1}]|n\rangle\langle n|.\end{aligned} \tag{23.3}$$

Noting that $\sum_{n=0}^{\infty}\zeta/(\zeta+1)^{n+1}=1$, it is seen that the density operator $\hat{\rho}$ can be expressed as a sum over the number states, with $p(n)=\zeta/(\zeta+1)^{n+1}$ being the probability of the number state $|n\rangle$. For this probability distribution, the average number of photons is readily evaluated as follows:

$$\langle n\rangle=\sum_{n=0}^{\infty}n\zeta/(\zeta+1)^{n+1}=\zeta\sum_{n=0}^{\infty}n(\zeta+1)^{-n-1}=-\zeta\frac{\mathrm{d}}{\mathrm{d}\zeta}\sum_{n=1}^{\infty}(\zeta+1)^{-n}=-\zeta\frac{\mathrm{d}}{\mathrm{d}\zeta}\zeta^{-1}=\zeta^{-1}. \tag{23.4}$$

Thus, the $P$-function of the mixed state is $(\zeta/\pi)\exp(-\zeta|\gamma|^2)=(\pi\langle n\rangle)^{-1}\exp(-|\gamma|^2/\langle n\rangle)$, while the probability of being in the number state $|n\rangle$ is $p(n)=\zeta/(\zeta+1)^{n+1}=\langle n\rangle^n/(1+\langle n\rangle)^{n+1}$.

**24. Concluding remarks**. This paper has showcased several instances where the application of Feynman's method to problems in classical optics yields well-known results, albeit in the context of single photons and their interaction with material media including simple optical components such as beam-splitters, mirrors, diffraction gratings, multilayer stacks, and quarter-wave plates. The fundamental underlying assumption has been that the incoming photon takes all the allowed paths through an optical system and, as a result, acquires a (complex) probability amplitude in connection with each such path. When the various paths taken by the photon are physically indistinguishable, one must add up their corresponding amplitudes to arrive at the overall amplitude for an observable event (or outcome). The probability of occurrence of an event will then be the squared magnitude of its overall probability amplitude. The amplitudes being complex numbers, they each come with a magnitude and a phase, which is why their superposition exhibits the interference phenomena that are so characteristic of EM waves encountered in all our conventional optical systems.

In addition to discussing several examples involving interference, scattering, and diffraction, we demonstrated the basic principles of reciprocity, time-reversal symmetry, the Ewald-Oseen extinction theorem, and the so-called optical theorem of classical optics, all based on the postulated behavior of single-mode wavepackets in the number state $|1\rangle$, while relying on the strong parallels between such single-photon states and the plane-waves of classical electrodynamics.

The concern of the latter part of the paper was multi-photon states such as the number state $|n\rangle$, the coherent state $|\gamma\rangle$, and a mixed thermal state. We examined the splitting and also combining of these states via beam-splitters. In particular, we applied the Feynman method to the number states $|n_1\rangle$ and $|n_2\rangle$ arriving at the entrance ports 1 and 2 of a beam-splitter, and derived the probability amplitudes of the various number states emerging from its exit ports 3 and 4. The operator algebra was subsequently brought in to confirm these results and also to extend them to other situations. In Sec.19, we introduced the operators $\hat{\boldsymbol{E}}(\boldsymbol{r},t)$ for the electric field, $\hat{\boldsymbol{B}}(\boldsymbol{r},t)$ for the magnetic field, $\hat{\boldsymbol{A}}(\boldsymbol{r},t)$ for the vector potential, and $\hat{\boldsymbol{S}}(\boldsymbol{r},t)$ for the Poynting vector, and used them to relate the expected values of the corresponding quantized fields to their classical counterparts.

A topic that was skipped over in Sec.19 is the quantum treatment of the angular momentum of wavepackets containing one or more photons. The angular-momentum-density operator is $\hat{\boldsymbol{\mathcal{L}}}(\boldsymbol{r},t) = \boldsymbol{r}\times\hat{\boldsymbol{S}}(\boldsymbol{r},t)/c^2$, with $\hat{\boldsymbol{S}} = \hat{\boldsymbol{E}}\times\hat{\boldsymbol{H}}$ being the Poynting vector operator. Here, one should consider a wavepacket as a superposition of $(\omega,\boldsymbol{k},\hat{\boldsymbol{e}})$ modes in free space, all having the same $\omega$ and perhaps the same RCP or LCP polarization state, but a distribution over a range of $k$-vector orientations $\hat{\boldsymbol{\kappa}}$ and the number states supported within these modes. Integrating the expected value of the angular momentum density over the volume $V$ of the wavepacket would then yield its average angular momentum. While such a treatment would be mathematically tedious, its essential idea is not too far from the way optical angular momentum is generally handled within the classical theory.[29,30]

## Appendix A
## Details of vector diffraction calculations leading to Eq.(14.9)

In the system of Fig.15(a), denoting by $E^{(\pm)}$ the RCP and LCP amplitudes of a diffracted plane-wave along the direction of $\boldsymbol{\sigma} = \sigma_x\hat{\boldsymbol{x}} + \sigma_y\hat{\boldsymbol{y}} + \sigma_z\hat{\boldsymbol{z}} = (\cos\theta)\hat{\boldsymbol{x}} + (\sin\theta\cos\varphi)\hat{\boldsymbol{y}} + (\sin\theta\sin\varphi)\hat{\boldsymbol{z}}$, we use the geometry of Fig.6 and Eqs.(6.1)-(6.3) to write the superposition of the two waves as

$$[E^{(+)} + E^{(-)}](-\cos\theta\cos\varphi\,\hat{\boldsymbol{x}} - \cos\theta\sin\varphi\,\hat{\boldsymbol{y}} + \sin\theta\,\hat{\boldsymbol{z}}) + \mathrm{i}[E^{(+)} - E^{(-)}](\sin\varphi\,\hat{\boldsymbol{x}} - \cos\varphi\,\hat{\boldsymbol{y}}). \quad \text{(A1)}$$

Now, in the system of Fig.15(a), the $x$-axis of Fig.6 is called $y$, while the $z$-axis of Fig.6 is called $x$. Consequently, the $z$ and $x$ components of Eq.(A1) become $\tilde{E}_x$ and $\tilde{E}_y$ of Fig.15(a), namely,

$$\tilde{E}_x = [E^{(+)} + E^{(-)}]\sin\theta \quad \rightarrow \quad E^{(+)} + E^{(-)} = \tilde{E}_x/(1-\sigma_x^2)^{½}. \quad \text{(A2)}$$

$$\tilde{E}_y = -[E^{(+)} + E^{(-)}]\cos\theta\cos\varphi + \mathrm{i}[E^{(+)} - E^{(-)}]\sin\varphi. \quad \text{(A3)}$$

Substituting from Eq.(A2) into Eq.(A3), and using the identities $\sin\varphi = \sigma_z/(1-\sigma_x^2)^{½}$ and $\cos\varphi = \sigma_y/(1-\sigma_x^2)^{½}$, we find

$$E^{(+)} - E^{(-)} = -\mathrm{i}\sigma_x\sigma_y\tilde{E}_x/\sigma_z(1-\sigma_x^2)^{½} - \mathrm{i}(1-\sigma_x^2)^{½}\tilde{E}_y/\sigma_z. \quad \text{(A4)}$$

Equations (A2) and (A4) may now be solved to arrive at $E^{(+)}$ and $E^{(-)}$ of Eq.(14.9); that is,

$$E^{(\pm)} = [(\sigma_z \mp \mathrm{i}\sigma_x\sigma_y)\tilde{E}_x \mp \mathrm{i}(1-\sigma_x^2)\tilde{E}_y]/2\sigma_z(1-\sigma_x^2)^{½}. \quad \text{(A5)}$$

## Appendix B
## Orthogonality of polarization states

A single mode of the electromagnetic field in free space is specified as $(\omega, \boldsymbol{k}, \hat{\boldsymbol{e}})$, where $\omega$ is the oscillation frequency, $\boldsymbol{k} = (\omega/c)\hat{\boldsymbol{\kappa}}$ is the wave-vector along the direction of the unit-vector $\hat{\boldsymbol{\kappa}}$, and $\hat{\boldsymbol{e}} = \boldsymbol{e}' + \mathrm{i}\boldsymbol{e}''$ is a complex unit-vector describing an arbitrary (generally elliptical) polarization state. In general, $|\hat{\boldsymbol{e}}|^2 = \hat{\boldsymbol{e}} \cdot \hat{\boldsymbol{e}}^* = \boldsymbol{e}' \cdot \boldsymbol{e}' + \boldsymbol{e}'' \cdot \boldsymbol{e}'' = 1$, and both vectors $\boldsymbol{e}'$ and $\boldsymbol{e}''$ are perpendicular to $\boldsymbol{k}$, so that $\hat{\boldsymbol{\kappa}} \cdot \hat{\boldsymbol{e}} = 0$. While it is convenient to assume that $\boldsymbol{e}' \cdot \boldsymbol{e}'' = 0$, such a restriction is by no means necessary; that is, the real and imaginary components of $\hat{\boldsymbol{e}}$ are not required to be perpendicular to each other. Each $(\omega, \boldsymbol{k})$ can be associated with two independent (or "mutually orthogonal") states of polarization $\hat{\boldsymbol{e}}_1 = \boldsymbol{e}_1' + \mathrm{i}\boldsymbol{e}_1''$ and $\hat{\boldsymbol{e}}_2 = \boldsymbol{e}_2' + \mathrm{i}\boldsymbol{e}_2''$, where

$$|\hat{\boldsymbol{e}}_1|^2 = \hat{\boldsymbol{e}}_1 \cdot \hat{\boldsymbol{e}}_1^* = \boldsymbol{e}_1' \cdot \boldsymbol{e}_1' + \boldsymbol{e}_1'' \cdot \boldsymbol{e}_1'' = 1, \tag{B1}$$

$$|\hat{\boldsymbol{e}}_2|^2 = \hat{\boldsymbol{e}}_2 \cdot \hat{\boldsymbol{e}}_2^* = \boldsymbol{e}_2' \cdot \boldsymbol{e}_2' + \boldsymbol{e}_2'' \cdot \boldsymbol{e}_2'' = 1. \tag{B2}$$

For the polarization states $\hat{\boldsymbol{e}}_1$ and $\hat{\boldsymbol{e}}_2$ of a given $(\omega, \boldsymbol{k})$ mode to be mutually orthogonal, it is necessary as well as sufficient to have $\hat{\boldsymbol{e}}_1 \cdot \hat{\boldsymbol{e}}_2^* = 0$. The rationale for this constraint is rooted in the fact that, in classical electrodynamics, the electric-field and the magnetic-field energy densities (and also the Poynting vector) are proportional to $\hat{\boldsymbol{e}} \cdot \hat{\boldsymbol{e}}^*$. Now, for any linear superposition of the two modes $(\omega, \boldsymbol{k}, \hat{\boldsymbol{e}}_1)$ and $(\omega, \boldsymbol{k}, \hat{\boldsymbol{e}}_2)$ with arbitrary complex coefficients $c_1$ and $c_2$, we will have

$$\hat{\boldsymbol{e}} \cdot \hat{\boldsymbol{e}}^* = (c_1\hat{\boldsymbol{e}}_1 + c_2\hat{\boldsymbol{e}}_2) \cdot (c_1\hat{\boldsymbol{e}}_1 + c_2\hat{\boldsymbol{e}}_2)^* = |c_1|^2|\hat{\boldsymbol{e}}_1|^2 + |c_2|^2|\hat{\boldsymbol{e}}_2|^2 + 2Re(c_1c_2^*\hat{\boldsymbol{e}}_1 \cdot \hat{\boldsymbol{e}}_2^*). \tag{B3}$$

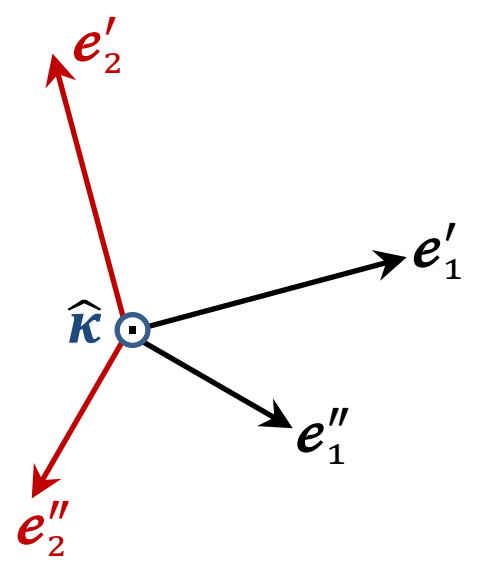

It is seen that, for all $c_1$ and $c_2$, the energy content of the superposed mode will be free of cross terms provided that $\hat{\boldsymbol{e}}_1 \cdot \hat{\boldsymbol{e}}_2^* = 0$. This is tantamount to requiring that $\boldsymbol{e}_1' \cdot \boldsymbol{e}_2' + \boldsymbol{e}_1'' \cdot \boldsymbol{e}_2'' = 0$ and also $\boldsymbol{e}_1' \cdot \boldsymbol{e}_2'' - \boldsymbol{e}_1'' \cdot \boldsymbol{e}_2' = 0$. Suppose now that, for a given mode $(\omega, \boldsymbol{k}, \hat{\boldsymbol{e}}_1)$, it is desired to identify the corresponding orthogonal polarization vector $\hat{\boldsymbol{e}}_2$. One can take $\boldsymbol{e}_1'$ and rotate it around $\hat{\boldsymbol{\kappa}}$ by 90°, say, counterclockwise, to arrive at $\boldsymbol{e}_2'$. One then takes $\boldsymbol{e}_1''$ and rotates it around $\hat{\boldsymbol{\kappa}}$ by 90°, this time clockwise, to arrive at $\boldsymbol{e}_2''$. This construction guarantees that $\hat{\boldsymbol{e}}_1 \cdot \hat{\boldsymbol{e}}_2^* = 0$ and that, therefore, the two modes $(\omega, \boldsymbol{k}, \hat{\boldsymbol{e}}_1)$ and $(\omega, \boldsymbol{k}, \hat{\boldsymbol{e}}_2)$ are mutually orthogonal.

The problem of identifying $\hat{\boldsymbol{e}}_2$ for a given $\hat{\boldsymbol{e}}_1$ is generally solved by writing $\hat{\boldsymbol{e}}_1 = e_{1x}\hat{\boldsymbol{x}} + e_{1y}\hat{\boldsymbol{y}}$, $\hat{\boldsymbol{e}}_2 = e_{2x}\hat{\boldsymbol{x}} + e_{2y}\hat{\boldsymbol{y}}$, then setting $\hat{\boldsymbol{e}}_1 \cdot \hat{\boldsymbol{e}}_2^* = e_{1x}e_{2x}^* + e_{1y}e_{2y}^* = 0$ and $\hat{\boldsymbol{e}}_2 \cdot \hat{\boldsymbol{e}}_2^* = |e_{2x}|^2 + |e_{2y}|^2 = 1$. The first equation gives $e_{2y}^*/e_{2x}^* = -e_{1x}/e_{1y}$ and, therefore, $\varphi_{2x} - \varphi_{2y} = \varphi_{1x} - \varphi_{1y} + \pi$. We also get $|e_{2y}|/|e_{2x}| = |e_{1x}|/|e_{1y}|$, which, upon substitution into the second equation, yields $|e_{2x}| = |e_{1y}|$ and $|e_{2y}| = |e_{1x}|$. Thus, aside from an arbitrary phase, say, $\varphi_{2x}$, the problem has a unique solution.

## Appendix C
## Thermal light passing through a beam-splitter

When a single mode $(\omega, \boldsymbol{k}, \hat{\boldsymbol{e}})$ of thermal radiation arrives at a beam-splitter whose Fresnel reflection and transmission coefficients are $\rho$ and $\tau$, respectively, the probability that none of the photons are reflected at the splitter will be

$$p(0) = \sum_{n=0}^{\infty}\left(1 - e^{-\hbar\omega/k_BT}\right)e^{-n\hbar\omega/k_BT}|\tau|^{2n} = \left(1 - e^{-\hbar\omega/k_BT}\right)/\left(1 - |\tau|^2e^{-\hbar\omega/k_BT}\right). \tag{C1}$$

Similarly, the probability $p(m)$ that $m \geq 1$ photons are reflected at the splitter is found to be

$$p(m) = \sum_{n=m}^{\infty}(1 - e^{-\hbar\omega/k_BT})e^{-n\hbar\omega/k_BT}\binom{n}{m}|\rho|^{2m}|\tau|^{2(n-m)}$$

$$= (1 - e^{-\hbar\omega/k_BT})(|\rho/\tau|^{2m}/m!)\sum_{n=m}^{\infty} n(n-1)\cdots(n-m+1)(|\tau|^2 e^{-\hbar\omega/k_BT})^n. \quad \text{(C2)}$$

Upon defining $x = |\tau|^2 e^{-\hbar\omega/k_BT}$, the sum in the above equation is computed as follows:

$$\sum_{n=m}^{\infty} n(n-1)\cdots(n-m+1)x^n = x^m \frac{\mathrm{d}^m}{\mathrm{d}x^m}\sum_{n=0}^{\infty} x^n = x^m \frac{\mathrm{d}^m}{\mathrm{d}x^m}(1-x)^{-1} = m!\,x^m/(1-x)^{m+1}. \quad \text{(C3)}$$

Substitution into Eq.(C2) now yields

$$p(m) = (1 - e^{-\hbar\omega/k_BT})\left(|\rho|^2 e^{-\hbar\omega/k_BT}\right)^m / \left(1 - |\tau|^2 e^{-\hbar\omega/k_BT}\right)^{m+1}. \quad \text{(C4)}$$

The expression for $p(m)$ in Eq.(C4) is seen to reduce to $p(0)$ of Eq.(C1) when $m$ is set to zero. Thus, Eq.(C4) is the general expression for the probability of $m \ge 0$ photons being reflected at the beam-splitter. Recalling that $|\rho|^2 + |\tau|^2 = 1$, it is now trivial to verify that $\sum_{m=0}^{\infty} p(m) = 1$. One may also write Eq.(C4) as $p(m) = \zeta/(\zeta+1)^{m+1}$, with $\zeta = (e^{\hbar\omega/k_BT} - 1)/|\rho|^2$. This alternative expression of the photon-number probability distribution clearly shows that the beam reflected at the splitter (reflectance $= |\rho|^2$) retains the characteristic Bose-Einstein distribution of the incident single mode thermal radiation, albeit one whose average number of photons is now given by $\langle n \rangle = \zeta^{-1} = |\rho|^2/(e^{\hbar\omega/k_BT} - 1)$. Using a similar argument, one can show that any subsequent passage of the reflected light through beam-splitters will preserve its Bose-Einstein character.

**Acknowledgement**. The author is grateful to Ewan Wright for numerous helpful discussions.